%% file: paper.tex
\documentclass[prd,onecolumn,superscriptaddress, nofootinbib, preprintnumbers]{revtex4}
\usepackage{amsmath, graphicx, hyperref, float, bm, epsfig}
\pdfoutput=1 

\usepackage{color}
\usepackage[dvipsnames]{xcolor}
\usepackage{multirow}
\usepackage{cleveref} 

\newcommand{\be}{\begin{equation}}
\newcommand{\ee}{\end{equation}}
\newcommand{\dd}{\text{d}}
\newcommand{\dec}{\text{dec}}
\newcommand{\MeV}{\text{MeV}}

\newcommand{\zdec}{z_\text{dec}}
\newcommand{\dzdec}{\Delta z_\text{dec}}
\newcommand{\Neff}{N_\text{eff}}
\newcommand{\Neffint}{N_\text{eff,int}}
\newcommand{\Nefffs}{N_\text{eff,fs}}
\newcommand{\meff}{\sum m}

\begin{document}

\preprint{YITP-SB-2020-41}

\title{Self-interacting neutrinos, the Hubble parameter tension, and the Cosmic Microwave Background}

\author{Thejs Brinckmann\footnote{E-mail: thejs.brinckmann@gmail.com}}
\affiliation{C.~N.~Yang Institute for Theoretical Physics and Department of Physics \& Astronomy, \\ Stony Brook University, Stony Brook, NY 11794, USA}
\author{Jae Hyeok Chang}
\affiliation{Maryland Center for Fundamental Physics, University of Maryland, College Park, MD 20742, USA}
\affiliation{Department of Physics and Astronomy, Johns Hopkins University, Baltimore, MD 21218, USA}
\author{Marilena LoVerde}
\affiliation{C.~N.~Yang Institute for Theoretical Physics and Department of Physics \& Astronomy, \\ Stony Brook University, Stony Brook, NY 11794, USA}

\begin{abstract}
We perform a comprehensive study of cosmological constraints on non-standard neutrino self-interactions using cosmic microwave background (CMB) and baryon acoustic oscillation data. We consider different scenarios for neutrino self-interactions distinguished by the fraction of neutrino states allowed to participate in self-interactions and how the relativistic energy density, $\Neff$, is allowed to vary. Specifically, we study cases in which: all neutrino states self-interact and $\Neff$ varies; two species free-stream, which we show alleviates tension with laboratory constraints, while the energy in the additional interacting states varies; and a variable fraction of neutrinos self-interact with either the total $\Neff$ fixed to the Standard Model value or allowed to vary. In no case do we find compelling evidence for new neutrino interactions or non-standard values of $\Neff$. In several cases we find additional modes with neutrino decoupling occurring at lower redshifts $\zdec \sim 10^{3-4}$. We do a careful analysis to examine whether new neutrino self-interactions solve or alleviate the so-called $H_0$ tension and find that, when all Planck 2018 CMB temperature and polarization data is included, none of these examples ease the tension more than allowing a variable $\Neff$ comprised of free-streaming particles. Although we focus on neutrino interactions, these constraints are applicable to any light relic particle.
\end{abstract}

\maketitle

\input{introduction.tex}
\input{SInu.tex}
\input{parameterization.tex}
\input{method.tex}
\input{results.tex}
\input{conclusions.tex}

\section*{Acknowledgements}
We thank Peizhi Du and Christina Kreisch for useful discussions. We also thank Daniel Green and the referee for very useful feedback on the first version of the paper. We especially thank Kevin J. Kelly and Sam McDermott for providing the data in \cite{Blinov:2019gcj}. TB and ML are supported by DOE Grant DE-SC0017848. JHC is supported by NSF Grant PHY-1914731, the Maryland Center for Fundamental Physics (MCFP), and the Johns Hopkins University Joint Postdoc Fund. JHC was also supported in part by DoE Grant DE-SC0017938. Results in this paper were obtained using the high-performance computing system at the Institute for Advanced Computational Science at Stony Brook University.

\input{appendix.tex}

\clearpage

\bibliography{references}

\end{document}

%% file: introduction.tex
\section{Introduction}
Neutrinos are among the least understood particles in the Standard Model. The origin of neutrino mass is unknown, as is their Dirac or Majorana nature. Moreover, a range of anomalies persist in laboratory neutrino experiments (for a review, see, e.g. \cite{Abazajian:2012ys, Zyla:2020zbs}). Cosmological datasets, which are sensitive to the gravitational effects of neutrinos throughout cosmic history, offer complementary information about neutrinos and may therefore shed light on these neutrino puzzles. In this paper we will generalize the phenomenological description of neutrinos as pertains to cosmological datasets to determine constraints on a variety of non-standard neutrino scenarios. 

In the standard cosmology, neutrinos were in thermal equilibrium with the rest of the Standard Model particles at temperatures $\gg 2$MeV. As the Universe expanded and cooled, neutrinos ceased to scatter frequently, a process referred to as {\em neutrino decoupling}. Neutrinos contribute a substantial fraction to the energy budget of the early universe, comprising roughly $40\%$ of the radiation density at epochs probed by the cosmic microwave background (CMB). CMB anisotropies are sensitive to both the total energy in neutrinos, through their contribution to the energy density and therefore the expansion rate, as well is inhomogeneities in the neutrino energy density (for a review, see, e.g. \cite{Abazajian:2016yjj}). 

The cosmological epochs probed by the CMB anisotropies are well after neutrino decoupling. That is, from the perspective of CMB data, standard neutrinos are {\em free-streaming} particles. While the total energy in neutrinos is unaffected by the decoupling transition, the behavior of neutrino perturbations changes qualitatively. If neutrinos scatter frequently, neutrino perturbations behave as a relativistic fluid and will participate in acoustic oscillations along with photons and baryons. After neutrino decoupling, neutrinos free-stream to cosmological distances, sourcing large anisotropic stress, which in turn modifies the behavior of the photon-baryon fluid \cite{Bashinsky:2003tk}. For standard neutrinos, neutrinos are free-streaming for the entire epoch probed by the CMB and the decoupling transition leaves no impact. On the other hand, if neutrinos have additional self-interactions, neutrino-neutrino scattering can persist until late enough times to have an observable impact on CMB data. In this paper, we will study CMB constraints on the decoupling of neutrino self-interactions. From now on, we refer to neutrino decoupling from the photon bath as  {\em standard neutrino decoupling} to distinguish from the decoupling of neutrino self-interactions. 

Before proceeding let us review related literature. The assumption of free-streaming neutrinos at CMB times has been relaxed to study a variety of specific non-standard neutrino scenarios (see, e.g. \cite{BialynickaBirula:1964zz, Raffelt:1987ah, Berkov:1987pz, Berkov:1988sd, Belotsky:2001fb, Chacko:2003dt, Hannestad:2004qu, Hannestad:2005ex, Bell:2005dr, Friedland:2007vv, Basboll:2008fx, Archidiacono:2013dua, Archidiacono:2014nda, Forastieri:2017oma, Oldengott:2017fhy,DiValentino:2017oaw,Song:2018zyl,Barenboim:2019tux,Esteban:2021ozz,Du:2021idh}). Other works have modified the behavior of neutrino perturbations by introducing a viscosity parameter to quantify the anisotropic stress and put constraints on that parameter with CMB data (e.g. \cite{Trotta:2004ty, Sawyer:2006ju, Smith:2011es, Gerbino:2013ova, Audren:2014lsa, Ade:2015xua}). A general framework for studying the impact of neutrino self-interactions and their decoupling on CMB data, along with constraints, were presented in \cite{Cyr-Racine:2013jua} and the subsequent work \cite{Lancaster:2017ksf}. Recently, neutrino self-interactions have been proposed as a solution to the Hubble tension \cite{Kreisch:2019yzn}, which has since been studied by \cite{Ghosh:2019tab,Escudero:2019gvw,He:2020zns,Berbig:2020wve,Mazumdar:2020ibx,Das:2020xke, Choudhury:2020tka}. Related work studies self-interacting dark-radiation, which will have similar consequences on CMB observables (e.g. \cite{Jeong:2013eza, Baumann:2015rya, Choi:2018gho,Blinov:2020hmc,Choi:2020pyy}). 

In this paper, we go beyond the previous works in several ways. First, we use the latest Planck 2018 data for our constraints. Second, we consider several different implementations of new neutrino interactions. In addition to studying interactions among all species of neutrinos, with a free total number of neutrino states, as in \cite{Cyr-Racine:2013jua,Lancaster:2017ksf, Kreisch:2019yzn} (our {\em Case 1}), we consider a scenario with two free-streaming neutrinos states and free number of self-interacting neutrino states ({\em Case 2}), which we will see alleviates some of the tension with current experimental constraints on new neutrino interactions \cite{Ng:2014pca, Blinov:2019gcj}. For {\em Case 3}, we fix the early universe energy density of neutrinos to the Standard Model value ($\Neff = 3.046$) and produce constraints on the fraction of those neutrinos that can have self-interactions. Finally, for {\em Case 4}, we allow both the total energy in relativistic neutrinos to vary, and the self-interacting fraction. We remind the reader that while we use the term ``neutrino" to describe the particles we are constraining, the physical effects of these particles on CMB and BAO data are purely gravitational and therefore the constraints on the energy density (parameterized by $\Neff$) and decoupling epoch described in this paper apply to any light relic particle (for a review see, e.g. \cite{Green:2019glg}). 

A second motivation for our work is the existence of tensions between different cosmological datasets. In recent years,  increasingly precise measurements of the Hubble parameter have led to a statistically significant tension between direct measurements of the Hubble expansion rate using supernovae calibrated with the distance ladder (e.g.~\cite{Riess:2011yx,Riess:2016jrr,Riess:2019cxk}), which find a high value of the Hubble parameter, and a host of alternative methods that do not make use of the distance ladder (see e.g.~\cite{Verde:2019ivm} for a review). These include: supernovae (i.e. the Pantheon~\cite{Scolnic:2017caz} or Dark Energy Survey (DES) samples~\cite{Macaulay:2018fxi}) calibrated by alternative means, e.g. using the Baryon Acoustic Oscillation (BAO) scale (e.g. measured by DES~\cite{Macaulay:2018fxi}); the so-called inverse distance ladder approach, as well as inferences of the Hubble parameter using early-time probes, such as the CMB (e.g. from Planck~\cite{Ade:2015xua,Aghanim:2018eyx} or ACT~\cite{Aiola:2020azj}); or, independently from the CMB, from the BAO scale in combination with measurements of the abundance of primordial elements from Big Bang Nucleosynthesis (BBN), either the two alone (see e.g.~\cite{Blomqvist:2019rah,Cuceu:2019for,Schoneberg:2019wmt}), or with galaxy clustering and weak lensing measurements (e.g. including DES~\cite{Abbott:2017smn}).

Similarly, recent measurements of the amplitude of matter fluctuations (quantified in this case by $\sigma_8$, the root mean square amplitude of fluctuations within 8 Mpc spheres, or by $S_8 = \sigma_8 (\Omega_m/0.3)^{0.5}$, a parameter that folds in the matter density in the universe, $\Omega_m$) in the late universe has seen a notable discrepancy between early and late time measurements. Specifically, a discrepancy exists between values inferred from Planck CMB data and late time cosmic shear measurements from e.g. KiDS+VIKING-450 alone~\cite{Hildebrandt:2018yau} and with DES~\cite{Joudaki:2019pmv,Asgari:2019fkq}, as well as the new cosmic shear and galaxy clustering results from KIDS-1000~\cite{2020arXiv200715632H,Asgari:2020wuj}, although note some analyses find a larger value for $S_8$ that is closer to Planck, e.g. KiDS-450+GAMA~\cite{vanUitert:2017ieu}, HSC SSP~\cite{Hamana:2019etx}, and DES-Y3~\cite{Abbott:2021bzy}. This tension is not as severe as that of the Hubble tension, but it behooves us to keep it in mind when searching for solutions to latter, as many natural solutions to the Hubble tension (e.g. simply increasing the amount of free-streaming relativistic species in the early universe) will worsen the aforementioned tension thereby making those models not viable candidates for alleviating the Hubble tension.

These tensions between cosmological datasets have led to a number of models being proposed to resolve or alleviate one or both of them. Examples of these include the introduction of extra dark radiation coupled to dark matter~\cite{Buen-Abad:2015ova,Lesgourgues:2015wza,Buen-Abad:2017gxg,Krall:2017xcw,Archidiacono:2019wdp,Becker:2020hzj}, which has the potential to alleviate both tensions, a phase of so-called Early Dark Energy~\cite{Poulin:2018cxd}, which can largely solve the Hubble tension (but may worsen the discrepancy related to the amplitude of matter fluctuations~\cite{Hill:2020osr,Ivanov:2020ril,DAmico:2020ods}, although attempts are being made to develop similar models that avoid this problem, see e.g.~\cite{Niedermann:2019olb,Niedermann:2020dwg,Niedermann:2020qbw}), and sterile neutrino secret interactions~\cite{Archidiacono:2014nda,Archidiacono:2015oma,Archidiacono:2016kkh,Archidiacono:2020yey}. Notably, the introduction of self-interactions between active neutrinos and extra relativistic species has been posited as a way to alleviate both tensions~\cite{Kreisch:2019yzn}, which is a topic we attempt to address in this work.

Finally, before proceeding we will mention two closely related papers that were posted while this manuscript was in preparation. Reference \cite{Das:2020xke} uses CMB, BAO, and $f\sigma_8$ data to constrain self-interacting neutrinos, primarily in cases where a fixed number of neutrino states self-interact (one, two, or three neutrino states) for a fixed $\Neff = 3.046$ and assuming massless neutrinos. The analysis in reference \cite{Choudhury:2020tka} considers all species of neutrinos to be self-interacting and varying both $\Neff$ and $\meff$, matching our Case 1. The particular dataset combinations in \cite{Choudhury:2020tka} differ somewhat from our choices (for instance, we always include CMB lensing in our analyses). Despite this, our results for Case 1 are in qualitative agreement.

This paper is organized as follows. In Sec.~\ref{sec:SInu}, we review self-interacting neutrinos using the example of an interaction mediated by a Majoran, and present relationships between the Majoran-neutrino coupling, an effective neutrino self-interaction parameters, and the associated redshift at which neutrino self-interactions will decouple. We also review experimental constrains on neutrino interactions. In Sec. ~\ref{sec:parameterization}, we describe our phenomenological parameterization of self-interacting neutrinos, which builds on \cite{Choi:2018gho}, how this is implemented in the CLASS code \cite{Lesgourgues:2011re}, and illustrate the changes to CMB temperature, polarization, and lensing power spectra induced by a self-interacting neutrino component. In Sec. \ref{sec:method}, we present our analysis method and choice of datasets. The results of our analyses are presented in Sec.~\ref{sec:results}. Conclusions are presented in Sec.~\ref{sec:discussion}. Details of the computation of neutrino opacity functions discussed in Sec.~\ref{sec:SInu} are given in Appendix \ref{app:opacity}. A study of the sensitivity of our analyses to the assumed duration of neutrino decoupling is in Appendix~\ref{app:decoupling_width}. And, complete parameter constraints plots for all scenarios are given in Appendix~\ref{app:full_plots}. 

%% file: SInu.tex
\section{Self-interacting neutrinos}
\label{sec:SInu}

In this section we review self-interacting neutrinos using the example of an interaction mediated via a massive scalar $\phi$, called the Majoron, the Goldstone boson associated with spontaneous breaking of neutrino flavor symmetry \cite{Chikashige:1980ui, Chikashige:1980qk}. Our goal here is to connect the neutrino interaction parameters with the redshift and duration of neutrino decoupling, which impact the CMB power spectra.  The Lagrangian for the Majoron reads,
\begin{equation}
\mathcal{L} = \frac{1}2 \partial_\mu \phi \partial^\mu \phi - \frac{1}2 m_\phi^2  \phi^2 + \frac{1}2 g_{ij} \bar{\nu}_i \nu_j \phi,
\end{equation}
where $m_\phi$ is the Majoron mass, $\nu_i$ is a mass eigenstate of neutrinos, and $g_{ij}$ are the coupling constants between neutrinos and the Majoron. We assume interactions among mass eigenstates so that when we study interactions among only a fraction of the mass eigenstates the number of interacting neutrinos is not changed by neutrino oscillations. We assume neutrinos are Majorana fermions because the Dirac neutrino case is widely constrained from $\Delta N_\text{eff}$ during Big Bang-Nucleosynthesis (BBN) \cite{Blinov:2019gcj}. For simplicity, we choose a diagonal and universal coupling, $g_{ij} \equiv g_\phi \delta_{ij}$. Neutrinos can interact with each other by exchanging the Majoron, and the effective Lagrangian for neutrino self-interaction with a heavy enough $\phi$ can be written as \cite{Lyu:2020lps},
\begin{equation}
\mathcal{L} = \frac{1}{8} G_{\nu} \bar{\nu}_i \nu_i \bar{\nu}_j \nu_j,
\end{equation}
where $G_{\nu} \equiv g_\phi^2/m_\phi^2$. If $G_\nu \gg G_F$, neutrinos continue to interact with each other even after they decouple from the Standard Model thermal bath at $T \sim 2$ MeV. 

\subsection{Decoupling of neutrino self-interactions}
The neutrino self-interaction rate drops as the Universe expands and the number density of neutrino decreases, eventually ceasing entirely. We define the decoupling redshift $z_\text{dec}$, as the redshift when neutrinos decouple from self-interactions (more concretely, when the neutrino opacity drops to $1/2$). The decoupling occurs roughly at $\Gamma_\nu \sim H$, where the $\Gamma_\nu$ is the neutrino self-interaction rate and $H$ is the Hubble parameter. From dimensional analysis, one finds $\Gamma_\nu \sim G_\nu^2 T^5$ and $H \sim T^2/m_{pl}$, which gives $z_\text{dec} \sim (G_\nu^2 T_0^3 m_{pl})^{-1/3}$, where $T_0$ is the CMB temperature today, and $m_{pl}$ is the Planck mass. As we shall see, this estimate gives a correct $z_\text{dec}$ up to an $\mathcal{O}(1)$ factor.

We will now study $z_\text{dec}$ in more detail. The exact form of $\Gamma_\nu$ for a neutrino $\nu_i$ with energy $E_1$ for the process $\nu_i(p_1) + \nu_j(p_2) \rightarrow \nu_k (p_3) +\nu_l(p_4)$ is
\begin{equation}
\Gamma_{\nu}(E_1) = \frac{1}{2E_1} \int d\Pi_2 d\Pi_3 d\Pi_4 f_\nu(E_2) (1-f_\nu(E_3))(1-f_\nu(E_4)) |\mathcal{M}|^2(2\pi)^4\delta^{(4)}(p_1+p_2-p_3-p_4)\,,
\end{equation}
where $d\Pi_i = \frac{g_\nu d^3 p_i}{(2\pi)^3 2 E_i}$ with the spin degeneracy $g_\nu=2$, $f_\nu(E) = \frac{1}{\exp(E/T_\nu)+1}$ is the Fermi-Dirac distribution, and $|\mathcal{M}|^2$ is the spin-averaged matrix element of the process, which is
\begin{eqnarray}
\label{eq:Mtot}
|\mathcal{M}_{\nu \nu \rightarrow \nu \nu}|^2 &=& \frac{1}{2}|\mathcal{M}_{\nu_i \nu_i \rightarrow \nu_i \nu_i}|^2 + \frac{1}{2} \times 2|\mathcal{M}_{\nu_i \nu_i \rightarrow \nu_j \nu_j}|^2\label{totalamp} + 2 |\mathcal{M}_{\nu_i \nu_j \rightarrow \nu_i \nu_j}|^2\,,
\end{eqnarray}
where,
\begin{eqnarray}
|\mathcal{M}_{\nu_i \nu_i \rightarrow \nu_i \nu_i}|^2 &=& G_{\nu}^2 (s^2 +s t +t^2)\,, \\
|\mathcal{M}_{\nu_i \nu_i \rightarrow \nu_j \nu_j}|^2 &=& G_{\nu}^2 s^2 \qquad (i \neq j)\,,\\
|\mathcal{M}_{\nu_i \nu_j \rightarrow \nu_i \nu_j}|^2 &=& G_{\nu}^2 t^2 \qquad (i \neq j)\,.
\end{eqnarray}

Here, $s=(p_1+p_2)^2$ and $t=(p_3-p_1)^2$ are Mandelstam variables, and the factor of $\frac{1}2$ in the first and the second terms in Eq.~\ref{eq:Mtot} accounts for the symmetric factor for identical outgoing particles. Note, since we are discussing neutrino scattering at $T \gg m_\nu$, we ignore neutrino masses throughout this section. If we ignore the Pauli-blocking factors $(1-f_\nu)$, $\Gamma_\nu$ reduces to,
\begin{eqnarray}
\Gamma_{\nu}(E_1) &=& \int \frac{d^3p_2}{(2\pi)^3} g_\nu f_\nu(E_2) \frac{s}{2 E_1 E_2}\sigma_{\nu \nu \rightarrow \nu \nu}\\
&=& \frac{35\pi}{1728} G_\nu^2 E_1 T_\nu^4
\end{eqnarray}
where $\sigma_{\nu \nu \rightarrow \nu \nu} = \int d\cos\theta_\textrm{CM} \frac{|\mathcal{M}_{\nu \nu \rightarrow \nu \nu}|^2}{32 \pi s}$ is the neutrino self-interaction cross section, and $\theta_\text{CM}$ is the angle between incoming and outgoing particles in the center of momentum frame. Neglecting the Pauli blocking factors gives $\mathcal{O}(10\%)$ errors on the rate, but we have checked this does not change the results of our computations of $z_\text{dec}$ significantly. This is discussed further in Appendix~\ref{app:opacity}.

Now we define the neutrino opacity function for scattering rate $\Gamma_\nu(E_1)$ as,
\begin{eqnarray}
O(T,E_1) &=& 1- \exp \left[- \int_{t(T)}^{t_0} \Gamma_\nu \dd t \right ] \,,\label{opacityt}\\
&=& 1- \exp \left[- \int_{T_0}^{T} \frac{\Gamma_\nu}{H(T') T'} \dd T' \right ]\,.\label{opacityT}
\end{eqnarray}
In this expression $T$ is the temperature of the photon bath and $H(T)$ is the Hubble parameter at temperature $T$\footnote{We assume the scale factor is inversely proportional to the temperature to get Eq.~\ref{opacityT} from Eq.~\ref{opacityt}, so Eq.~\ref{opacityT} is only exact for temperatures after electron-positron annihilations.}. We take the neutrino temperature to be $T_\nu = \left( \frac{4}{11} \right)^{1/3} T$. The opacity function averaged over neutrino energies is,
\begin{equation}
\langle O(T) \rangle  = \frac{1}{n_\nu} \int \frac{d^3 p_1}{(2\pi)^3} g_\nu f_\nu(E_1) \, O(T,E_1)\,,
\end{equation}
where $n_\nu = \frac{3 \zeta(3)}{2\pi^2}T_\nu^3$ is the number density of each neutrino species. 

\begin{figure}[tb]
	\centering
	\includegraphics[width=8cm]{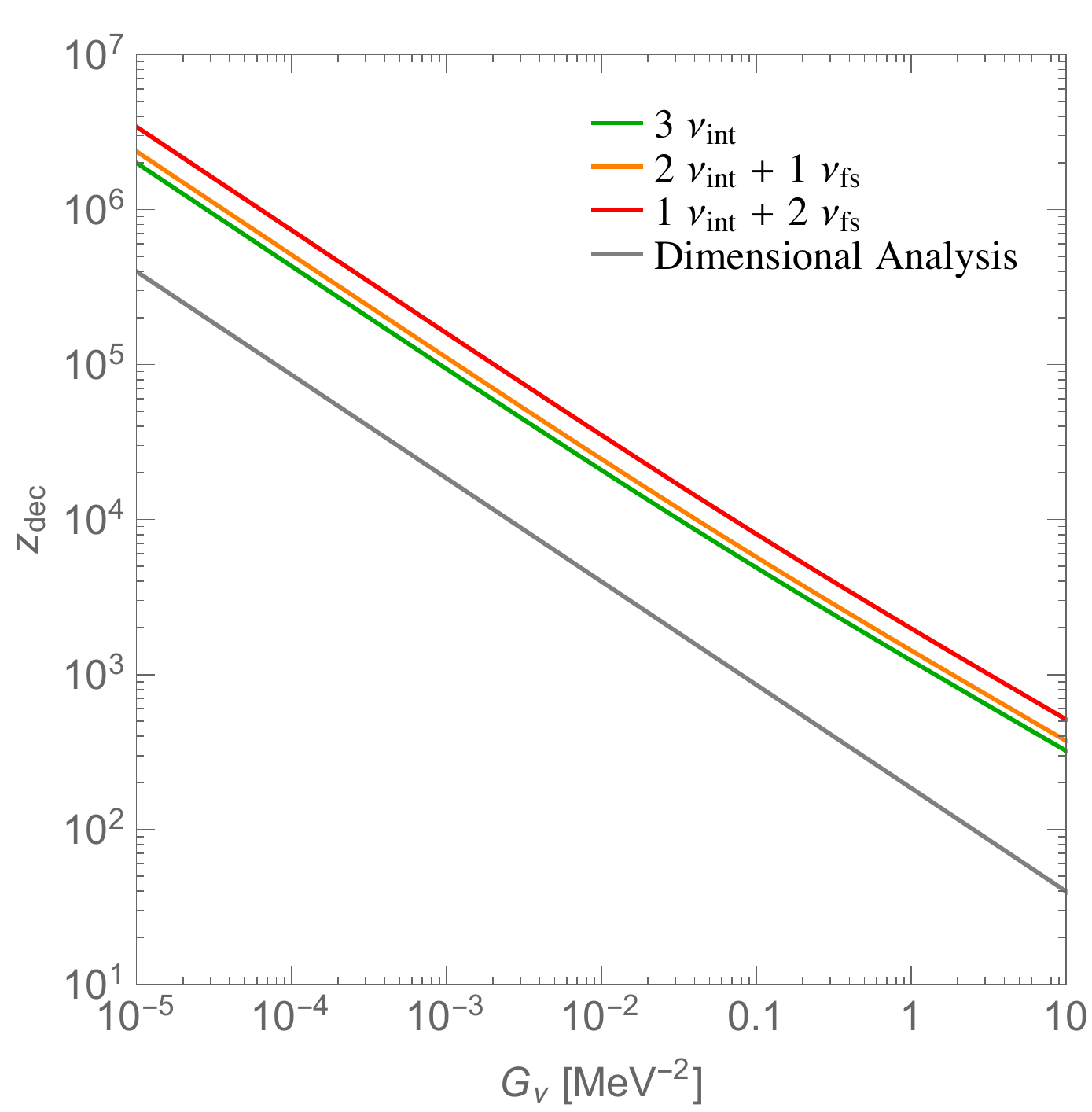}
	\caption{The relationship between the redshift of neutrino decoupling, $z_\text{dec}$, and the effective neutrino self-interaction strength $G_\nu$. Plotted is the relationship for different numbers of interacting neutrino species. The red, orange, and green lines correspond to 1, 2, and 3 self-interacting neutrino species, respectively. The gray line is result from dimensional analysis, $z_\text{dec} \sim (G_\nu^2 T_0^3 m_{pl})^{-1/3}$.}
	\label{fig:zdecvsGnu}
\end{figure}

To find the decoupling redshift $z_\text{dec}$, we fit the opacity function in terms of the redshift $O_z(z) = \langle O((1+z)T_0) \rangle$ to the transition function from~\cite{Choi:2018gho},
\begin{equation}\label{eq:transitionfunc}
\mathcal{T}(z) = \frac{1}{2} \left( \tanh \left(\frac{z-z_{\dec}}{\Delta z_{\dec}} \right)+1 \right).
\end{equation}
We show the decoupling redshifts in terms of $G_\nu$ in Figure~\ref{fig:zdecvsGnu}. In Figure~\ref{fig:zdecvsGnu}, we also show the results for the case of partially interacting neutrinos as uniform couplings for neutrino self-interactions are widely constrained by terrestrial experiments, yet these constraints can be weakened by assuming only certain species of neutrinos are self-interacting \cite{Blinov:2019gcj} (see Section.~\ref{subsec:expconstraint} for more detail). The amplitudes for one and two interacting neutrino species are,
\begin{eqnarray}
|\mathcal{M}_{\nu \nu \rightarrow \nu \nu}^\text{1-int}|^2 &=& \frac{1}{2}|\mathcal{M}_{\nu_i \nu_i \rightarrow \nu_i \nu_i}|^2\,,\\
|\mathcal{M}_{\nu \nu \rightarrow \nu \nu}^\text{2-int}|^2 &=& \frac{1}{2}|\mathcal{M}_{\nu_i \nu_i \rightarrow \nu_i \nu_i}|^2 + \frac{1}{2} |\mathcal{M}_{\nu_i \nu_i \rightarrow \nu_j \nu_j}|^2\label{totalamp} + |\mathcal{M}_{\nu_i \nu_j \rightarrow \nu_i \nu_j}|^2\,.
\end{eqnarray}
And the scattering rates are, 
\begin{eqnarray}
\Gamma_{\nu}^\text{1-int}(E_1) &=& \frac{7\pi}{1728} G_\nu^2 E_1 T_\nu^4\,,\\
\Gamma_{\nu}^\text{2-int}(E_1) &=& \frac{7\pi}{576} G_\nu^2 E_1 T_\nu^4\,.
\end{eqnarray}
Note neutrino oscillation does not change the number of interacting neutrino species since we assume diagonal couplings in the mass eigenstates.

We find $\Delta \zdec \sim 0.4 \zdec$ is a good description of  the Majoron case. See Appendix~\ref{app:opacity} for comparison with the actual opacity function. This is a generic feature of decoupling from a dimension-6 operator with the number density of particles changing only from the expansion. Precise values for $\Delta \zdec$ vary with $\zdec$, but we have checked that this approximation is enough for the purpose of the work. A study of the (in)sensitivity of our results to the assumed decoupling width is presented in Appendix~\ref{app:decoupling_width}. 

\subsection{Experimental constraints on neutrino self-interactions}
\label{subsec:expconstraint}
Experimental constraints on neutrino self-interactions have been studied in previous works \cite{Bardin:1970wq,Bilenky:1992xn,Bilenky:1999dn,Blinov:2019gcj,Lyu:2020lps,Brdar:2020nbj}, and we review most relevant constraints on the model with the Majoron, mostly following \cite{Blinov:2019gcj}.\footnote{Note we ignore terms from the UV completion considered in \cite{Lyu:2020lps}. As pointed out in \cite{Blinov:2019gcj}, UV completion of self-interacting neutrinos cannot be a minimal see-saw mechanism, but needs separate seesaw mechanisms for the neutrino masses and the Majoron coupling.} Reference \cite{Blinov:2019gcj} discusses the experimental constraints on the coupling between the Majoron and the neutrino flavor eigenstates. The strongest constraints on the coupling to  $\nu_e$ come from Kaon decay ($K \rightarrow e \nu \phi$) and neutrinoless double-beta decay, while the coupling to $\nu_\mu$ is constrained from Kaon decay to muon ($K \rightarrow \mu \nu \phi$) and coupling to $\nu_\tau$ from $\tau$ decay ($\tau \rightarrow \ell \nu \nu \phi$). In this work, we consider the Majoron coupling to be diagonal to the neutrino mass eigenstates instead of neutrino flavor eigenstates as in \cite{Blinov:2019gcj}, we translate the constraints with the Pontecorvo–Maki–Nakagawa–Sakata (PMNS) matrix, where we take the values in the PMNS matrix from \cite{Esteban:2018azc}.

\begin{figure}[tb]
	\centering
	\includegraphics[width=8cm]{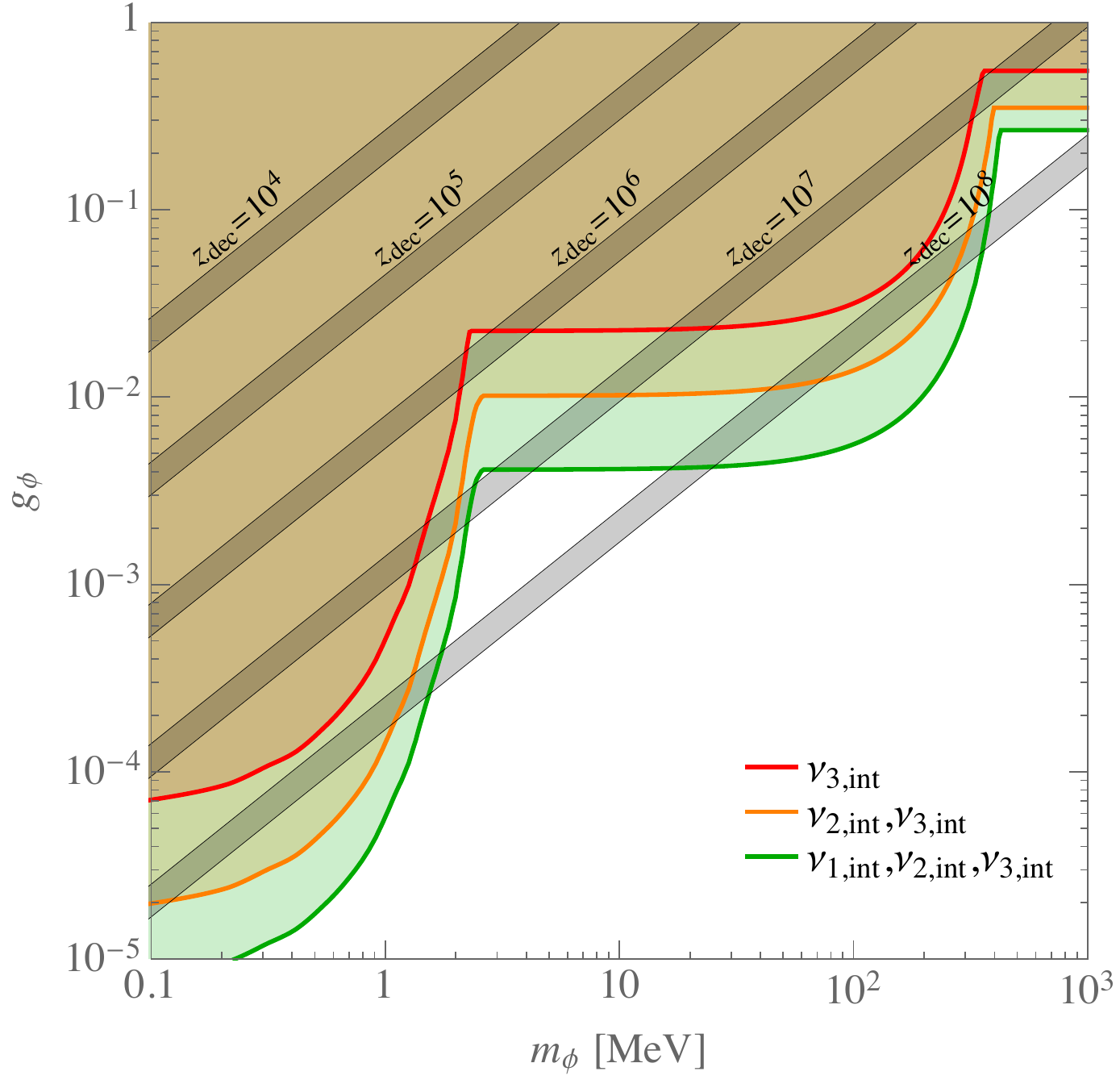}
	\caption{Experimental constraints for the Majoron coupled to neutrino mass eigenstates in the $g_\phi-m_\phi$ plane translated from the constraints on the coupling to neutrino flavor eigenstates in \cite{Blinov:2019gcj}. We show these constraints translated under different assumptions about the number of interacting neutrinos. The gray bands show different values of $\zdec$ in this parameter space with the upper lines of the bands corresponding to the 1-interacting neutrino case and the lower lines to the 3-interacting neutrino case.}
	\label{fig:app:gphiconst}
\end{figure}

The Majoron couplings to mass eigenstates can be converted to couplings to flavor eigenstates with the PMNS matrix $U$ by
\begin{equation}
g_{\alpha \beta}=U_{\alpha i} g_{ij} (U^\dagger)_{j \beta}.
\end{equation}
For the all interacting neutrino case (our {\em Case 1}) where the Majoron couplings to neutrino mass eigenstates are diagonal and universal, $g_{ij} = g_\phi \delta_{ij}$, we have $g_{ij} = g_{\alpha \beta}$ hence the constraints on $g_\phi$ are same as the universal case in \cite{Blinov:2019gcj}. For partially interacting cases (our {\em Cases 2-4}), we calculate $g_{\alpha \beta}$ in terms of $g_\phi$ and demand that each component obeys the constraints in $g_{\alpha \beta}$. We show the results in Fig.~\ref{fig:app:gphiconst}. For partially interacting neutrinos, we choose the cases that are minimally constrained. Since $\nu_\tau$ is the least constrained flavor eigenstate, choosing the mass eigenstates that contain more $\nu_\tau$ gives the desired combination. As a result, we consider $\nu_3$ for 1-interacting neutrino, and  $\nu_2$ and $\nu_3$ for 2-interacting neutrinos. As we shall see, current CMB data is only able to probe neutrino decoupling that occurs at redshifts below $\zdec \lesssim 10^{6}$ so Fig.~\ref{fig:app:gphiconst} demonstrates that laboratory constraints on new neutrino interactions are stronger than CMB constraints, with the possible exception of interactions mediated exclusively through $\nu_3$.

%% file: parameterization.tex
\section{Parameterization, Implementation, and Effect on CMB Power Spectra}
\label{sec:parameterization}
To model the impact of self-interacting neutrinos on CMB observables, we extend the decoupling redshift approach of~\cite{Choi:2018gho}. Neutrino self-interactions have the effect of suppressing higher moments of the Boltzmann hierarchy  (i. e. moments of the perturbation to the neutrino distribution function, $F_{\nu,\ell}$, are suppressed for $\ell >2$), causing neutrino perturbations to evolve as a relativistic fluid.  After decoupling, neutrinos free-stream allowing the higher moments of the Boltzmann hierarchy to take nonzero values. The transition between these two epochs is imposed manually with the transition function in Eq.~\ref{eq:transitionfunc}. In this paper, we extend the implementation in ~\cite{Choi:2018gho} to allow for massive self-interacting species in combination with ordinary, massive, free-streaming neutrinos. Where~\cite{Choi:2018gho} considered a near-instantaneous decoupling width $\Delta \zdec = 0.01 \zdec$, we approximately match the opacity function for the case of neutrino decoupling with the Majoron as stated in Section~\ref{sec:SInu}, which is closer to $\Delta \zdec = 0.4 \zdec$. This approach also gives CMB power spectra that are in excellent agreement with those in~\cite{Cyr-Racine:2013jua,Lancaster:2017ksf,Kreisch:2019yzn,Park:2019ibn} and, as is shown in Appendix~\ref{app:decoupling_width}, our final results are relatively insensitive to the precise value of $\Delta \zdec$ for $\Delta \zdec/\zdec$ in the range $[0.1, 0.8]$. 

\subsection{Parameterization}
\label{ssec:parameters}

As usual, the radiation density in the early universe is parameterized by 
\begin{equation}
\rho_{\text{rad}}(T \lesssim 1{\rm MeV}) = \rho_{\gamma}\left[1+ \frac{7}{8}\left(\frac{4}{11}\right)^{4/3}\Neff\right]\,,
\end{equation}
where $\rho_\gamma = (\pi^2/15)T_\gamma^4$ is the energy density in CMB photons and $T_\gamma = 2.725K$ today (see e.g. \cite{Ade:2015xua}). With this definition, $\Neff = 3$ corresponds to the radiation energy density expected from three Standard Model neutrinos that decouple instantaneously. In the standard cosmology $\Neff \approx 3.046$, due to residual heating of neutrinos from electron-positron annihilation \cite{Mangano:2001iu, Mangano:2005cc, Gnedin:1997vn, Hannestad:1995rs, Heckler:1994tv, Dolgov:1992qg, Dolgov:2002wy, Dolgov:1997mb, Dodelson:1992km, Rana:1991xk, Esposito:2000hi,deSalas:2016ztq,Akita:2020szl}. 

We further split the $\Neff$ parameter into two components: $\Nefffs$, the effective number of free-streaming relativistic species and $\Neffint$ the effective number of interacting relativistic species. The total energy in relativistic species is then given by
\be
\Neff = \Nefffs + \Neffint
\ee
the interacting species $\Neffint$ is assumed to decouple at a redshift $\zdec$ with a duration $\Delta \zdec$ that is fixed to $\Delta \zdec = 0.4 \zdec$ (see Sec.~\ref{sec:SInu} and Appendix~\ref{app:decoupling_width}). The effective mass of all species is given by $\meff = \displaystyle \sum_i N_{\rm{eff},i} \times m_{i}$. We will always use the degenerate neutrino mass approximation where all interacting or free-streaming species are presumed to have the same mass.\footnote{This has been shown to be sufficiently accurate for all current and most future cosmological analyses, see e.g.~\cite{Lesgourgues:2006nd,DiValentino:2016foa,Vagnozzi:2017ovm,Archidiacono:2020dvx}}

In this paper we will consider four cases of parameter choices,  designed to mimic different scenarios for interacting neutrinos. 
\begin{itemize}
	\item {\em Case 1: All species interacting} \\
	In this example, all neutrino species are presumed to participate in the new self-interactions, which decouple at $\zdec$, a free parameter. The total energy in self-interacting neutrinos, $\Neffint$, and the masses of self-interacting neutrinos are both allowed to vary. This is implemented by allowing variable $\Neffint$, $\zdec$, and $\meff$ with $\Nefffs = 0$.
	
	\item {\em Case 2: Two free-streaming species plus interacting species}\\
	In this example, we force two neutrino states to be free-streaming, as a way to account for laboratory constraints on new neutrino self-interactions (see Section~\ref{subsec:expconstraint}). This is implemented by fixing $\Nefffs =2$ and allowing the additional interacting relativistic degrees of freedom, characterized by $\Neffint$ to vary\footnote{Technically, because of the way \texttt{CLASS} computes $\Neff$ we set the number of extra free-streaming species to $\Nefffs=2.0328$ so that we get $\Neff=3.046$ if we add exactly one massive extra relativistic species.}. We assume $\Nefffs$ degrees of freedom are massless and the masses of $\Neffint$ are $\meff/\Neffint$.
	\item {\em Case 3: Fixed number of relativistic species and varying fraction of interacting species}\\
	In this example, we fix the number of relativistic species to $\Neff=3.046$, but vary the fraction that is self-interacting. The mass sum of all relativistic species is also allowed to vary. This is implemented by allowing variable $\Neffint$, $\zdec$, and $\meff$ while fixing $\Neff = \Nefffs + \Neffint = 3.046$.
	\item {\em Case 4: Varying fraction of interacting species}:\\
	Finally, we consider an example where both the free-streaming and self-interacting degrees of freedom are allowed to vary. This is implemented by allowing variable $\Neffint$, $\Nefffs$, and $\zdec$. This is equivalent to treating the total relativistic degrees of freedom ($\Neff = \Neffint+ \Nefffs$) as a free parameter, as well as the interacting fraction.  For simplicity, in this case we  fixed to $\meff = 0.11~$eV, by setting the individual mass of the interacting and the free-streaming species to $\meff / \Neff$.
	\item {\em Reference Cosmologies} \\
	We have one reference cosmology per case, which uses a standard implementation of variable $\Neff$, where all degrees of freedom contributing to $\Neff$ are free-streaming. Masses are arranged as in Cases 1-4, i.e. for Case 4 all relativistic species are massless, for Case 1 and 3 all species are massive, and for Case 2 only species in excess of $\Neff = 2.0328$ are massive. In either cases with a massive relativistic species, the mass is characterized by $\meff$. Specifically, we allow variable $\Nefffs$ (except reference Case 3) and $\meff$ (except reference Case 4), and fix $\Neffint = 0$.
\end{itemize}

For Cases 2, 3, 4, we also study the pure fluid-like limit, equivalent to setting $\zdec$ to a value in the future so that $\Neffint$ remains interacting through today.  Case 1 is analogous to that considered in \cite{Kreisch:2019yzn}, with the exception that we treat the neutrinos as having degenerate masses, rather than putting all the mass associated with $\sum m_\nu$ into a single mass state. 

\subsection{Implementation in CLASS}
We generalize the implementation of a decoupling non-cold dark matter species from~\cite{Choi:2018gho} by modifying \texttt{CLASS v2.7} to add a new species \texttt{ddec}, which is similar to the existing \texttt{ncdm} species. This separation allows us to control all aspects of the \texttt{ddec} species, while maintaining the current implementation of the \texttt{ncdm} species and allows for flexible computation options with both active (or sterile) neutrinos and/or a new class of self-interacting species, each with their own precision parameters and settings. We verified that the implementation is still valid when considering massive species and a wider decoupling width\footnote{Since~\cite{Choi:2018gho} only considered massless species and instantaneous decoupling.}, tuning the precision parameters where needed\footnote{This primarily involves turning off the fluid approximation and increasing the precision requirements for the other \texttt{ddec} precision parameters (which match the \texttt{ncdm} ones), see Appendix~\ref{app:precision_settings} for details.}. The neutrino self-interaction is added as a function modifying the Boltzmann hierarchy, so that the species behaves as a perfect fluid prior to decoupling and is free-streaming after decoupling, with some intermediate region defined by the decoupling width. For this work we define the decoupling width as 40\% of the decoupling redshift, $\Delta \zdec = 0.4 \zdec$, since we find this to be a good approximation for an effective self-interaction (see Sec. \ref{sec:SInu}). In Appendix~\ref{app:decoupling_width}, we discuss how this choice affects bounds on the time of decoupling.

\subsection{Effects on CMB Power Spectra}
\begin{figure}[tb]
	\centering
	\includegraphics[width=\textwidth]{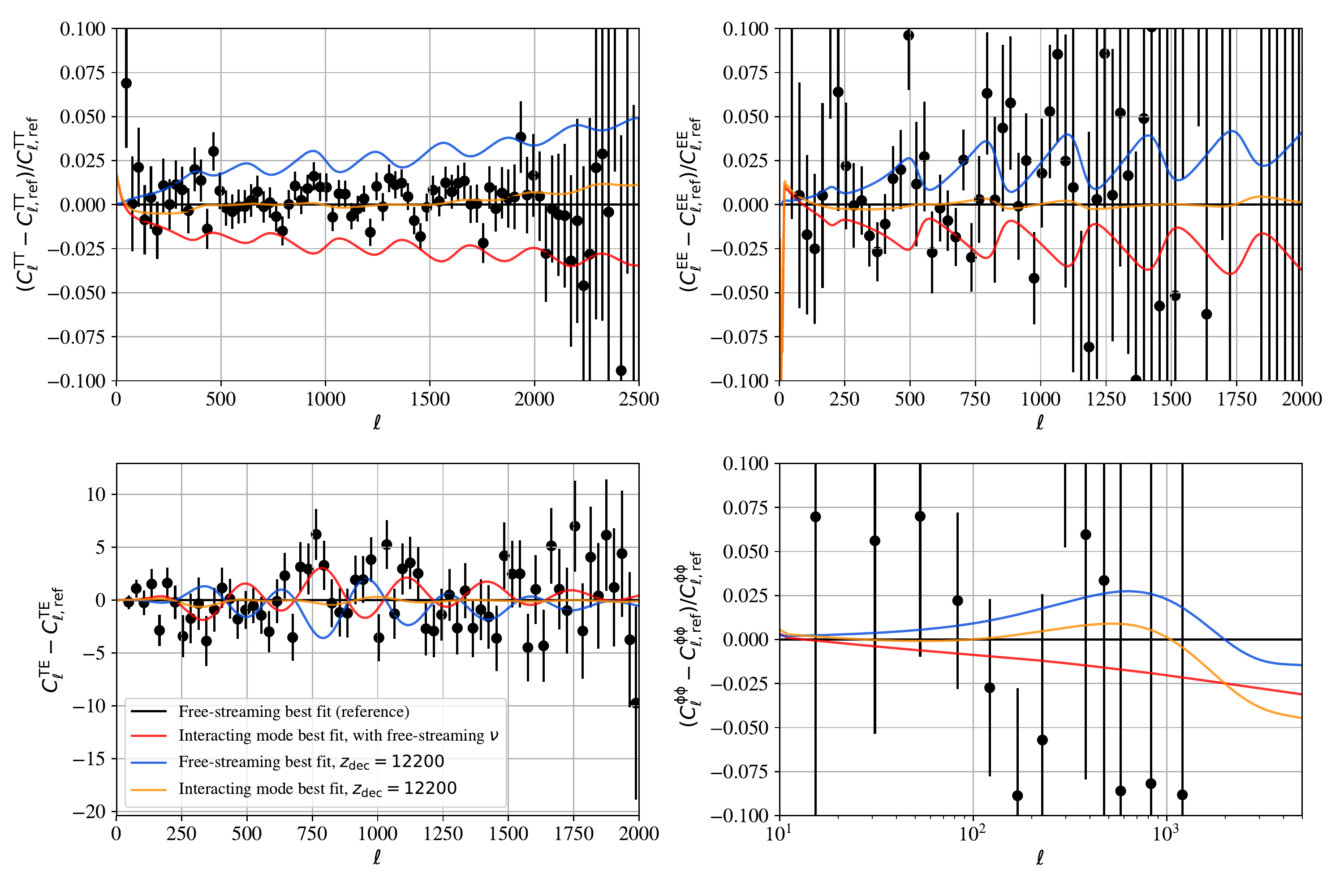}
	\caption{Fractional differences between CMB temperature (top left) and polarization (top right) auto-correlation and cross-correlation (bottom left, showing difference rather than fractional difference) angular power spectra, and the CMB lensing power spectrum (bottom right). In all panels, the Planck 2018 binned data~\cite{Aghanim:2019ame,Aghanim:2018oex} are shown in black, for comparison. Fractional differences are computed with respect to a best-fit free-streaming neutrinos comparison case. The orange curves show power spectra in a model with both free-streaming and self-interacting neutrinos ({\em Case 2}), computed at the best fit cosmological parameters for that model with a decoupling redshift of $\zdec=\textrm{12,200}$. In red the same cosmological parameters are assumed as for the orange curve, but the the power spectra are computed assuming only free-streaming neutrinos. In blue, the power spectra are computed in a cosmology with interacting neutrinos assuming $\zdec = \textrm{12,200}$ using the best-fit cosmological parameters for the free-streaming reference case. See Sec.~(\ref{ssec:parameters}) for a description of the reference and interacting neutrino models and Table \ref{results:tab:2FSz6m} (P18 +lens +BAO) for the precise parameter choices.}
	\label{fig:2FSb_spectra}
\end{figure}

Let us now review the effects of neutrino self-interactions on CMB power spectra. Neutrino self-interactions qualitatively change the evolution of neutrino perturbations. Perturbations in  free-streaming neutrinos propagate at the speed of light $c$, while perturbations in a relativistic fluid of self-interacting neutrinos propagate at smaller speed $c_s \approx c/\sqrt{3}$. This  difference leads to changes in the phase and amplitude of acoustic oscillations in the photon-baryon fluid in the early universe (see, e.g. \cite{Bashinsky:2003tk,Hou:2011ec,Baumann:2015rya,Choi:2018gho}). In what follows, we illustrate the changes to the CMB power spectra caused by neutrino self-interactions that decouple at different epochs, using example parameter choices from our results in Sec.~\ref{sec:results}, paying particular attention to how these changes can be mimicked by changing other cosmological parameters. 

\begin{figure}[tb]
	\centering
	\includegraphics[width=\textwidth]{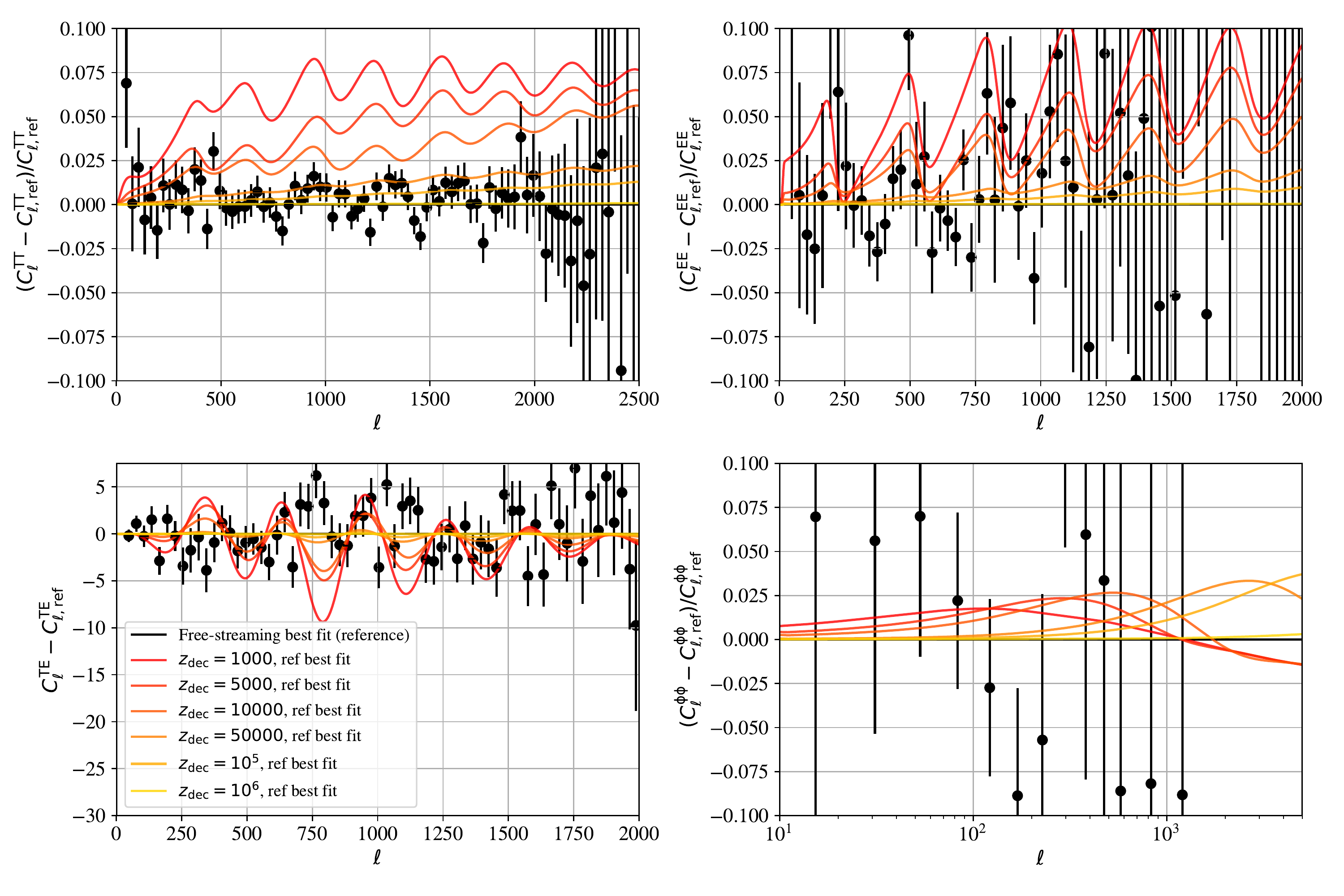}
	\caption{Fractional differences between the CMB power spectra computed in a model with self-interacting and free-streaming neutrinos (Case 2, see Sec. \ref{ssec:parameters}), as compared to a model with only free-streaming neutrinos. Each curve shows a different assumed value of $\zdec$, with all other parameters held fixed. Shown are CMB temperature (top left) and polarization (top right) auto-correlation and cross-correlation (bottom left, showing difference rather than fractional difference) angular power spectra, and the CMB lensing power spectrum (bottom right). In all panels, the Planck 2018 binned data~\cite{Aghanim:2019ame,Aghanim:2018oex} are shown in black, for comparison.}
	\label{fig:varying_zdec}
\end{figure}

In Figure~\ref{fig:2FSb_spectra}, we show the effect of neutrino self-interactions on the CMB temperature and polarization auto-correlation ($C_{\ell}^{TT}$, $C_{\ell}^{EE}$) and cross-correlation ($C_{\ell}^{TE}$) angular power spectra and the CMB lensing power spectrum ($C_{\ell}^{\phi\phi}$). We compare power spectra computed assuming different cosmologies with either free-streaming or a combination of free-streaming and self-interacting neutrinos (e.g. Case 2 from \ref{ssec:parameters}). The reference (``ref") cosmology is the best-fit parameters assuming a free $\Neff$, all free-streaming ($\Lambda$CDM+$\nu_\text{fs}$, see P18 +lens +BAO on Table \ref{results:tab:2FSz6m} for the precise parameter values). In orange we show the best-fit of a $\Lambda$CDM+$\nu_\text{fs}$+$\nu_\text{int}$ case (Case 2 in Section~\ref{ssec:parameters} above) for $\zdec = \textrm{12,200}$ (the low-z decoupling mode discussed later in Section~\ref{sec:results:two_fs})). In red we show the power spectra for a free-streaming cosmology, but computed using the best fit cosmological parameters for the interacting scenario (including the total $\Neff$ value). In blue we show the converse: the power spectra for an interacting neutrino cosmology with $\zdec = \textrm{12,200}$, but computed using the best-fit parameters from the free-streaming comparison case. In all panels, the Planck 2018 binned data~\cite{Aghanim:2019ame,Aghanim:2018oex} are shown in black, for comparison. As was noted in~\cite{Lancaster:2017ksf, Kreisch:2019yzn}, a significant change in cosmological parameters is nearly offset by changing the decoupling redshift (or equivalently, neutrino interaction strength), $\zdec$. This is illustrated by the orange curve, which is also approximately the sum of the blue and red curves, where for most scales the difference compared to the free-streaming comparison case is sub-percent, with percent-level differences at very small scales (high multipole, $\ell$).

\begin{figure}[tb]
	\centering
	\includegraphics[width=\textwidth]{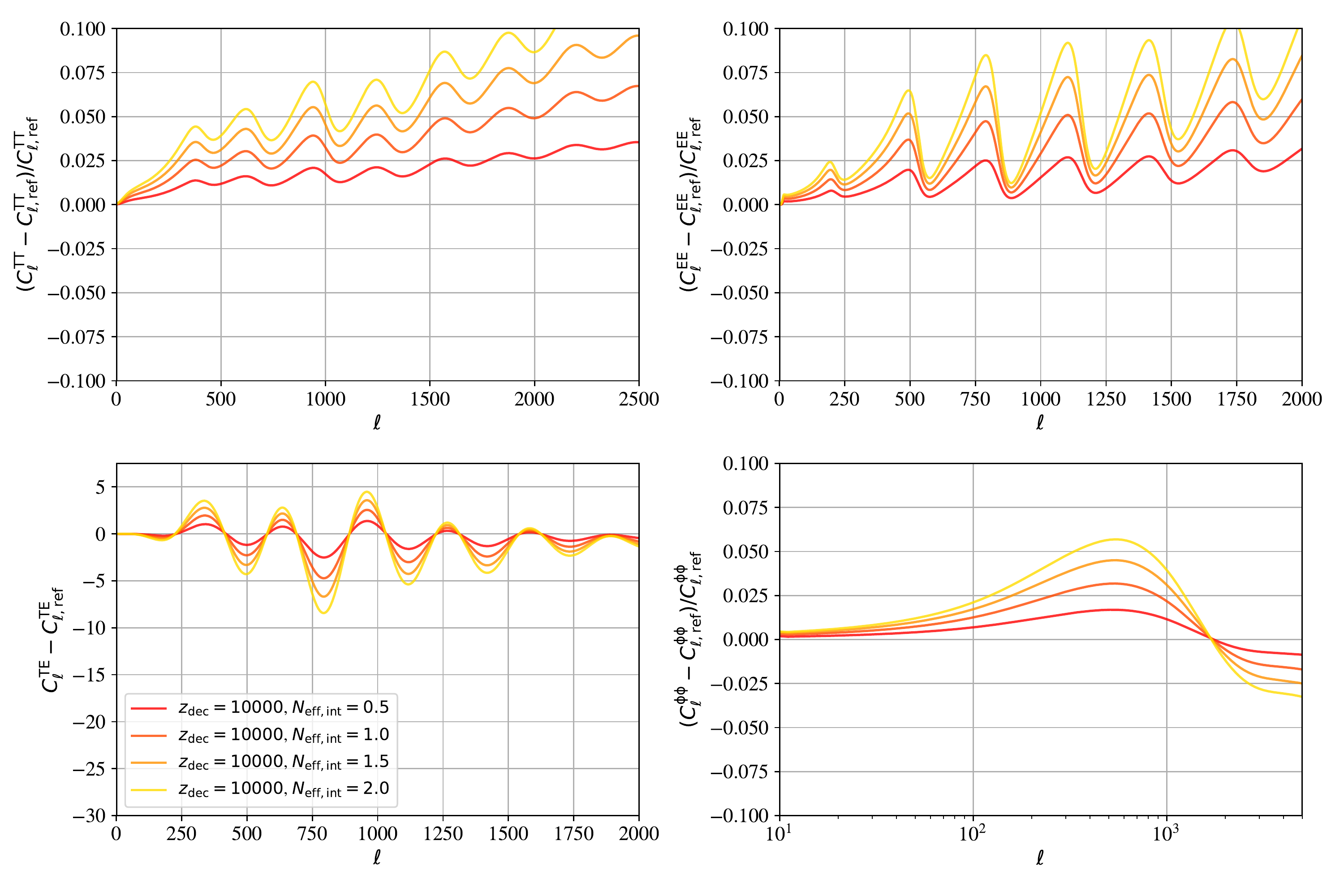}
	\caption{The fractional differences between CMB power spectra computed in a scenario with a varying fraction of interacting neutrinos that decouple from self-interactions at $\zdec = \textrm{10,000}$, as compared to those computed in a cosmology with the same total $\Neff$ comprised of all free-streaming neutrinos (all other parameters are also held fixed). Shown are the CMB temperature (top left) and polarization (top right) auto-correlation and cross-correlation (bottom left, showing difference rather than fractional difference) angular power spectra, and the CMB lensing power spectrum (bottom right).}
	\label{fig:varying_Neff}
\end{figure}

In Figure~\ref{fig:varying_zdec}, we isolate the effect of varying the decoupling redshift $\zdec$, holding all other parameters fixed to the best fit free-streaming ($\Lambda$CDM+$\nu_\text{fs}$) comparison case (``ref"). We can see that varying $\zdec$ does not just shift the phase and overall amplitude of the acoustic peaks but introduces more subtle changes, depending on the value of $\zdec$:
\begin{itemize}
	\item The amplitude always increases across all scales with lower $\zdec$, but not in a scale-independent way (this is easier to see in the $C_{\ell}^{TT}$ plot, but the effect persists in the $C_{\ell}^{EE}$ plot).
	\item Going from later decoupling redshifts to earlier, between $\zdec = \textrm{10,000}$ and $\zdec = 1000$ we see that at higher values the spectra experience a fairly regular scale-dependent amplitude shift roughly corresponding to a larger effect with smaller $\zdec$, but at low $\zdec$ values it is not this simple.
	\item At $\ell > 800$ the amplitude increases across all scales in roughly the same way as between $\zdec = 1000$ and $\zdec = \textrm{10,000}$, except for a small change of the damping tail, where the amplitude increases more slowly with lower $\zdec$ and goes from increasing with $\ell$ in a scale-dependent way for $\zdec = \textrm{10,000}$ to a roughly flat amplitude increase at $\zdec = 1000$.
	\item At $\ell < 800$ the amplitude increases steadily up until around $\zdec = \textrm{10,000}$ and then barely changes through $\zdec = 5000$ before going back to changing in a regular way at $\zdec = 1000$. This change coincides with the transition from radiation to matter domination.
\end{itemize}

Finally, in  Fig.~\ref{fig:varying_Neff}, we isolate the effect of varying the total $\Neffint$ with fixed $\zdec = \textrm{10,000}$, in comparison to a predictions for power spectra with the same total $\Neff$. Precisely, we consider power spectra computed in a cosmology with two free-streaming massless neutrinos and variable amount of $\Neffint$ ranging from $\Neffint = 0.5$ to $\Neffint=2$. We compare those power spectra to ones computed a cosmology with only free-streaming neutrinos with the same abundances set by $\Nefffs = 2 + \Neffint$, where $\Neffint$ takes the same values ranging from $0.5$ to $2$. In this case, the binned Planck data is omitted. Since we do not use a universal reference case (with $\Nefffs$ varying), the Planck residuals would get shifted for each curve. From Fig. \ref{fig:varying_Neff}, we see that the overall shape for $C_{\ell}^{TT}$ and $C_{\ell}^{EE}$ is changed by increasing the self-interacting fraction. The fractional difference between the $C_{\ell}^{TE}$ are very similar to that shown in the varying $\zdec$ plot. The changes to $C_{\ell}^{\phi\phi}$ with varying interacting fraction are instead purely an amplitude shift with the same peak value and zero crossing point irrespective of $\Neffint$. 

%% file: method.tex
\section{Method and datasets}
\label{sec:method}
For our analyses we use the Cosmological sampling package \texttt{MontePython v3.2}\footnote{Get the new \texttt{MontePython v3.4} at https://github.com/brinckmann/montepython\_public}~\cite{Audren:2012wb,Brinckmann:2018cvx}, interfaced with a modified version of the Boltzmann Solver \texttt{CLASS v2.7}\footnote{Get the current \texttt{CLASS v2.9} at https://github.com/lesgourg/class\_public}~\cite{Blas:2011rf,Lesgourgues:2011re,Lesgourgues:2011rh} and with the MultiNest sampler~\cite{Feroz:2007kg,Feroz:2008xx,Feroz:2013hea} via the PyMultiNest wrapper~\cite{Buchner:2014nha} (see Appendix~\ref{app:precision_settings} for the MultiNest sampling settings used for the runs in this paper).\\
\\
In the following we use shorthand notation to refer to the following datasets:
\begin{itemize}
	\item P18: Planck 2018 CMB temperature and polarization auto- and cross-correlation, both high-$\ell$ and low-$\ell$~\cite{Aghanim:2019ame}.
	\item TT: Planck 2018 CMB temperature auto-correlation, both high-$\ell$ and low-$\ell$~\cite{Aghanim:2019ame}.
	\item lowEE: Planck 2018 CMB polarization auto-correlation, low-$\ell$~\cite{Aghanim:2019ame}.
	\item lens: Planck 2018 CMB lensing~\cite{Aghanim:2018oex}.
	\item BAO: 6dFGS ($z=0.106$)~\cite{Beutler:2011hx}, SDSS DR7 MGS ($z=0.15$)~\cite{Ross:2014qpa}, and BOSS DR12 ($z=0.38, 0.51, 0.61$) three redshift bin sample~\cite{Alam:2016hwk} (formerly the CMASS and LOWZ galaxy samples~\cite{Anderson:2013zyy})\footnote{Note that since neutrino self-interactions introduce a phase shift in the acoustic peaks it is possible constraints using the standard BAO approach are biased. However,~\cite{Bernal:2020vbb} studied the reliability of this approach when confronted with some beyond $\Lambda$CDM cosmologies, including a model with interactions between dark matter and neutrinos, which exhibit a similar phase shift. As such, for current data the self-interacting neutrino model is unlikely to result in large biases from the BAO analysis, especially when BAO data is combined with other datasets, such as CMB data. However, for future data it would be prudent to ideally analyse the BAO data consistently or at least to redo the analysis of~\cite{Bernal:2020vbb} for the self-interacting neutrino case in question in order to ensure that any possible bias is sufficiently small.}.
	\item R19: Prior on the Hubble parameter today, $H_{0}$, from Riess et al. 2019~\cite{Riess:2019cxk}.
\end{itemize}
Note we make use of the ``lite" version of the Planck likelihoods in order to speed up rate of convergence, as the full set of nuisance parameters are expected to have only a small effect on cosmological constraints for most models. Although not an ideal choice, the use of MultiNest requires we restrict the total number of parameters to make the analysis feasible and we checked that the resulting bias is less than about 0.2$\sigma$ shifts in all parameters (with $\Neff$ biased towards lower values). We consider P18 +lens and P18 +lens +BAO as our baseline configurations (we always include lensing), but sometimes add R19 to explore whether the model in question helps alleviate the Hubble tension. We also consider the case without high-$\ell$ polarization with and without the R19 prior for the model with all neutrinos self-interacting, in order to compare to previous work.

\subsection{A discussion of tensions and analysis choices}
\label{sec:tensions}
Cosmology has seen a number of tensions between datasets grow in recent years, most notably the $H_0$ and $S_8$ tensions. Aside from the possibility of unresolved systematics, these tensions could arise from the assumption of an incorrect model, as (nearly) all cosmological analyses assume a model to conduct the analysis. As a result, many works have striven to address this tension by changing assumptions within the cosmological model e.g. on the nature of dark matter or dark energy. Before proceeding, we discuss our philosophy on exercising great caution when combining discrepant datasets when performing parameter inference analyses. 

When combining two discrepant datasets, we will generally expect to find a result between the two measurements, e.g. between Planck CMB extrapolations of the $H_0$ value and late-time cepheid calibrated supernovae measurements of the same. \textit{This does not mean a model alleviates the tension}, as it is merely a consequence of the statistical analysis. In cases where a model does not resolve (or at least significantly alleviate) the tension in question, a combination of discrepant datasets is not consistent and these dataset combinations should be avoided (e.g.~\cite{Lemos:2019txn,Handley:2020hdp}).

In the following, we will test our models by combining datasets that are discrepant within $\Lambda CDM$, specifically by including a prior on the Hubble parameter from Riess et al.~\cite{Riess:2019cxk}
\footnote{Note that it was recently pointed out that rather than including a prior on $H_0$, from e.g. the SH0ES collaboration~\cite{Riess:2019cxk}, it is more appropriate to include the Pantheon supernovae sample~\cite{Scolnic:2017caz} along with a prior on the absolute peak magnitude $M_B$ from SH0ES~\cite{Benevento:2020fev,Camarena:2021jlr,Efstathiou:2021ocp} (see those works for details). While the authors would encourage doing so in the future irrespective of the cosmological model being studied (or indeed to not include a prior from a discrepant dataset at all, as discussed in this section), we stress that for a change to early time cosmology, such as the self-interacting neutrinos studied in this work, the inferred value for $H_0$ by SH0ES would be expected to correct (with all the usual caveats) and the use of a prior on $H_0$ should not differ significantly from a prior on $M_B$. However, for models that change the late time evolution of the Universe correct use of the $M_B$ prior is crucial.}.
We evaluate whether the addition of an $H_0$ prior appears to be reasonable, by comparing to a control case (a free-streaming only model) that does not help resolve the $H_0$ tension (see e.g.~\cite{Knox:2019rjx}) and for which, therefore, the combination of datasets is not consistent. In many cases, the analysis will show that our model does not help alleviate the $H_0$ tension beyond what we find for the free-streaming control case. As such, the combination of discrepant datasets is suspect. We include these null-results in order to further the discussion of which types of models work to resolve the tensions, but also what does \textit{not} work to resolve the $H_0$ tension. In all cases, the validity of this combination for a particular model will always be discussed in the text.

%% file: results.tex
\newpage
\section{Results}
\label{sec:results}
In this section, we present the results of our analyses. As introduced in Sec.~\ref{ssec:parameters}, we consider four scenarios for interacting neutrinos and a corresponding reference example with free-streaming neutrinos. 

\subsection{Case 1: All species interacting}
\label{sec:results:all_interacting}

\subsubsection{Baseline data configurations}
\begin{figure}[tb]
	\centering
	\includegraphics[width=8.4cm]{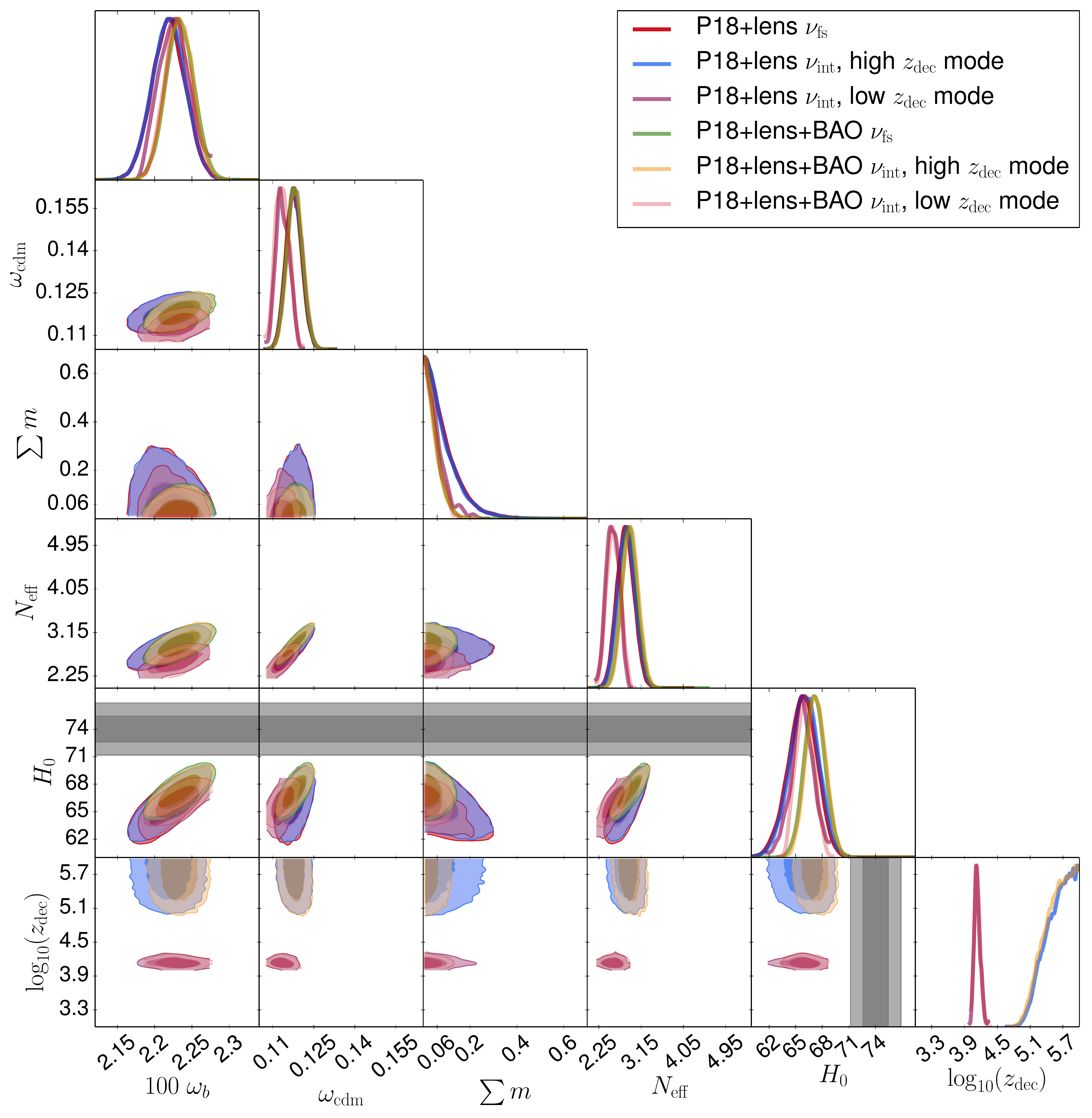}
	\hspace{0.1cm}\includegraphics[width=8.4cm]{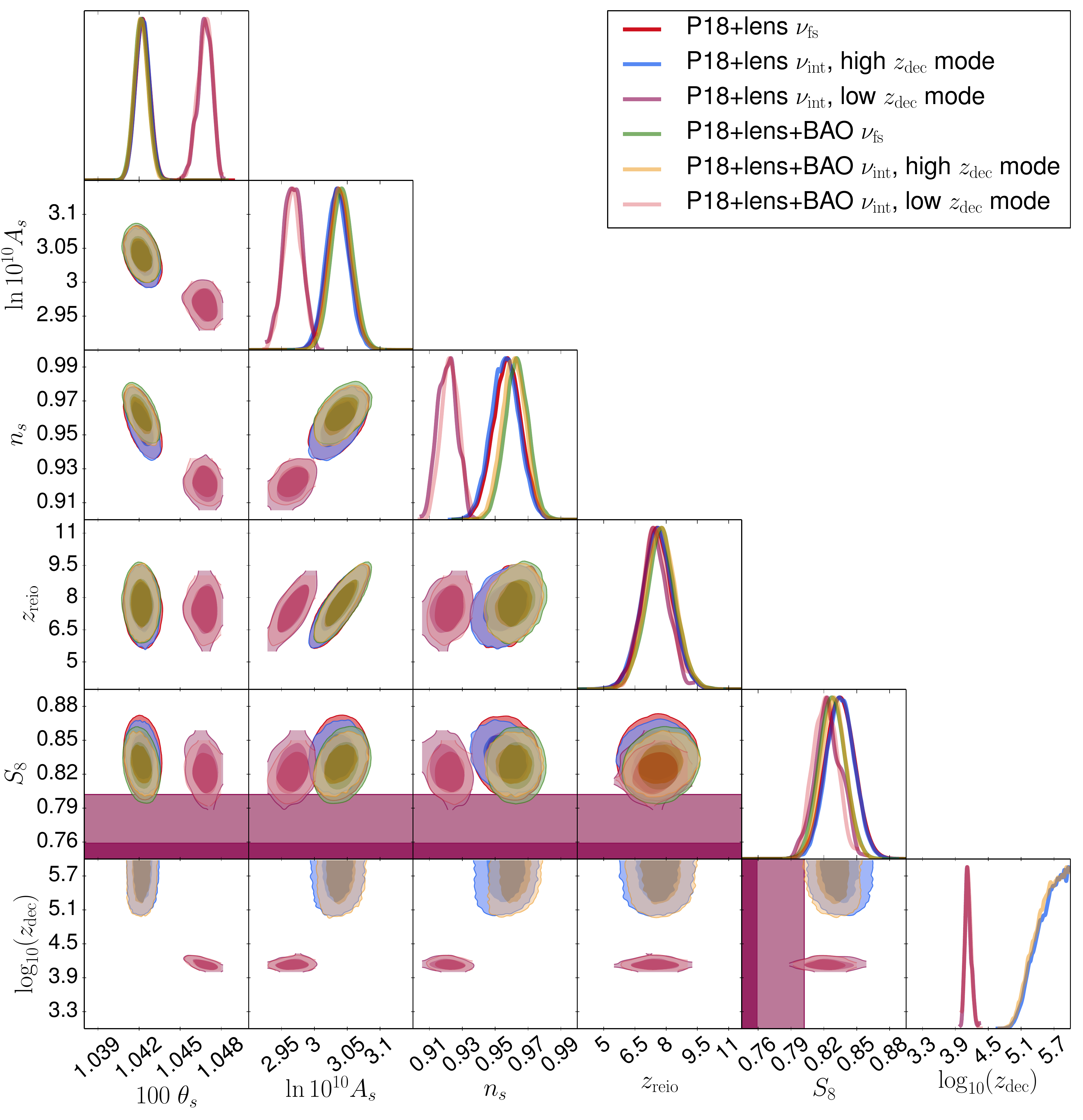}
	\caption{\textbf{Case 1: all species interacting.} Plotted are parameter constraints from models with all free-streaming neutrinos (red and green) vs all species interacting separated into high $\zdec$ (blue and yellow) and low $\zdec$ modes (purple and pink) for Planck-only (red, blue and purple) and Planck + BAO (green, yellow and pink). The grey bands correspond to the $H_0$ measurement from Riess et al.~\cite{Riess:2019cxk}, while the purple band is the $S_8$ measurement from KiDS+VIKING-450~\cite{Wright:2020ppw}. See Appendix \ref{app:full_plots}, Figure \ref{fig:app:P_Pb_R1z6m} for the full parameter space plot. \textbf{Left and right panel:} both show the same cases, but different cosmological parameters. \textbf{Left panel}: energy density parameters. \textbf{Right panel:} power spectrum shape and amplitude parameters. Note that beyond $\zdec > 10^6$ the 1-d posterior flattens out and is very slowly increasing until a peak near standard neutrino decoupling. In several cases it can be hard to see some contours as they are neatly overlapping, e.g. red and blue as well as green and yellow, which in these cases signify minimal difference between the free-streaming comparison case and the high $\zdec$ mode. The same is also true to a lesser degree for the two strongly interacting modes in purple and pink. This figure shows that enabling self-interactions for all neutrinos is not a solution to the Hubble tension when including all Planck primary anisotropies and lensing. This can be seen from the self-interacting cases closely overlapping with the free-streaming cases for $H_0$ (left panel), i.e. there is no improvement over free-streaming neutrinos.}
	\label{fig:results:P_Pb_R1z6m}
\end{figure}

\begin{figure}[tb]
	\centering
	\includegraphics[width=8.4cm]{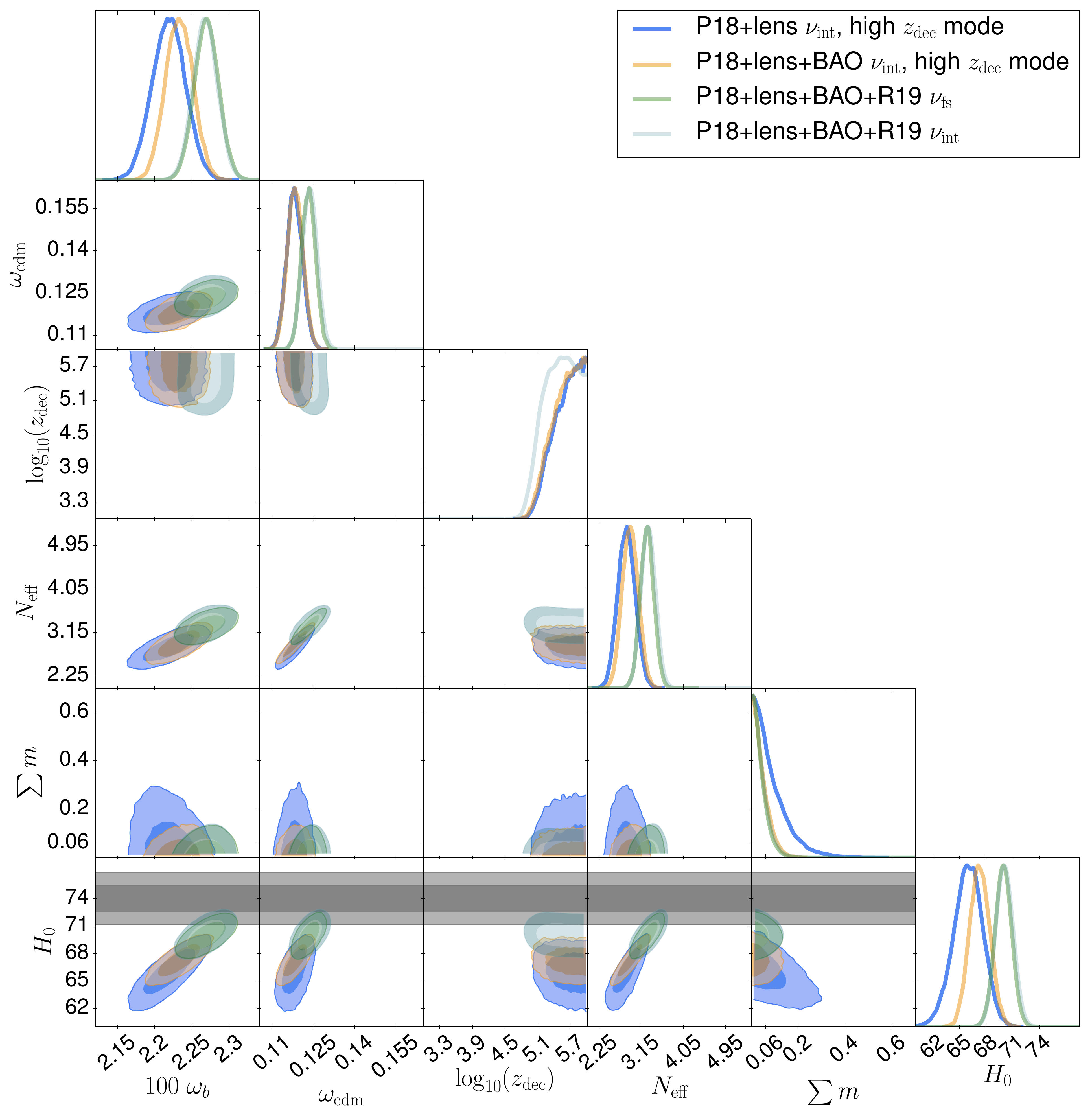}
	\hspace{0.1cm}\includegraphics[width=8.4cm]{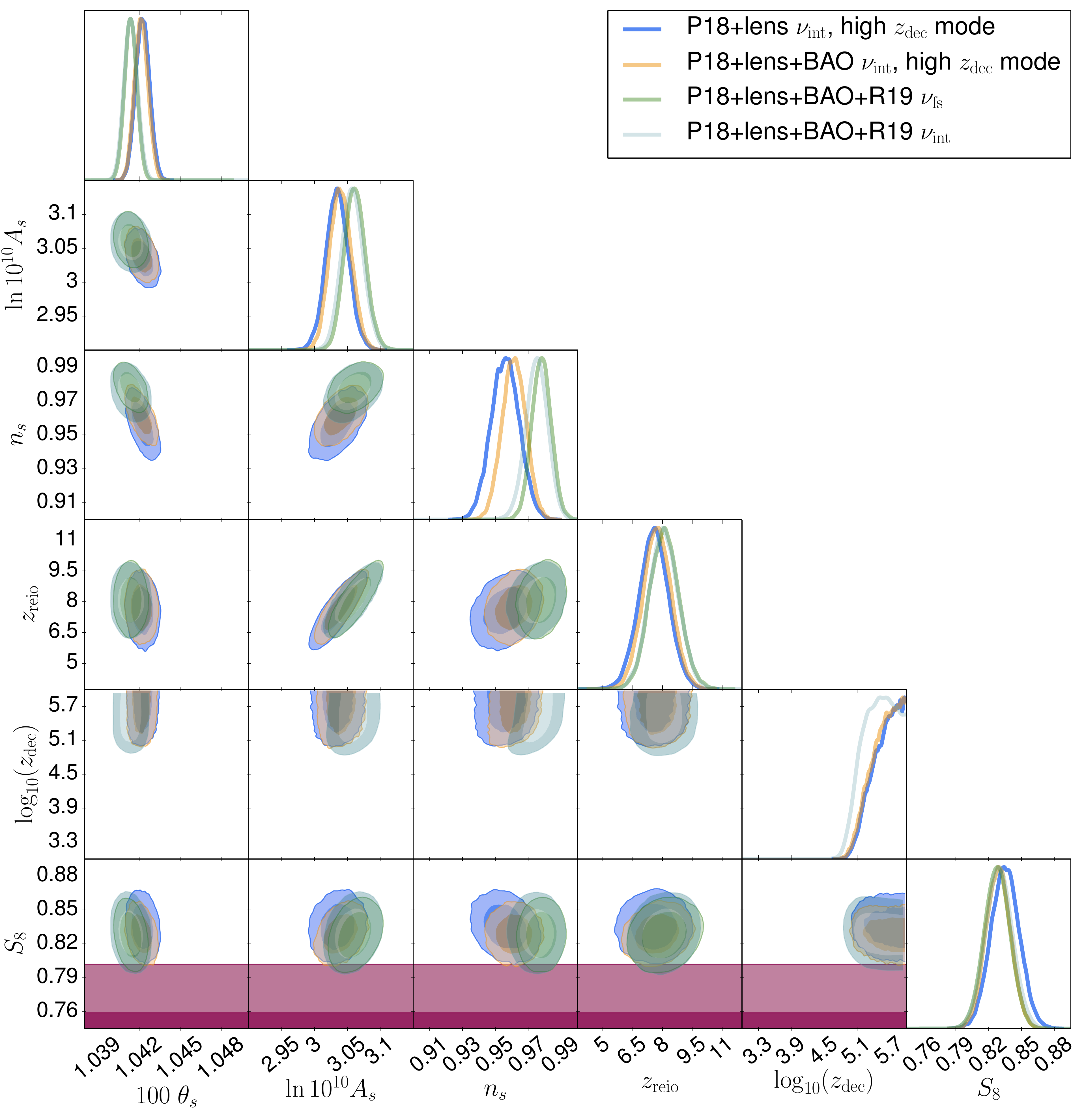}
	\caption{\textbf{Case 1: all species interacting, including $H_0$ prior.} Parameter constraints comparing models with all free-streaming neutrinos (light green) vs all species interacting separated (cyan) for Planck + BAO + R19. For comparison, we also show the high $\zdec$ modes from Figure \ref{fig:results:P_Pb_R1z6m} for Planck-only (blue) and Planck + BAO (yellow). The grey bands correspond to the $H_0$ measurement from Riess et al.~\cite{Riess:2019cxk}, while the purple band is the $S_8$ measurement from KiDS+VIKING-450~\cite{Wright:2020ppw}. See Appendix \ref{app:full_plots}, Figure \ref{fig:app:Pbr_R1z6m} for the full parameter space plot. \textbf{Left and right panel:} both show the same cases, but different cosmological parameters. \textbf{Left panel}: energy density parameters. \textbf{Right panel:} power spectrum shape and amplitude parameters. Note that beyond $\zdec > 10^6$ the 1-d posterior is approximately flat up to standard neutrino decoupling. This figure shows that enabling self-interactions for all neutrinos is not a solution to the Hubble tension when including all Planck primary anisotropies and lensing, as well as a prior on $H_0$. This can be seen from the self-interacting case (cyan) neatly overlapping with the free-streaming case (green) for $H_0$ (left panel), i.e. there is no improvement over free-streaming neutrinos. Note that this makes the inclusion of the $H_0$ prior suspect and those results should be regarded with caution.}
	\label{fig:results:Pbr_R1z6m}
\end{figure}

In order to compare to~\cite{Kreisch:2019yzn} we include a case with all neutrinos interacting. Note that in our case the mass is distributed across all of the neutrino mass states in a degenerate mass hierarchy instead of one massive and the rest massless like in~\cite{Kreisch:2019yzn}. In this section, we present our baseline data configurations for Case 1 (all species interacting), consisting of Planck primary anisotropies plus lensing, alone and with BAO data (see Figure~\ref{fig:results:P_Pb_R1z6m}). Additionally, we add a prior corresponding to the Hubble constant measurement from Riess et al.~\cite{Riess:2019cxk}, in order to see if this decoupling model alleviates the Hubble tension and whether this data combination is reasonable for this decoupling model (see Figure~\ref{fig:results:Pbr_R1z6m}).
\newpage
\noindent In this section, we refer to these figures, datasets, and configurations:
\begin{itemize}
	\item Figure \ref{fig:results:P_Pb_R1z6m}. P18 +lens. Red (all free-streaming), blue (all interacting, high $\zdec$ mode), and purple (all interacting, low $\zdec$ mode).
	\item Figure \ref{fig:results:P_Pb_R1z6m}. P18 +lens +BAO. Green (all free-streaming), yellow (all interacting, high $\zdec$ mode), and pink (all interacting, low $\zdec$ mode).
	\item Figure \ref{fig:results:Pbr_R1z6m}. P18 +lens +BAO +R19. Light green (all free-streaming) and cyan (all interacting).
	\item Figure \ref{fig:results:Pbnhr_R1z6m}. TT +lowEE +lens +BAO (no high-$\ell$ polarization). Red (all free-streaming) and blue (all interacting).
	\item Figure \ref{fig:results:Pbnhr_R1z6m}. TT +lowEE +lens +BAO +R19 (no high-$\ell$ polarization). Green (all free-streaming), yellow (all interacting, high $\zdec$ mode), and pink (all interacting, low $\zdec$ mode).
\end{itemize}

When considering the full set of Planck primary anisotropies and lensing (see Figure \ref{fig:results:P_Pb_R1z6m}, purple and pink), referred to here as P18 +lens, we find the data allows for a low $\zdec$ mode with $\log_{10}(\zdec)=4.14 \pm 0.58$ ($\log_{10}(\zdec)=4.14 \pm 0.56$ when including BAO). Note, however, that this mode is disfavored by the data. In itself, the low $\zdec$ mode is a slightly worse fit to the data compared to the free-streaming and high $\zdec$ cases (see Table~\ref{results:tab:P_Pb_R1z6m}), which is unfortunate considering that we have added a free parameter. Additionally, once we consider the parameter volume effects using Bayesian evidences this mode is further disfavored: we have a large allowed parameter space for $\zdec$, where the posterior for the high $\zdec$ mode (Figure \ref{fig:results:P_Pb_R1z6m}, blue and yellow) flattens out above $\zdec \sim 10^{6}$ and is very slowly increasing all the way to standard neutrino decoupling ($z \sim 10^{9}$). This is because variations in $\zdec$ values above $\zdec \gtrsim 10^{6}$ do not have distinguishable effects on observables so that values of $\zdec \sim 10^6$ produce power spectra that closely resemble those for free-streaming neutrinos. As such, the high $\zdec$ mode gives us a bound of $\log_{10}(\zdec) > 5.2$ for P18 +lens and $\log_{10}(\zdec) > 5.1$ for P18 +lens +BAO (both 68\%CL, note that these bounds depend weakly on the prior range as pointed out in the Table~\ref{results:tab:P_Pb_R1z6m} caption) and has a cosmology fairly similar to a free-streaming one and the data shows no preference for it over a free-streaming case.

\begin{table}[t]
	\scriptsize
	\centering
	\begin{tabular}{ l | c c | c c c c }
		& \multicolumn{2}{ c | }{Free-streaming} & \multicolumn{4}{ c }{Self-interacting (Case 1)} \\
		& P18 +lens & +BAO & \multicolumn{2}{ c }{P18 +lens} & \multicolumn{2}{ c }{+BAO} \\
		& & & mode 1 & mode 2 & mode 1 & mode 2  \\ \hline
		$\omega_{b}$ & $0.02219 \pm 0.00022$ & $0.02234 \pm 0.00019$ & $0.02219 \pm 0.00022$ & $0.02226 \pm 0.00020$ & $0.02233 \pm 0.018$ & $0.02230 \pm 0.0017$ \\
		$\omega_{cdm}$ & $0.1177 \pm 0.0029$ & $0.1179 \pm 0.0028$ & $0.1180 \pm 0.0029$ & $0.1136 \pm 0.0024$ & $0.1182 \pm 0.0029$ & $0.1135 \pm 0.0025$ \\
		$100 \times \theta_s$ & $1.04226 \pm 0.00051$ & $1.04217 \pm 0.00050$ & $1.04225 \pm 0.00051$ & $1.04679 \pm 0.00055$ & $1.04217 \pm 0.00049$ & $1.04678 \pm 0.00055$ \\
		$\ln({10}^{10} A_s)$ & $3.037 \pm 0.017$ & $3.042 \pm 0.017$ & $3.035 \pm 0.017$ & $2.967 \pm 0.014$ & $3.040 \pm 0.016$ & $2.967 \pm 0.014$ \\
		$n_s$ & $0.9573 \pm 0.0085$ & $0.9631 \pm 0.0071$ & $0.9560 \pm 0.0085$ & $0.9209 \pm 0.0061$ & $0.9613 \pm 0.0071$ & $0.9226 \pm 0.0055$ \\
		$z_{reio}$ & $7.57 \pm 0.76$ & $7.75 \pm 0.73$ & $7.56 \pm 0.76$ & $7.45 \pm 0.67$ & $7.74 \pm 0.72$ & $7.48 \pm 0.65$ \\
		$\log_{10}$$(\zdec)$ & --- & --- & $> 5.2$ (95\%CL) & $ 4.14 \pm 0.058$ & $> 5.1$ (95\%CL) & $4.14 \pm 0.056$ \\
		$\Neff$ & $ 2.82 \pm 0.18 $ & $2.90 \pm 0.17$ & $ 2.84 \pm 0.18$ & $ 2.55 \pm 0.14$ & $2.92 \pm 0.17$ & $ 2.57 \pm 0.14$ \\
		$\meff$ & $< 0.227$ (95\%CL) & $< 0.108$ (95\%CL) & $< 0.225$ (95\%CL) & $< 0.160$ (95\%CL) & $< 0.107$ (95\%CL) & $< 0.108$ (95\%CL) \\
		$H_0$ $\left[\frac{\textrm{(km/s)}}{\textrm{Mpc}}\right]$ & $ 65.8 \pm 1.6 $ & $67.2 \pm 1.1$ & $65.9 \pm 1.7$ & $65.7 \pm 1.3$ & $67.3 \pm 1.1$ & $66.1 \pm 1.0$ \\
		$S_8$ & $ 0.835 \pm 0.013 $ & $0.828 \pm 0.012$ & $0.835 \pm 0.013$ & $0.825 \pm 0.013$ & $0.829 \pm 0.011$ & $0.821 \pm 0.011$ \\ [0.5mm] \hline
		ln($E$) & $-0.5282 \times 10^{3}$ & $-0.5320 \times 10^{3}$ & $-0.5333 \times 10^{3}$ & $-0.5388 \times 10^{3}$ & $-0.5370 \times 10^{3}$ & $-0.5418 \times 10^{3}$ \\
		$E_{\rm int}/E_{\rm fs}$ & --- & --- & $6.1 \times 10^{-3}$ & $2.5 \times 10^{-5}$ & $6.7 \times 10^{-3}$ & $5.5 \times 10^{-5}$ \\ [0.5mm] \hline
		\multicolumn{7}{ c }{Best fit} \\ [0.5mm] \hline
		$\Neff$ & 2.846 & 2.922 & 2.859 & 2.572 & 2.819 & 2.519 \\
		$\log_{10}$$(\zdec)$ & --- & --- & 5.953 & 4.119 & 5.997 & 4.126 \\
		$\chi^2_{\rm{eff}}$ & 1011.08 & 1016.72 & 1011.67 & 1018.35 & 1016.94 & 1023.39 \\
		$\Delta \chi^2_{\rm{eff}}$& --- & --- & +0.59 & +7.27 & +0.22 & +6.67
	\end{tabular}
	\caption{Case 1: statistical information for our baseline configurations. $\chi^2_{\rm{eff}} = -2 \ln \mathcal{L}$ is the minimum effective chi square, $\Delta \chi^2_{\rm{eff}}$ is with regards to the corresponding free-streaming case, ln($E$) is the Bayesian evidence, and $E_{\rm int}/E_{\rm fs}$ is the Bayesian evidence ratio with regards to the corresponding free-streaming case. All credibility intervals are 68\%CL centered around the mean unless otherwise noted. Note that because the 1-d marginalized posterior distribution above $\zdec>10^{6}$ is not quite flat the bound on $\zdec$ for mode 1 (high $\zdec$ mode) is somewhat prior dependent. The results listed in the table were derived with a prior bound of $10^{2} < \zdec < 10^{6}$ as the posterior distrubtion is close to flat beyond this range, this was done in order to speed up sampling while properly resolving the low $\zdec$ mode. If we allow sampling up to around standard neutrino decoupling ($\zdec < 10^{9}$) the bound for P18 +lens and P18 +lens +BAO are both $\log_{10}(\zdec)> 5.5$ (95\%CL), so slightly larger than the values listed in the table. The other bounds are largely unaffected by this prior choice and the shape of the posterior for $\zdec$ shown in Fig.~\ref{fig:results:P_Pb_R1z6m} is also insensitive to the allowed range of $\zdec$.}
	\label{results:tab:P_Pb_R1z6m}
\end{table}
\begin{table}[t]
	\scriptsize
	\centering
	\begin{tabular}{ l | c  c | c }
		& \multicolumn{2}{ c |}{Free-streaming} & \multicolumn{1}{ c }{Self-interacting (Case 1)} \\
		& P18 +lens +BAO & P18 +lens +BAO +R19 & \multicolumn{1}{ c }{P18 +lens +BAO + R19} \\ \hline
		$\omega_b$ & $0.02234 \pm 0.00019$ & $ 0.02270 \pm 0.00016$ & $0.2268 \pm 0.0016$ \\
		$\omega_{cdm}$ & $0.1179 \pm 0.0028$ & $0.1231 \pm 0.0026$ & $0.1235 \pm 0.0027$ \\
		$100 \times \theta_s$ & $1.04217 \pm 0.00050$ & $1.04139 \pm 0.00043$ & $1.04139 \pm 0.00045$ \\
		$\ln({10}$$^{10} A_s)$ & $3.042 \pm 0.017$ & $3.062 \pm 0.016$ & $3.058 \pm 0.016$ \\
		$n_s$ & $0.9631 \pm 0.0071$ & $0.9780 \pm 0.0058$ & $0.9751 \pm 0.0066$ \\
		$z_{reio}$ & $7.75 \pm 0.73$ & $8.10 \pm 0.74$ & $8.07 \pm 0.74$ \\
		$\log_{10}$$(\zdec)$ & --- & --- & $> 5.3$ (68\%CL) \\
		$\Neff$ & $2.90 \pm 0.17$ & $3.28 \pm 0.14$ & $3.30 \pm 0.15$ \\
		$\meff$ & $< 0.108$ (95\%CL) & $< 0.0965$ (95\%CL) & $< 0.102$ (95\%CL) \\
		$H_0$ $\left[\frac{\textrm{(km/s)}}{\textrm{Mpc}}\right]$ & $67.2 \pm 1.1$ & $69.93 \pm 0.92$ & $70.05 \pm 0.94$ \\
		$S_8$ & $0.828 \pm 0.012$ & $0.828 \pm 0.012$ & $0.829 \pm 0.012$ \\ [0.5mm] \hline
		ln($E$) & $-0.5320 \times 10^{3}$ & $-0.5393 \times 10^{3}$ & $-0.5439 \times 10^{3}$ \\
		$E_{\rm int}/E_{\rm fs}$ & --- & --- & $1.0 \times 10^{-2}$ \\ [0.5mm] \hline
		\multicolumn{4}{ c }{Best fit} \\ [0.5mm] \hline
		$\Neff$ & 2.922 & 3.209 & 3.254 \\
		$\log_{10}$$(\zdec)$ & --- & --- & 5.571 \\
		P18 highTTTEEE & 583.23 & 588.54 & 589.17 \\
		P18 lowTT & 23.45 & 21.95 & 22.17 \\
		P18 lowEE & 396.01 & 396.43 & 396.26 \\
		P18 lensing & 8.73 & 9.08 & 9.08 \\
		P18 total & 1011.4 & 1016.0 & 1016.7 \\
		BAO & 5.30 & 5.69 & 5.78 \\
		R19 & --- & 9.14 & 8.17 \\
		$\chi^2_{\rm{eff}}$ & 1016.72 & 1030.83 & 1030.64 \\
		$\Delta \chi^2_{\rm{eff}}$ & --- & --- & -0.19
	\end{tabular}
	\caption{Case 1: statistical information including a prior on $H_0$ from Riess et al.~\cite{Riess:2019cxk}. $\chi^2_{\rm{eff}} = -2 \ln \mathcal{L}$ is the minimum effective chi square, $\Delta \chi^2_{\rm{eff}}$ is with regards to the corresponding free-streaming case, ln($E$) is the Bayesian evidence, and $E_{\rm int}/E_{\rm fs}$ is the Bayesian evidence ratio with regards to the corresponding free-streaming case. All credibility intervals are 68\%CL centered around the mean unless otherwise noted. The baseline case withour R19 is included for comparison, to show how adding R19 degrades the fit to the CMB.}
	\label{results:tab:Pbr_R1z6m}
\end{table}

The low $\zdec$ mode has a significantly different cosmology, as reported by e.g.~\cite{Kreisch:2019yzn}, with wildly different $\theta_s$, $A_s$ and $n_s$ values compared to the free-streaming neutrino comparison case (Figure \ref{fig:results:P_Pb_R1z6m}, red and green) and the high $\zdec$ mode (Figure \ref{fig:results:P_Pb_R1z6m}, blue and yellow). There is, however, virtually no change in $H_0 = 65.7 \pm 1.3$ (km/s)/Mpc for P18 +lens ($H_0 = 66.1 \pm 1.0$ (km/s)/Mpc for P18 +lens +BAO), compared to $H_0 = 65.8 \pm 1.6$ (km/s)/Mpc for P18 +lens ($H_0 = 67.2 \pm 1.1$ (km/s)/Mpc for P18 +lens +BAO) for the free-streaming case, as any increase of $H_0$ allowed by the self-interactions is neatly off-set by a lower value for $\Neff = 2.55 \pm 0.14$ for P18 +lens ($\Neff = 2.57 \pm 0.14$ for P18 +lens +BAO) compared to the free-streaming value of $\Neff=2.82 \pm 0.18$ for P18 +lens ($\Neff=2.90 \pm 0.17$ for P18 +lens +BAO). On its own, with the baseline data configurations, all neutrinos self-interacting does not help alleviate the $H_0$ tension.

Since this model has been proposed as a solution to the Hubble tension~\cite{Kreisch:2019yzn}, let us examine if that picture changes if we include a prior on the Hubble parameter from Riess et al.~\cite{Riess:2019cxk} (R19) of $H_0 = 74.04 \pm 1.42$ (km/s)/Mpc (see Figure \ref{fig:results:Pbr_R1z6m}), and if such a combination is consistent in the first place.

Perhaps surprisingly, once we include the $H_0$ prior, the low $\zdec$ mode is ruled out by the data (Figure \ref{fig:results:Pbr_R1z6m}, cyan).
We already saw a hint this might happen from Figure \ref{fig:results:P_Pb_R1z6m}. In order to accommodate a larger $H_0$ value, we need to increase the effective number of relativistic species $\Neff$. However, because the effect on observables when all species are interacting is so strong, the data does not allow for a large amount of it, and instead the mode is ruled out and we are left with a bound $\log_{10}(\zdec) > 5.3$ (68\% CL). For the remaining high $\zdec$ mode (for which the posterior is approximately flat from the edge of the plotted parameter space to standard neutrino decoupling), we find $\Neff=3.30 \pm 0.15$, which is comparable to the free-streaming value of $\Neff=3.28 \pm 0.14$ and is much larger than the low $\zdec$ value without R19 from before of $\Neff = 2.55 \pm 0.14$ for P18 +lens ($\Neff = 2.57 \pm 0.14$ for P18 +lens +BAO).

Finally, comparing the self-interacting case (cyan) and free-streaming comparison case (light green) from Figure \ref{fig:results:Pbr_R1z6m}, it is clear that neutrino self-interactions are not a solution to the Hubble parameter tension. The $H_0=70.05 \pm 0.94$ (km/s)/Mpc value inferred in the self-interacting case is almost identical to that for the free-streaming case $H_0=69.93 \pm 0.92$ (km/s)/Mpc (the curves are neatly on top of one another, making it hard to tell them apart). The higher value compared to the cases without R19 is simply due to combining discrepant datasets, where the value for $H_0$ is increased at the expense of worsening the fit to the CMB data (see Table~\ref{results:tab:Pbr_R1z6m}, where the best-fit chi square contribution from Planck is 1011.4 for the free-streaming case without R19, compared to 1016.0 and 1016.7 when including R19 for the free-streaming and interacting cases, respectively). Moreover, neutrino self-interactions do not help improve the consistency with the Riess et al. $H_0$ value compared to simply adding free-streaming neutrinos. For any case where this is true we recommend viewing any such combination of discrepant datasets with caution.

\subsubsection{Removing high-$\ell$ polarization}
\begin{figure}[tb]
	\centering
	\includegraphics[width=8.4cm]{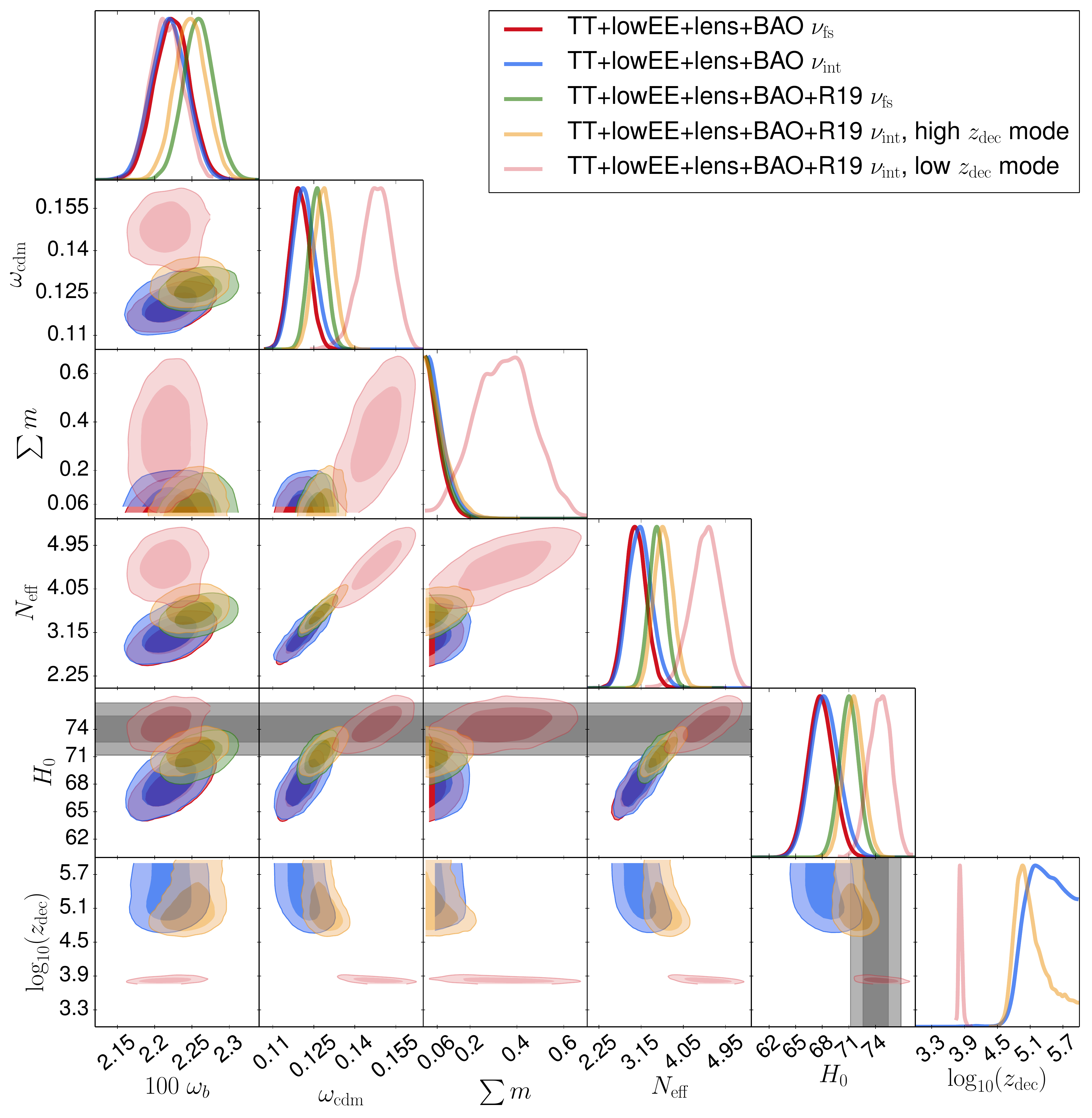}
	\hspace{0.1cm}\includegraphics[width=8.4cm]{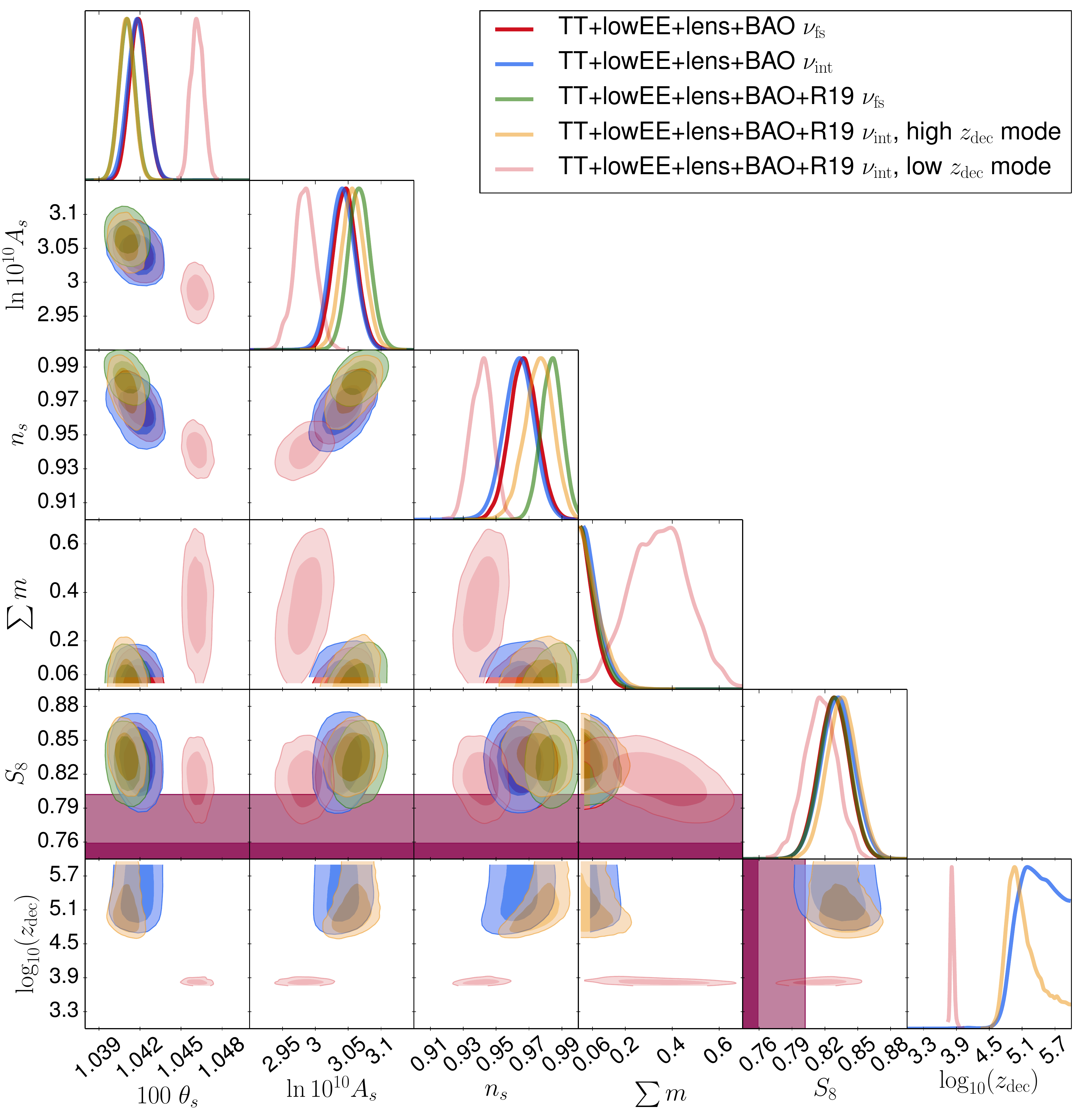}
	\caption{\textbf{Case 1: all species interacting, no high-$\ell$ polarization.} All free-streaming (red and green) vs all species interacting separated into high $\zdec$ (blue and yellow) and low $\zdec$ modes (pink) for Planck-only with no high-$\ell$ polarization (red and blue) and the same plus BAO (green, yellow and pink). The grey bands correspond to the $H_0$ measurement from Riess et al.~\cite{Riess:2019cxk}, while the purple band is the $S_8$ measurement from KiDS+VIKING-450~\cite{Wright:2020ppw}. See Appendix \ref{app:full_plots}, Figure \ref{fig:app:Pbnhr_R1z6m} for the full parameter space plot. \textbf{Left and right panel:} both show the same cases, but different cosmological parameters. \textbf{Left panel}: energy density parameters. \textbf{Right panel:} power spectrum shape and amplitude parameters. Note that beyond $\zdec > 10^6$ the 1-d posterior is approximately flat up to standard neutrino decoupling. It can be hard to see some contours, e.g. red and blue, as they are nearly overlapping as there is only a small difference between the free-streaming comparison case and the high $\zdec$ mode. In agreement with~\cite{Kreisch:2019yzn}, once we exclude high-$\ell$ polarization and include a prior on $H_0$ we get a dramatically different cosmology for the low $\log_{10}(\zdec)=3.83 \pm 0.03$ mode (pink), including a very large neutrino mass sum $\meff=0.35 \pm 0.13$ (68\%CL), a very large number of extra relativistic species $\Neff=4.53 \pm 0.32$ (68\% CL), a low $S_8=0.816 \pm 0.015$ value closer to those of late time weak lensing surveys such as KiDS and DES, and a very high $H_0=74.5 \pm 1.2$ (km/s)/Mpc in perfect agreement with Riess et al.~\cite{Riess:2019cxk} (in fact, the value of $H_0$ we find is larger and in even better agreement with Riess et al. than the one reported in \cite{Kreisch:2019yzn}).}
	\label{fig:results:Pbnhr_R1z6m}
\end{figure}
In order to fully compare to~\cite{Kreisch:2019yzn}, we also produce parameter constraints removing high-$\ell$ polarization data. This is the choice of datasets that produced the most convincing argument for neutrino self-interactions as a solution to the Hubble tension~\cite{Kreisch:2019yzn}. Note that although there have been questions about Planck high-$\ell$ polarization in the past  \cite{Aghanim:2015xee}, we do not have a convincing argument for excluding high-$\ell$ polarization from the analysis at present and leave the interpretation of that choice and these results up to the reader.\\
\\
\noindent In this subsection, we refer to these figures, datasets, and configurations:
\begin{itemize}
\item Figure \ref{fig:results:Pbnhr_R1z6m}. TT +lowEE +lens +BAO (no high-$\ell$ polarization). Red (all free-streaming) and blue (all interacting).
\item Figure \ref{fig:results:Pbnhr_R1z6m}. TT +lowEE +lens +BAO +R19 (no high-$\ell$ polarization). Green (all free-streaming), yellow (all interacting, high $\zdec$ mode), and pink (all interacting, low $\zdec$ mode).
\end{itemize}

High-$\ell$ polarization strongly constrains the number of relativistic degrees of freedom, $\Neff$, and the neutrino mass sum, $\meff$. Additionally, neutrino self-interactions leave a clear signature on the polarization power spectrum. Therefore, when removing high-$\ell$ polarization the picture changes dramatically (see Figure \ref{fig:results:Pbnhr_R1z6m}), as reported by~\cite{Kreisch:2019yzn}. In this section, we update their results for Planck 2018.

When all relativistic species are self-interacting and the prior on the Hubble parameter, $H_0$, from Riess et al.~\cite{Riess:2019cxk} is included (Figure \ref{fig:results:Pbnhr_R1z6m}, yellow and pink contours), the data allows for a very high value of $\Neff=4.53 \pm 0.32$ for the low $\zdec$ mode (\ref{fig:results:Pbnhr_R1z6m}, pink contours), which in turn leads to a high value for $H_0 = 74.5 \pm 1.2$ (km/s)/Mpc, which is larger and in even better agreement with Riess et al. than the one reported in \cite{Kreisch:2019yzn} and is even slightly larger than the Riess et al.~\cite{Riess:2019cxk} measurement of $H_0 = 74.04 \pm 1.42$ (km/s)/Mpc and completely eliminates the Hubble parameter tension and indicating the combination of datasets is valid (with the strong assumption that there is a problem with the Planck high-$\ell$ polarization).

These remarkably high values for $\Neff$ and $H_0$ are only allowed because of the strong neutrino self-interactions of the low $\zdec$ mode and are a consequence of the $H_0$ prior. In comparison, still with the $H_0$ prior is included, the high $\zdec$ mode (Figure \ref{fig:results:Pbnhr_R1z6m}, yellow contours) has much lower values of $\Neff=3.63 \pm 0.20$ and $H_0 = 71.5 \pm 1.1$ (km/s)/Mpc. Similarly, when all species are free-streaming we find $\Neff=3.50 \pm 0.18$ and $H_0 = 71.0 \pm 1.1$ (km/s)/Mpc (Figure \ref{fig:results:Pbnhr_R1z6m}, green contours). In both of these examples, the $H_0$ constraint is simply the result of combining two discrepant datasets making the combination suspect, as we are degrading the fit to the CMB data in order to accommodate a larger $H_0$ value. The degradation to the fit to CMB data can be seen from the individual $\chi_{\rm eff}^2$ contributions from the different datasets shown in Table~\ref{results:tab:Pbnhr_R1z6n}, where the fit to the CMB data worsens when R19 is added, in particular Planck high-$\ell$ temperature auto-correlation (P18 highTT) and Planck lensing, with a fairly large chi square contribution increase from all the Planck likelihoods for the best fit cosmology from 633.5 without R19 to 638.5 with R19 for the free-streaming case and to 636.4 with R19 for the high $\zdec$ mode. So even though the $H_0$ tension is alleviated, it comes at the expense of a worse fit to the CMB data.

For the high $\zdec$ mode we do find a better fit overall for the best fit cosmology, with a $\Delta\chi_{\rm eff}^2=-2.98$ compared to the free-streaming case including R19, but this is offset by the increased complexity of the model as seen from the Bayesian evidence ratio $E_{\rm int}/E_{\rm fs} = 2.7 \times 10^{-2}$, where a value less than 1 means the free-streaming comparison model is favored. However, for the low $\zdec$ mode it looks even worse. The $H_0$ tension is completely eliminated, but the fit to the CMB data is severely degraded with a chi square contribution from the Planck likelihoods of 647.5, resulting in an overall best fit $\Delta\chi_{\rm eff}^2=+3.98$ compared to the free-streaming comparison case (with R19), resulting in a lower Bayesian evidence ratio of $E_{\rm int}/E_{\rm fs} = 2.7 \times 10^{-4}$. So although this model is technically allowed by the data, it is disfavored compared to a free-streaming scenario. If we disregard the $H_0$ prior we instead find a bound on $\log_{10}(\zdec)>4.9$ (95\% CL) (Figure \ref{fig:results:Pbnhr_R1z6m}, blue contours) and lower values of $\Neff=3.13 \pm 0.25$ and $H_0 = 68.3 \pm 1.5$ (km/s)/Mpc for all extra relativistic species self-interacting and $\Neff=3.04 \pm 0.22$ and $H_0 = 67.8 \pm 1.4$ (km/s)/Mpc for all extra relativistic species free-streaming.

\begin{table}[htb]
	\scriptsize
	\centering
	\begin{tabular}{ l | c c | c c c }
		& \multicolumn{2}{ c | }{Free-streaming} & \multicolumn{3}{ c }{Self-interacting (Case 1)} \\
		& TT +lowEE +lens +BAO & +R19 & \multicolumn{1}{ c }{TT +lowEE +lens +BAO} & \multicolumn{2}{ c }{+R19} \\
		& & & & mode 1 & mode 2  \\ \hline
		$\omega_{b}$ & $0.02223 \pm 0.00023$ & $0.02259 \pm 0.00020$ & $0.02219 \pm 0.00024$ & $0.02248 \pm 0.021$ & $0.02216 \pm 0.0021$ \\
		$\omega_{cdm}$ & $0.1199 \pm 0.0036$ & $0.1263 \pm 0.0031$ & $0.1213 \pm 0.0041$ & $0.1286 \pm 0.0036$ & $0.1478 \pm 0.0057$ \\
		$100 \times \theta_s$ & $1.04189 \pm 0.00063$ & $1.04204 \pm 0.00054$ & $1.04103 \pm 0.00056$ & $1.04217 \pm 0.00049$ & $1.04617 \pm 0.00046$ \\
		$\ln({10}$$^{10} A_s)$ & $3.045 \pm 0.018$ & $3.067 \pm 0.017$ & $3.041 \pm 0.018$ & $3.057 \pm 0.018$ & $2.984 \pm 0.016$ \\
		$n_s$ & $0.9668 \pm 0.0085$ & $0.9839 \pm 0.0065$ & $0.9639 \pm 0.0092$ & $0.9760 \pm 0.0087$ & $0.9411 \pm 0.0067$ \\
		$z_{reio}$ & $7.74 \pm 0.76$ & $8.10 \pm 0.78$ & $7.69 \pm 0.77$ & $7.95 \pm 0.78$ & $7.63 \pm 0.87$ \\
		$\log_{10}$$(\zdec)$ & --- & --- & $> 4.9$ (95\%CL) & $ 5.15^{+0.40}_{-0.16}$ & $3.83 \pm 0.03$ \\
		$\Neff$ & $ 3.04 \pm 0.22 $ & $3.50 \pm 0.18$ & $ 3.13 \pm 0.25$ & $3.63 \pm 0.20$ & $ 4.53 \pm 0.32$ \\
		$\meff$ & $< 0.130$ (95\%CL) & $< 0.144$ (95\%CL) & $< 0.151$ (95\%CL) & $< 0.173$ (95\%CL) & $0.35 \pm 0.13$ \\
		$H_0$ $\left[\frac{\textrm{(km/s)}}{\textrm{Mpc}}\right]$ & $ 67.8 \pm 1.4 $ & $71.0 \pm 1.0$ & $68.3 \pm 1.5$ & $71.5 \pm 1.1$ & $74.5 \pm 1.2$ \\
		$S_8$ & $ 0.828 \pm 0.014 $ & $0.829 \pm 0.014$ & $0.832 \pm 0.015$ & $0.836 \pm 0.014$ & $0.816 \pm 0.015$ \\ [0.5mm] \hline
		ln($E$) & $-0.3419 \times 10^{3}$ & $-0.3468 \times 10^{3}$ & $-0.3462 \times 10^{3}$ & $-0.3504 \times 10^{3}$ & $-0.3550 \times 10^{3}$ \\
		$E_{\rm int}/E_{\rm fs}$ & --- & --- & $1.4 \times 10^{-2}$ & $2.7 \times 10^{-2}$ & $2.7 \times 10^{-4}$ \\ [0.5mm] \hline
		\multicolumn{6}{ c }{Best fit} \\ [0.5mm] \hline
		$\Neff$ & 2.971 & 3.494 & 3.123 & 3.591 & 4.653 \\
		$\log_{10}$$(\zdec)$ & --- & --- & 5.224 & 4.970 & 3.8208 \\
		P18 highTT & 205.34 & 211.24 & 204.95 & 208.11 & 216.01 \\
		P18 lowTT & 23.58 & 21.56 & 23.95 & 22.74 & 24.44 \\
		P18 highEE & --- & --- & --- & --- & --- \\
		P18 lowEE & 395.77 & 396.28 & 395.75 & 396.15 & 395.87 \\
		P18 lensing & 8.81 & 9.46 & 8.88 & 9.36 & 11.20 \\
		P18 total & 633.5 & 638.5 & 633.5 & 636.4 & 647.5 \\
		BAO & 5.40 & 6.54 & 5.25 & 6.53 & 4.96 \\
		R19 & --- & 3.75 & --- & 2.97 & 0.33 \\
		$\chi^2_{\rm{eff}}$ & 638.89 & 648.83 & 638.79 & 645.85 & 652.81 \\
		$\Delta \chi^2_{\rm{eff}}$& --- & --- & -0.10 & -2.98 & +3.98
	\end{tabular}
	\caption{Case 1: statistical information when excluding high-$\ell$ polarization. $\chi^2_{\rm{eff}} = -2 \ln \mathcal{L}$ is the minimum effective chi square, $\Delta \chi^2_{\rm{eff}}$ is with regards to the corresponding free-streaming case, ln($E$) is the Bayesian evidence, and $E_{\rm int}/E_{\rm fs}$ is the Bayesian evidence ratio with regards to the corresponding free-streaming case. All credibility intervals are 68\%CL centered around the mean unless otherwise noted.}
	\label{results:tab:Pbnhr_R1z6n}
\end{table}

\subsection{Case 2: Two free-streaming species plus interacting species}
\label{sec:results:two_fs}
Considering the bounds on electron and muon self-interactions pointed out by~\cite{Blinov:2019gcj}, we want to consider a case of two free-streaming neutrinos plus one interacting neutrino, while simultaneously allowing for additional interacting species. We use the normal hierarchy approximation for the neutrino masses\footnote{Known to not be reliable enough for precise neutrino mass sum estimates (see e.g.~\cite{Lesgourgues:2006nd,DiValentino:2016foa,Vagnozzi:2017ovm}), but nevertheless good enough for our purposes as the mass constraint is not the focus of our study.}, treating the two lighter mass states as massless. This should allow us to approximately equate a flavor state with a mass state, a translation that might otherwise be non-trivial. The extra relativistic species share the mass of the massive standard neutrino.

\begin{figure}[htb]
	\centering
	\includegraphics[width=8.4cm]{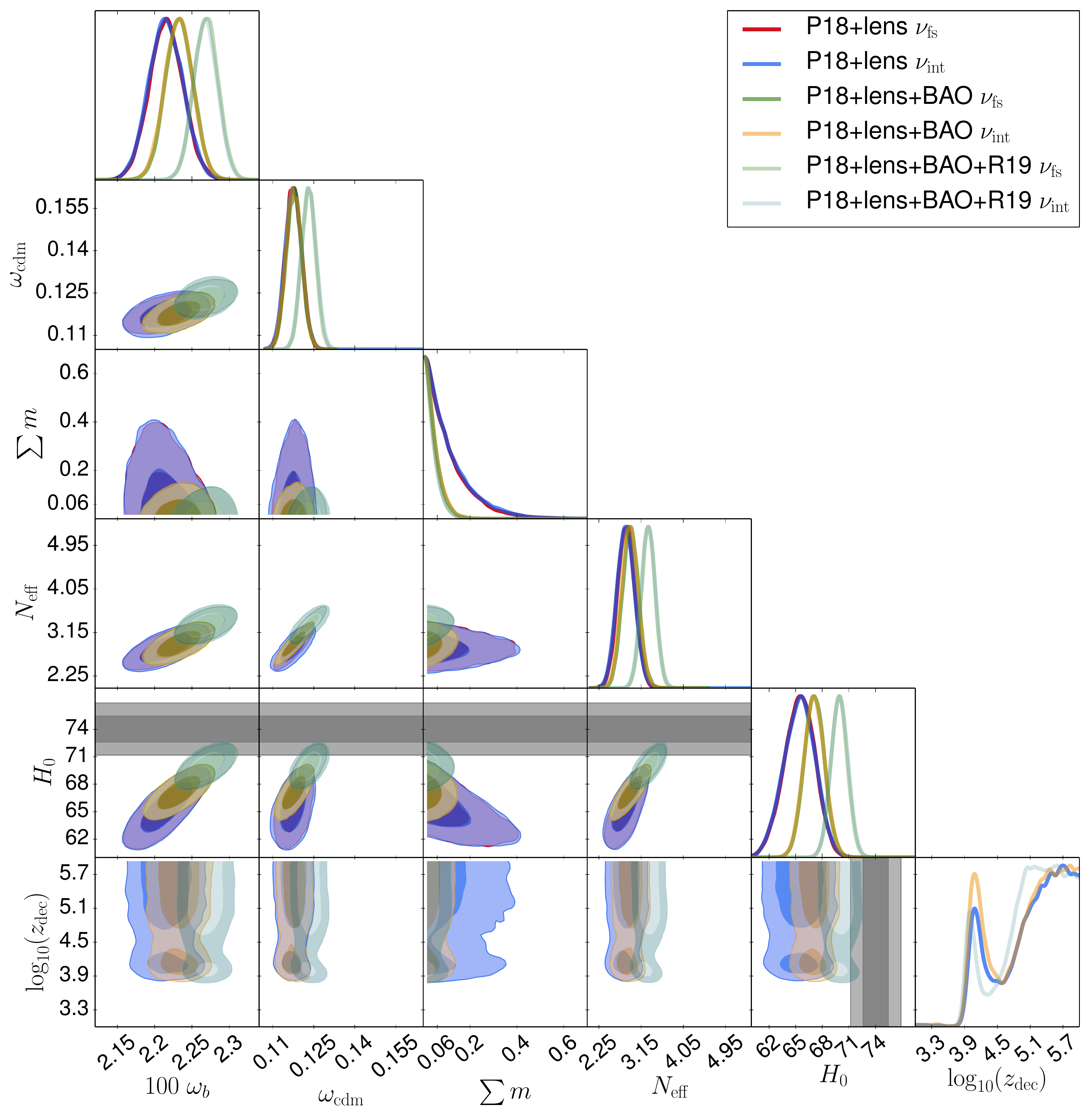}
	\hspace{0.1cm}\includegraphics[width=8.4cm]{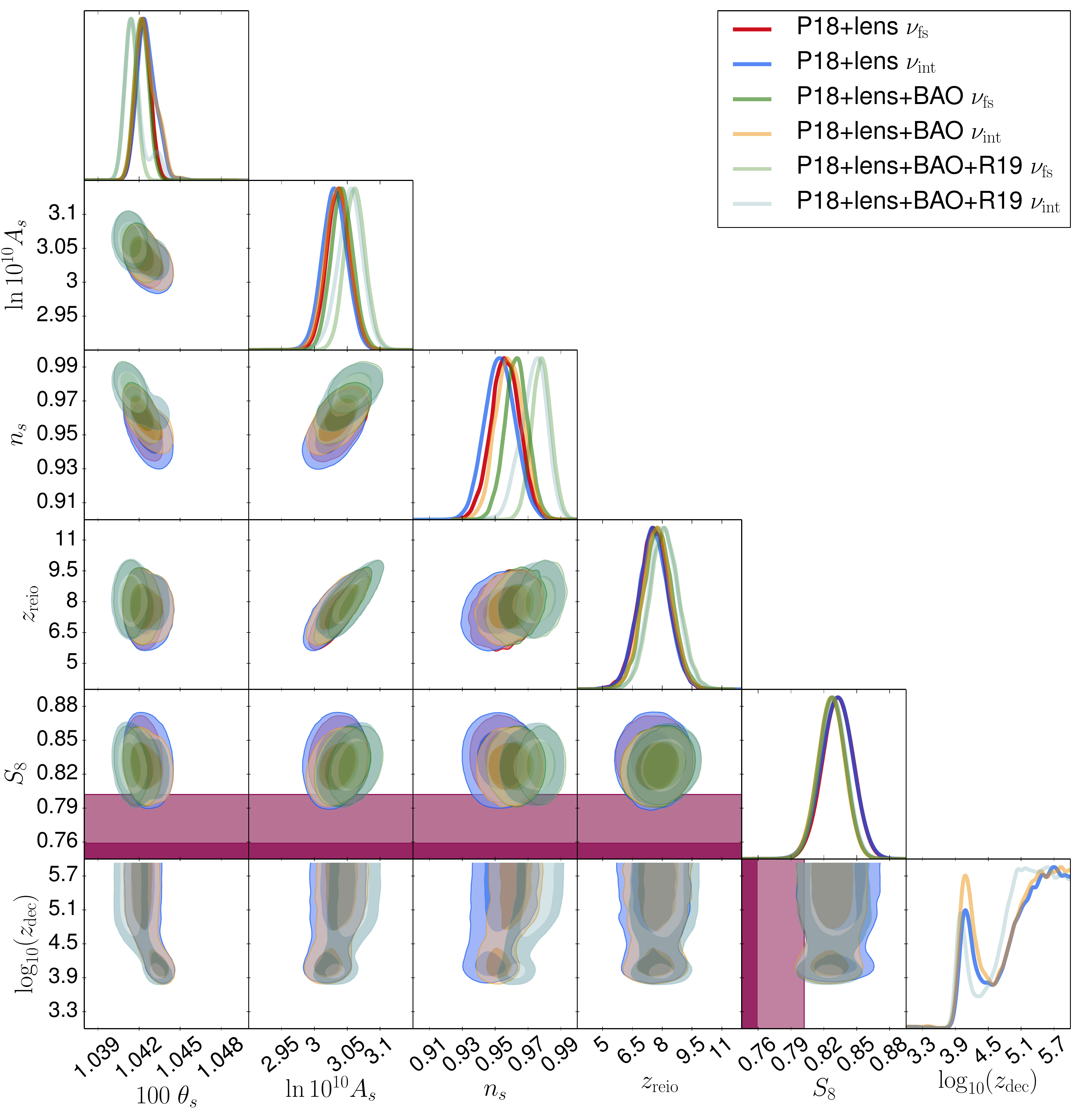}
	\caption{\textbf{Case 2: two free-streaming neutrinos plus self-interacting relativistic species.} All free-streaming (red, green and light green) vs self-interacting species (blue, yellow and cyan) for Planck-only (red and blue), Planck + BAO (green and yellow) and Planck + BAO + $H_0$ prior (light green and cyan). The grey bands correspond to the $H_0$ measurement from Riess et al.~\cite{Riess:2019cxk}, while the purple band is the $S_8$ measurement from KiDS+VIKING-450~\cite{Wright:2020ppw}. See Appendix \ref{app:full_plots}, Figure \ref{fig:app:2FSz6m} for the full parameter space plot. \textbf{Left and right panel:} both show the same cases, but different cosmological parameters. \textbf{Left panel}: energy density parameters. \textbf{Right panel:} power spectrum shape and amplitude parameters. Note that beyond $\zdec > 10^6$ the 1-d posterior is approximately flat up to standard neutrino decoupling. Several of the contours are hard to see as the self-interacting cases are nearly on top of the free-streaming ones, i.e. red and blue, green and yellow, as well as light green and cyan, indicating that this case is not a solution to the Hubble tension. Even though we find a low $\zdec$ mode, it shows no improvement compared to the free-streaming comparison case, even when including the $H_0$ prior (on the left panel the cyan contour is directly on top of the light green one). Indeed, this is further confirmed when considering the low $\zdec$ part of parameter space (bottom row), where we see the low $\zdec$ mode separated at greater than $1\sigma$ from the high $\zdec$ mode, but still showing a low value for $H_0$ in line with the free-streaming case (note that this makes the inclusion of the $H_0$ prior suspect and those results should be regarded with caution). Similarly to Case 1 (Section~\ref{sec:results:all_interacting}), the low $\zdec$ mode has a significantly different cosmology than the free-streaming comparison case, including a much larger $\theta_s$, related to the scale of the first peak, and lower values for the primordial power spectrum tilt, $n_s$, and amplitude, $A_s$.}
	\label{fig:results:2FSz6m}
\end{figure}
\begin{figure}[htb]
	\centering
	\includegraphics[width=8.4cm]{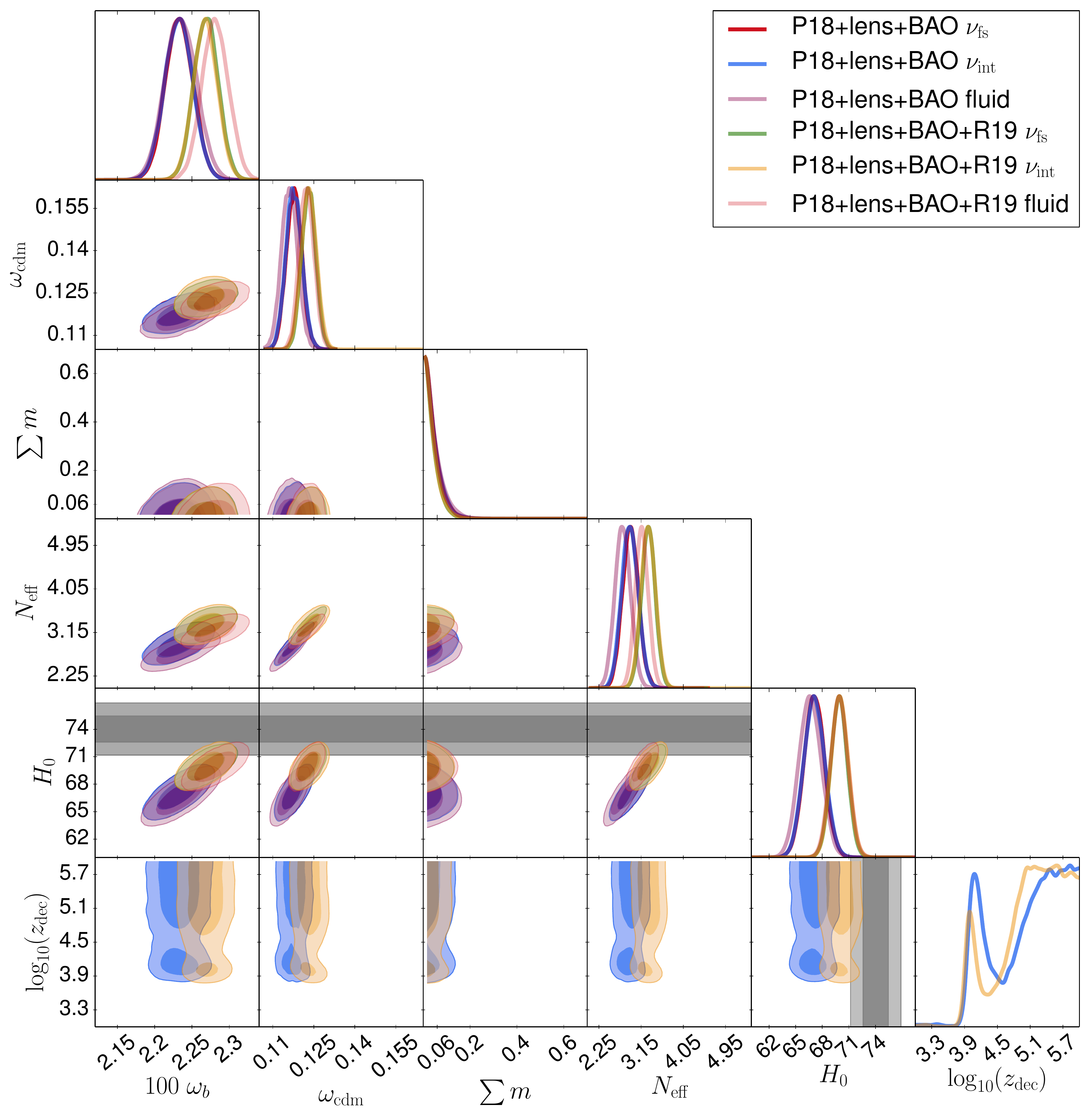}
	\hspace{0.1cm}\includegraphics[width=8.4cm]{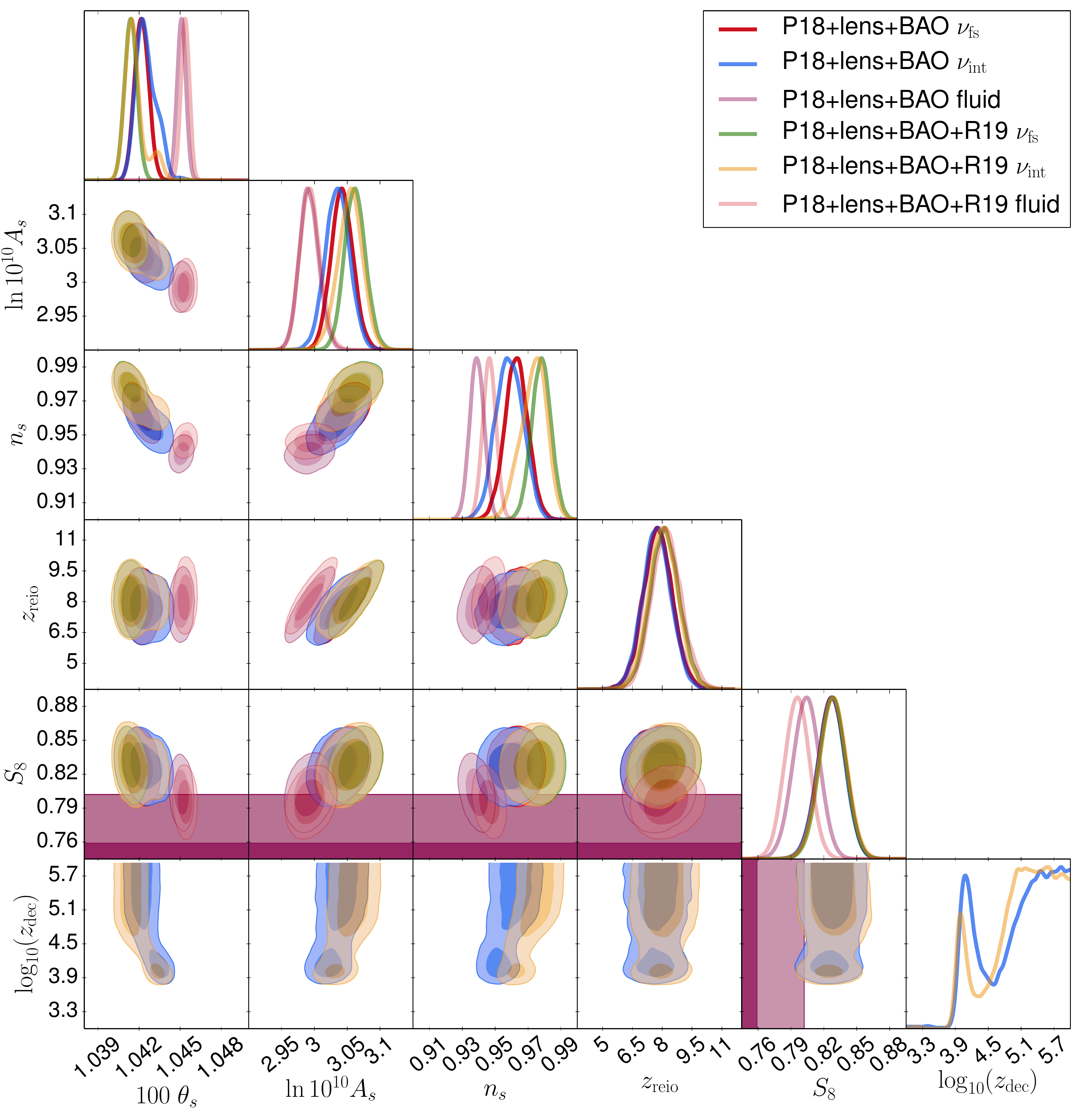}
	\caption{\textbf{Case 2: two free-streaming neutrinos plus self-interacting relativistic species, including fluid-like.} All free-streaming (red and green) vs self-interacting species (blue and yellow) and fluid-like (purple and pink) for Planck + BAO (red, blue and purple) and Planck + BAO + $H_0$ prior (green, yellow and pink). The grey bands correspond to the $H_0$ measurement from Riess et al.~\cite{Riess:2019cxk}, while the purple band is the $S_8$ measurement from KiDS+VIKING-450~\cite{Wright:2020ppw}. See Appendix \ref{app:full_plots}, Figure \ref{fig:app:2FSfl} for the full parameter space plot. \textbf{Left and right panel:} both show the same cases, but different cosmological parameters. \textbf{Left panel}: energy density parameters. \textbf{Right panel:} power spectrum shape and amplitude parameters. Note that beyond $\zdec > 10^6$ the 1-d posterior is approximately flat up to standard neutrino decoupling. Some contours are hard to see as they are neatly on top of each other, i.e. red and blue as well as green and yellow. In this figure we include a fluid-like case, corresponding to a species decoupling after recombination up until today, in the future, or never. The fluid-like cases reduces the $S_8$ tension to about $2\sigma$ for the case without the $H_0$ prior (note that although the largest improvement is for the case with the $H_0$ prior, this combination of datasets is suspect as this case does not alleviate the $H_0$ tension) and displays a different cosmology similar to the low $\zdec$ mode, but with parameter values for $\theta_s$, $n_s$, and $A_s$ even further away from the free-streaming case.}
	\label{fig:results:2FSfl}
\end{figure}

\newpage
\noindent In this section, we refer to these figures, datasets, and configurations:
\begin{itemize}
	\item Figure \ref{fig:results:2FSz6m}. P18 +lens. Red (all free-streaming) and blue (partially interacting).
	\item Figure \ref{fig:results:2FSz6m}. P18 +lens +BAO. Green (all free-streaming) and yellow (partially interacting).
	\item Figure \ref{fig:results:2FSz6m}. P18 +lens +BAO +R19. Light green (all free-streaming) and cyan (partially interacting).
	\item Figure \ref{fig:results:2FSfl}. P18 +lens +BAO. Red (all free-streaming), blue (partially interacting), and purple (partially fluid-like).
	\item Figure \ref{fig:results:2FSfl}. P18 +lens +BAO +R19. Green (all free-streaming), yellow (partially interacting), and pink (partially fluid-like).
\end{itemize}

We find the strongly interacting mode at $\log_{10}(\zdec) \approx 4.09$ from the previous Section~\ref{sec:results:all_interacting}, although with less significance than shown there (MultiNest does not consider it a separate mode as the intermediate $\zdec$ values are not ruled out). Note that beyond $\log_{10}(\zdec) \gtrsim 5.5$ the posterior continues to be approximately flat all the way to standard neutrino decoupling, as the data prefers free-streaming neutrinos and cannot distinguish between free-streaming and slightly-interacting species with such an early decoupling time. Given the shape of the posterior (see Figure~\ref{fig:results:2FSz6m}), highly non-Gaussian with two modes not clearly separated in parameter space, any attempt to derive bounds on $\zdec$ will naturally be prior dependent, so we instead refer the reader to the $\zdec$ posterior on Figure~\ref{fig:results:2FSz6m} and the discussion in the caption of Table~\ref{results:tab:2FSz6m}.

Although the low $\zdec$ mode does have a significantly different cosmology (most notably with larger $\theta_s$ and lower $A_s$ and $n_s$ values), this case does not help with the current cosmological tension related to $H_0$. For all combinations of datasets the free-streaming and interacting lines are neatly on top of each other in the 1-d plot for $H_0$ (left panel of Figure~\ref{fig:results:2FSz6m}), i.e. red and blue lines for Planck-only (P18 +lens), green and yellow for P18 +lens +BAO and cyan and light green for P18 +lens +BAO +R19. This is further confirmed when considering the 2-d plot for $H_0$-$\zdec$ and comparing the high and low $\zdec$ part of parameter space, where $H_0$ for the low $\zdec$ part of parameter space is actually very slightly lower. This can be understood by the self-interactions being slightly disfavored by the data and therefore allowing for a lower value of $\Neff$ (which reduces the effect of the interactions as illustrated by Figure~\ref{fig:varying_Neff}), which in turn leads to a smaller value for $H_0$. The fluid-like case (i.e. setting decoupling to today) similarly does not help with the $H_0$ tension (see Figure \ref{fig:results:2FSfl}, left panel). Note that~\cite{Blinov:2020hmc} also considered fluid-like radiation, but did not consider the same cases we do here. 
The $S_8$ tension, on the other hand, although is not improved at all for the marginalized 1-d posterior distribution for the self-interacting case, it is very slightly improved for the low $\zdec$ mode (Figure \ref{fig:results:2FSz6m}, right panel) and significantly improved for the fluid-like case (Figure \ref{fig:results:2FSfl}, right panel, purple and pink contours). Note, however, that a self-interaction this strong would appear to be strongly ruled out for the most massive standard neutrino mass state, unless we can construct a scenario where e.g. these early universe cosmological neutrinos are not exactly the same ones we measure on Earth today.

However, we do not consider late time large-scale structure data in the analysis, so it is hard to say if this shift in $S_8$ helps the model compared to free-streaming neutrinos and extra relativistic species. But when considering only Planck, with or without BAO, we find all of these cases are disfavored compared to free-streaming species (see Table~\ref{results:tab:2FSz6m}). The best fit cosmologies of the self-interacting modes have slightly better $\chi^2_{\rm{eff}}$ values than the free-streaming comparison cases, but this is outweighed by the added complexity of the model resulting in a Bayesian evidence ratio of less than 1.

\begin{table}[tb]
	\tiny
	\centering
	\begin{tabular}{ l | c c c | c c c | c c }
		& \multicolumn{3}{ c | }{Free-streaming} & \multicolumn{3}{ c }{Self-interacting (Case 2)} & \multicolumn{2}{ c }{Self-interacting (fluid-like)} \\
		& P18 +lens & +BAO & +R19 & P18 +lens & +BAO & +R19 & P18 +lens +BAO & +R19 \\ \hline
		$\omega_{b}$ & $0.02216 \pm 0.00023$ & $0.02233 \pm 0.00019$ & $0.02269 \pm 0.00016$ & $0.02215 \pm 0.00023$ & $0.02232 \pm 0.00019$ & $0.02268 \pm 0.00016$ & $0.02235 \pm 0.00021$ & $0.02281 \pm 0.00018$ \\
		$\omega_{cdm}$ & $0.1177 \pm 0.0029$ & $0.1178 \pm 0.0029$ & $0.1231 \pm 0.0026$ & $0.1175^{+0.0029}_{-0.0031}$ & $0.1177 \pm 0.0029$ & $0.1232^{+0.0028}_{-0.0027}$ & $0.1162 \pm 0.0030$ & $0.1224^{+0.0025}_{-0.0029}$\\
		$100 \times \theta_s$ & $1.04229 \pm 0.00051$ & $1.04219 \pm 0.00050$ & $1.04140 \pm 0.00044$ & $1.04257^{+0.00056}_{-0.00082}$ & $1.04254^{+0.00056}_{-0.00093}$ & $1.04173^{+0.00032}_{-0.00088}$ & $1.04510 \pm 0.00032$ & $1.04537 \pm 0.00030$ \\
		$\ln({10}^{10} A_s)$ & $3.036 \pm 0.017$ & $3.042 \pm 0.017$ & $3.062 \pm 0.016$ & $3.031^{+0.019}_{-0.018}$ & $3.035 \pm 0.018$ & $3.054^{+0.020}_{-0.018}$ & $2.992 \pm 0.015$ & $2.994 \pm 0.030$ \\
		$n_s$ & $0.9563 \pm 0.0087$ & $0.9629 \pm 0.0071$ & $0.9781 \pm 0.0059$ & $0.9530^{+0.0092}_{-0.0095}$ & $0.9588^{+0.0084}_{-0.0085}$ & $0.9735^{+0.0091}_{-0.0067}$ & $0.9389 \pm 0.0044$ & $0.9465 \pm 0.0040$ \\
		$z_{reio}$ & $7.58 \pm 0.77$ & $7.78 \pm 0.74$ & $8.12 \pm 0.74$ & $7.56 \pm 0.75$ & $7.73 \pm 0.72$ & $8.06^{+0.72}_{-0.74}$ & $7.81 \pm 0.73$ & $8.24 \pm 0.75$ \\
		$\Neff$ & $2.83 \pm 0.18$ & $2.92 \pm 0.17$ & $3.30 \pm 0.15$ & $2.83^{+0.18}_{-0.19}$ & $2.91^{+0.18}_{-0.17}$ & $3.30^{+0.15}_{-0.16}$ & $2.76 \pm 0.17$ & $3.16 \pm 0.14$\\
		$\meff$ & $< 0.301$ (95\%CL) & $< 0.108$ (95\%CL) & $< 0.095$ (95\%CL) & $<0.312$ (95\%CL) & $<0.110$ (95\%CL) & $<0.097$ (95\%CL) & $< 0.122$ (95\%CL) & $< 0.110$ (95\%CL) \\
		$H_0$ $\left[\frac{\textrm{(km/s)}}{\textrm{Mpc}}\right]$ & $65.4 \pm 1.7$ & $67.1 \pm 1.1$ & $69.9 \pm 0.9$ & $65.4^{+1.9}_{-1.7}$ & $ 67.1 \pm 1.2$ & $70.0 \pm 1.0$ & $66.7 \pm 1.2$ & $69.9 \pm 1.0$ \\
		$S_8$ & $0.833 \pm 0.014$ & $0.827 \pm 0.012$ & $0.828 \pm 0.012$ & $0.832^{+0.015}_{-0.014}$ & $0.827 \pm 0.012$ & $0.828 \pm 0.012$ & $0.804 \pm 0.012$ & $0.796 \pm 0.011$ \\ [0.5mm] \hline
		ln($E$) & $-0.5280 \times 10^{3}$ & $-0.5322 \times 10^{3}$ & $-0.5394 \times 10^{3}$ & $-0.5324 \times 10^{3}$ & $-0.5323 \times 10^{3}$ & $-0.5436 \times 10^{3}$ & $-0.5365 \times 10^{3}$ & $-0.5444 \times 10^{3}$ \\
		$E_{\rm int}/E_{\rm fs}$ & --- & --- & --- & $1.3 \times 10^{-2}$ & $0.86$ & $1.6 \times 10^{-2}$ & $1.4 \times 10^{-2}$ & $6.8 \times 10^{-3}$ \\ [0.5mm] \hline
		\multicolumn{9}{ c }{Best fit} \\ [0.5mm] \hline
		$\Neffint$ & --- & --- & --- & 0.834 & 0.787 & 1.239 & 0.646 & 1.153 \\
		$\log_{10}$$\zdec$ & --- & --- & --- & 5.442 & 4.085 & 5.163 & --- & --- \\
		$\chi^2_{\rm{eff}}$ & 1011.10 & 1016.79 & 1030.98 & 1011.24 & 1016.62 & 1030.85 & 1025.76 & 1041.15 \\
		$\Delta \chi^2_{\rm{eff}}$ & --- & --- & --- & +0.14 & -0.16 & -0.12 & +8.97 & +10.17 \\ [0.5mm] \hline
		\multicolumn{9}{ c }{Second mode best fit} \\ [0.5mm] \hline
		$\Neffint$ & --- & --- & --- & 0.687 & 0.822 & 1.127 & --- & --- \\
		$\log_{10}$$\zdec$ & --- & --- & --- & 4.118 & 5.456 & 4.002 & --- & --- \\
		$\chi^2_{\rm{eff}}$ & --- & --- & --- & 1011.43 & 1016.71 & 1031.39 & --- & --- \\
		$\Delta \chi^2_{\rm{eff}}$ & --- & --- & --- & +0.33 & -0.08 & +0.41 & --- & ---
	\end{tabular}
	\caption{Case 2: statistical information for our baseline configurations and the one including a prior on $H_0$. $\chi^2_{\rm{eff}} = -2 \ln \mathcal{L}$ is the effective chi square, $\Delta \chi^2_{\rm{eff}}$ is with regards to the corresponding free-streaming case, ln($E$) is the Bayesian evidence, and $E_{\rm int}/E_{\rm fs}$ is the Bayesian evidence ratio with regards to the corresponding free-streaming case. $\chi^2_{\rm{eff}}$ is presented for the best fit cosmology (giving the minimum effective chi square) and for the best fit cosmology of the second mode (note that the best fit cosmology does not always correspond to the same $\zdec$ mode). All credibility intervals are 68\%CL centered around the mean unless otherwise noted. Note that the best fit $\Neff = 2.0328 + \Neffint$, is distinct from the mean value of $\Neff$ in line 7. For the $\zdec$ parameter we refer the reader to Figure~\ref{fig:results:2FSz6m}, as the posterior distribution for $\zdec$ is multimodal and highly non-Gaussian, which means attempts to derive bounds on $\zdec$ naturally end up being prior dependent with $\log_{10}(\zdec) > 4.9$ at 68\%CL ($\log_{10}(\zdec) > 3.9$ at 95\%CL) when the sampling range is $10^{2} < \zdec < 10^{6}$, compared to $\log_{10}(\zdec) > 6.1$ at 68\%CL ($\log_{10}(\zdec) > 4.5$ at 95\%CL) when allowing $\zdec$ to vary freely up to around standard neutrino decoupling $\zdec < 10^9$. As the posterior distribution is nearly flat beyond the range $10^{2} < \zdec < 10^{6}$ we chose this prior range in order to accurately resolve the low $\zdec$ mode. The other bounds are largely unaffected by this prior choice.}
	\label{results:tab:2FSz6m}
\end{table}

\subsection{Case 3: Fixed number of relativistic species and varying fraction of interacting species}
\label{sec:results:fixed_Neff}
Simplifying to only including standard model neutrinos, we want to see whether having one or more species that are self-interacting is allowed by current cosmological data. This case includes free-streaming massive neutrinos plus massive interacting species. The total effective number of relativistic species is fixed to that expected from the Standard Model, $\Neff =3.046$, so this case amounts to varying the fraction that is self-interacting, while also varying the total mass sum $\meff$.

\begin{figure}[tb]
	\centering
	\includegraphics[width=8.4cm]{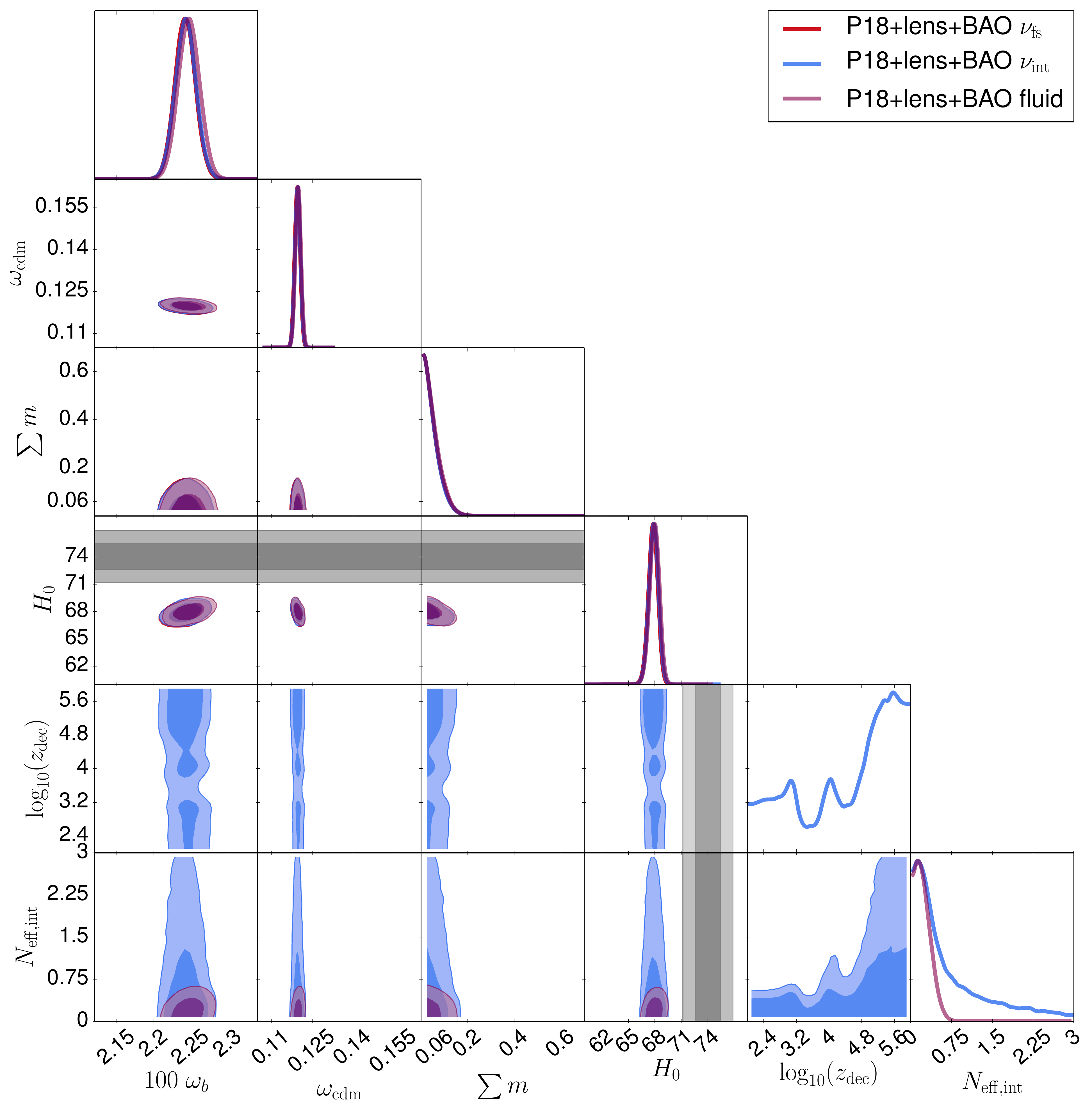}
	\hspace{0.1cm}\includegraphics[width=8.4cm]{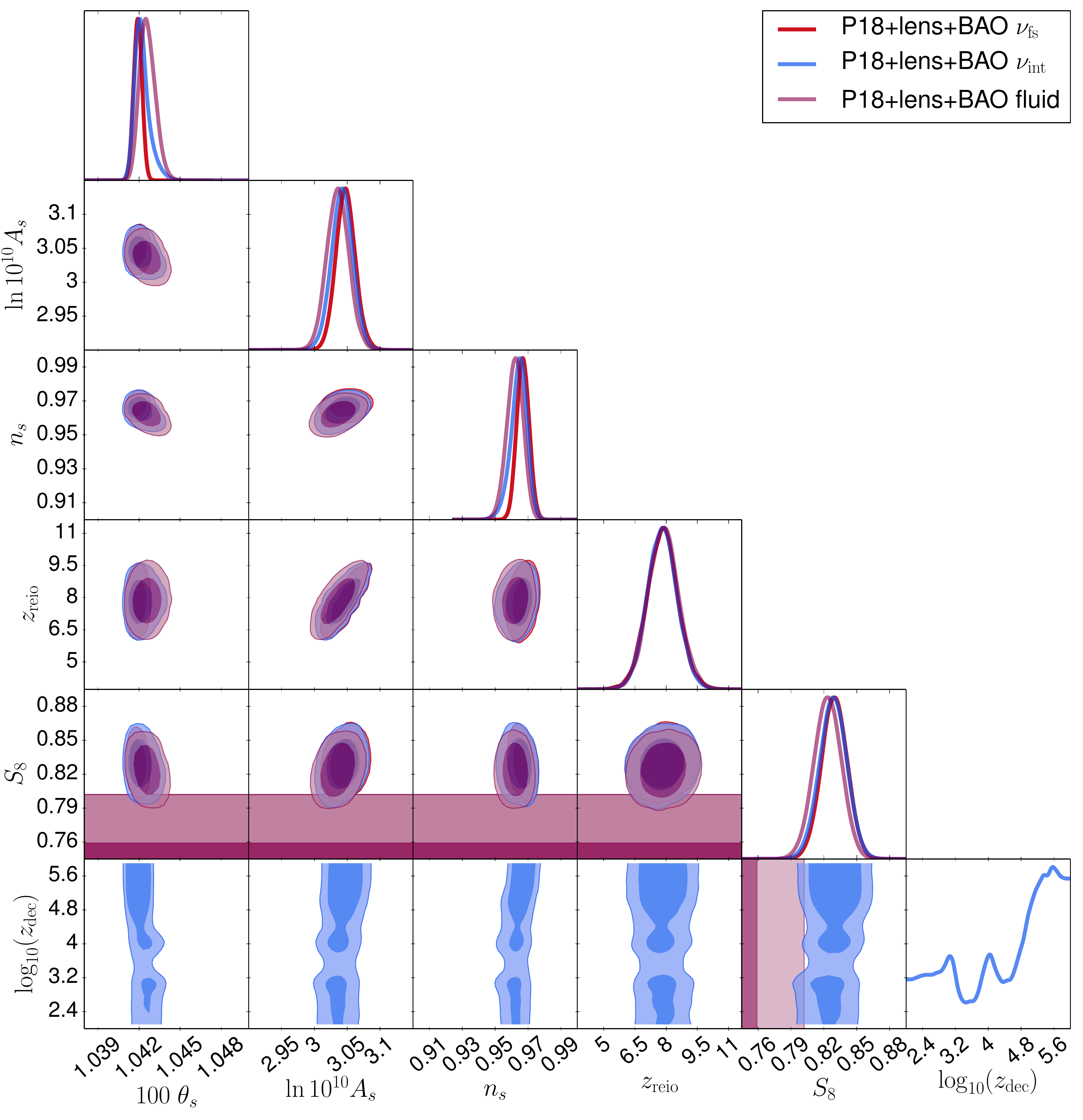}
	\caption{\textbf{Case 3: varying fraction of self-interacting relativistic species, including fluid, with fixed $\mathbf{\Neff=3.046}$.} All free-streaming (red) vs fraction of self-interacting (blue) and fraction of fluid-like (purple) for Planck + BAO. The grey bands correspond to the $H_0$ measurement from Riess et al.~\cite{Riess:2019cxk}, while the purple band is the $S_8$ measurement from KiDS+VIKING-450~\cite{Wright:2020ppw}. See Appendix \ref{app:full_plots}, Figure \ref{fig:app:PSIz6m} for the full parameter space plot. \textbf{Left and right panel:} both show the same cases, but different cosmological parameters. \textbf{Left panel}: energy density parameters. \textbf{Right panel:} power spectrum shape and amplitude parameters. Note that beyond $\zdec > 10^6$ up to standard neutrino decoupling and below $\zdec < 10^2$ the 1-d posterior is approximately flat. For this case any value of $\zdec$ is allowed but is limited to $\Neffint<0.79$ at 68\%CL ($\Neffint<2.34$ at 95\%CL), but high $\zdec$ values are preferred at about $1-\sigma$ (with a bound on $\zdec$ that is strongly prior dependent as discussed in the Table~\ref{results:tab:PSI} caption). Low values of $\zdec$ require a low number of self-interacting extra relativistic species of about $\Neffint \sim 0.5$ at $1\sigma$ (with the fluid-like case allowing for a smaller value of $\Neffint<0.28$ at 68\%CL or $\Neffint < 0.50$ at 95\%CL). Note that around $\zdec \sim \textrm{10,000}$ there is a local maximum that allows up to about $\Neffint \lesssim 1$ at $2\sigma$.}
	\label{fig:results:PSIz6m}
\end{figure}

\noindent In this section, we refer to the following figure, dataset, and configurations:
\begin{itemize}
	\item Figure \ref{fig:results:PSIz6m}. P18 +lens +BAO. Red (all free-streaming), blue (partially interacting), and purple (partially fluid-like).
\end{itemize}
We see a small local maximum at around $\zdec \sim \textrm{10,000}$ (Figure \ref{fig:results:PSIz6m}, left panel), roughly at the $\zdec$ of the strongly interacting mode of the other cases, that allows for about $\Neffint \lesssim 1$ at around $2\sigma$. However, it is clearly disfavored compared to higher values of $\zdec$ that are approximately free-streaming (again the 1-d posterior remains flat from $\zdec= 10^6$ to around standard neutrino decoupling), as $\zdec \gtrsim 10^{5}$ is preferred by the data (with $\Neffint \lesssim 1$, at about $1\sigma$). This picture does not change significantly depending on whether BAO data is included or when considering fluid-like vs strongly interacting, although the bounds on allowed $\Neffint$ values tighten slightly for the fluid-like case. Insofar that bounds were possible to derive they are summarized in Table~\ref{results:tab:PSI}.

\begin{table}[tb]
	\scriptsize
	\centering
	\begin{tabular}{ l | c | c | c }
		& Free-streaming & Self-interacting (Case 3) & Self-interacting (fluid-like) \\
		& P18 +lens +BAO & P18 +lens +BAO & P18 +lens +BAO \\ \hline
		$\omega_{b}$ & $0.02242 \pm 0.00014$ & $0.02243 \pm 0.00014$ & $0.02247 \pm 0.00015$ \\
		$\omega_{cdm}$ & $0.1196 \pm 0.0010$ & $0.1197 \pm 0.0010$ & $0.1199 \pm 0.0010$ \\
		$100 \times \theta_s$ & $1.04191 \pm 0.00032$ & $1.04220^{+0.00033}_{-0.00066}$ & $1.04261^{+0.00050}_{-0.0067}$ \\
		$\ln({10}^{10} A_s)$ & $3.043 \pm 0.015$ & $3.042 \pm 0.016$ & $3.036 \pm 0.017$ \\
		$n_s$ & $0.9669 \pm 0.0038$ & $0.9643^{+0.0055}_{-0.0041}$ & $0.9621 \pm 0.0050$ \\
		$z_{reio}$ & $7.82 \pm 0.73$ & $7.82 \pm 0.72$ & $7.87 \pm 0.73$ \\
		$\Neffint$ & --- & $<0.79$ (68\%CL) & $<0.28$ (68\%CL) \\
		$\meff$ & $< 0.115$ (95\%CL) & $<0.115$ (95\%CL) & $< 0.118$ (95\%CL) \\
		$H_0$ $\left[\frac{\textrm{(km/s)}}{\textrm{Mpc}}\right]$ & $67.8 \pm 0.5$ & $67.9^{+0.6}_{-0.5}$ & $68.0 \pm 0.6$ \\
		$S_8$ & $0.829 \pm 0.012$ & $0.829 \pm 0.013$ & $0.823 \pm 0.013$ \\ [0.5mm] \hline
		ln($E$) & $-0.5306 \times 10^{3}$ & $-0.5309 \times 10^{3}$ & $-0.5320 \times 10^{3}$ \\
		$E_{\rm int}/E_{\rm fs}$ & --- & 0.74 & 0.24 \\ [0.5mm] \hline
		\multicolumn{4}{ c }{Best fit} \\ [0.5mm] \hline
		$\Neffint$ & --- & 0.199 & 0.139 \\
		$\log_{10}$$(\zdec)$ & --- & 3.084 & --- \\
		$\chi^2_{\rm{eff}}$ & 1017.32 & 1016.65 & 1017.05 \\
		$\Delta \chi^2_{\rm{eff}}$ & --- & -0.67 & -0.28 \\ [0.5mm] \hline
		\multicolumn{4}{ c }{High $\zdec$ mode best fit} \\ [0.5mm] \hline
		$\Neffint$ & --- & 0.020 & --- \\
		$\log_{10}$$(\zdec)$ & --- & 5.077 & --- \\
		$\chi^2_{\rm{eff}}$ & --- & 1017.47 & --- \\
		$\Delta \chi^2_{\rm{eff}}$ & --- & +0.15 & ---
	\end{tabular}
	\caption{Case 3: statistical information for our baseline configurations. $\chi^2_{\rm{eff}} = -2 \ln \mathcal{L}$ is the effective chi square, $\Delta \chi^2_{\rm{eff}}$ is with regards to the corresponding free-streaming case, ln($E$) is the Bayesian evidence, and $E_{\rm int}/E_{\rm fs}$ is the Bayesian evidence ratio with regards to the corresponding free-streaming case. $\chi^2_{\rm{eff}}$ is presented for the best fit cosmology (giving the minimum effective chi square) and for the best fit of the high $\zdec$ mode. All credibility intervals are 68\%CL centered around the mean unless otherwise noted. Note that at 95\%CL $\Neffint < 2.34$ ($\Neffint < 0.50$ for fluid-like) and so although one or more interacting species is disfavored, it is not ruled out by cosmological data (however, it is hard to accommodate a fluid-like neutrino species). The $\zdec$ bound is not reported in the table as it is strongly prior dependent. We find $\log_{10}(\zdec)>4.0$ (68\%CL) for the decoupling case for our standard prior range $10^{2} < \zdec < 10^{6}$, which was chosen because the posterior is flat below and above this range and the narrower prior range allows us to accurately probe the intermediate $\zdec$ range, while keeping the case computationally feasible. With a prior $\zdec < 10^{9}$, we find $\log_{10}(\zdec)>5.4$ (68\%CL) and $\Neffint < 1.10$ at 68\%CL ($\Neffint < 2.51$ at 95\%CL). We instead refer to Figure~\ref{fig:results:PSIz6m} for the $\zdec$ constraints.}
	\label{results:tab:PSI}
\end{table}

\subsection{Case 4: Varying fraction of interacting species}
\label{sec:results:varying_fraction}
We want to open up parameter space and allow for a freely varying number of free-streaming effective degrees of freedom, while still varying the fraction that is interacting. In order to keep the total number of varying parameters fixed, we fix the effective mass to roughly the current $2\sigma$ upper bound on the neutrino mass sum~\cite{Aghanim:2018eyx}. This case has free-streaming massless neutrinos with varying $\Nefffs$, plus massive interacting species with a fixed mass sum $\meff = 0.11$ eV and varying $\Neffint$ and $\zdec$.

\begin{figure}[tb]
	\centering
	\includegraphics[width=8.4cm]{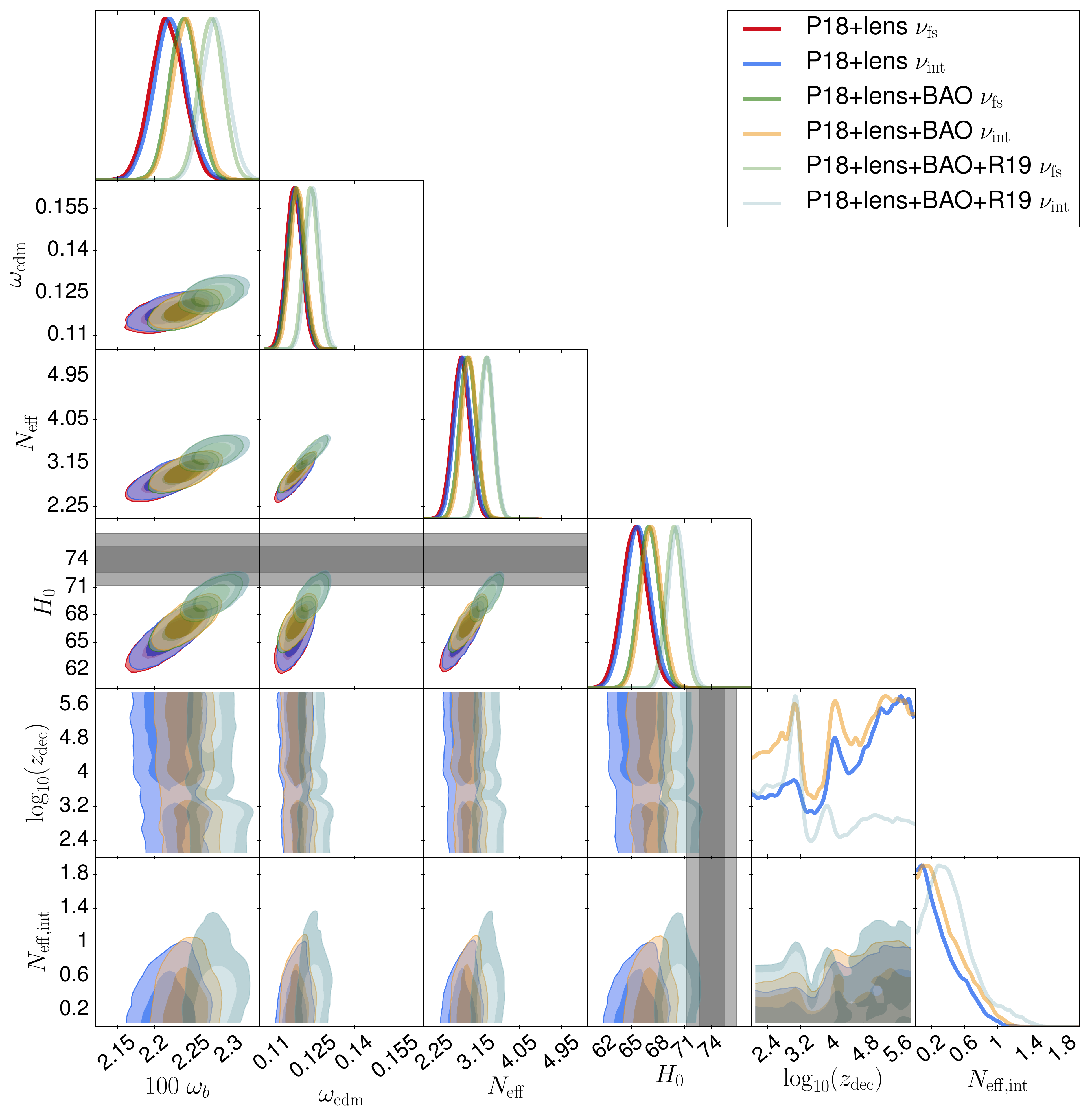}
	\hspace{0.1cm}\includegraphics[width=8.4cm]{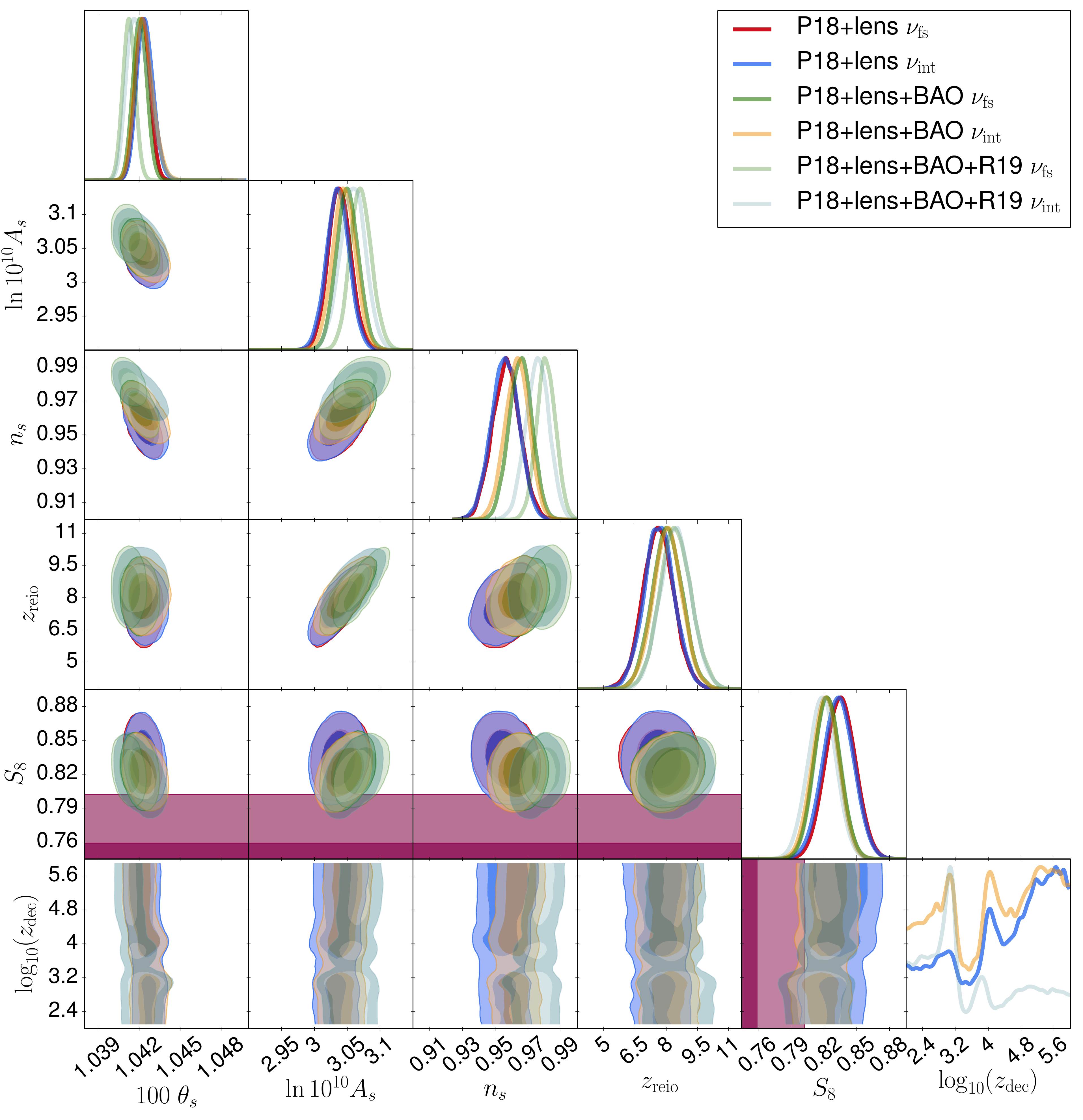}
	\caption{\textbf{Case 4: varying fraction of self-interacting relativistic species.} All free-streaming (red, green and light green) vs fraction of self-interacting (blue, yellow and cyan) for Planck-only (red and blue), Planck + BAO (green and yellow) and Planck + BAO + $H_0$ prior (light green and cyan). The grey bands correspond to the $H_0$ measurement from Riess et al.~\cite{Riess:2019cxk}, while the purple band is the $S_8$ measurement from KiDS+VIKING-450~\cite{Wright:2020ppw}. See Appendix \ref{app:full_plots}, Figure \ref{fig:app:urz6i} for the full parameter space plot. \textbf{Left and right panel:} both show the same cases, but different cosmological parameters. \textbf{Left panel}: energy density parameters. \textbf{Right panel:} power spectrum shape and amplitude parameters. Some contours are hard to see as the self-interacting cases are on top of the free-streaming comparison cases, e.g. red and blue, green and yellow, and light green and cyan. This is true for e.g. $H_0$, where there is negligible difference between the free-streaming and interacting cases except at low values of $\zdec \lesssim 2000$ in the fluid-like regime, making the combination of Planck and the $H_0$ prior suspicious. The parameter space for the decoupling redshift, $\zdec$, can be roughly separated into three regions, with slightly disfavored regions inbetween: 1) nearly free-streaming at $\zdec \gtrsim \textrm{60,000}$ (note the posterior is approximately flat up to standard neutrino decoupling), 2) a strongly interacting mode at around $\zdec \sim \textrm{10,000}$, and 3) a fluid-like region from just before recombination until today (the posterior is approximately flat below $\zdec < 10^2$), with a peak in the posterior shortly before recombination. At low $\zdec \lesssim 2000$, in the fluid-like regime, we see $H_0$ and $S_8$ tensions are slightly alleviated compared to the free-streaming comparison cases (see $\zdec$-$H_0$ and $\zdec$-$S_8$ 2-d plots for $\zdec \lesssim 2000$), while larger values for $\zdec$ shows no improvement compared to the free-streaming cases.}
	\label{fig:results:urz6i}
\end{figure}

\begin{figure}[tb]
	\centering
	\includegraphics[width=8.4cm]{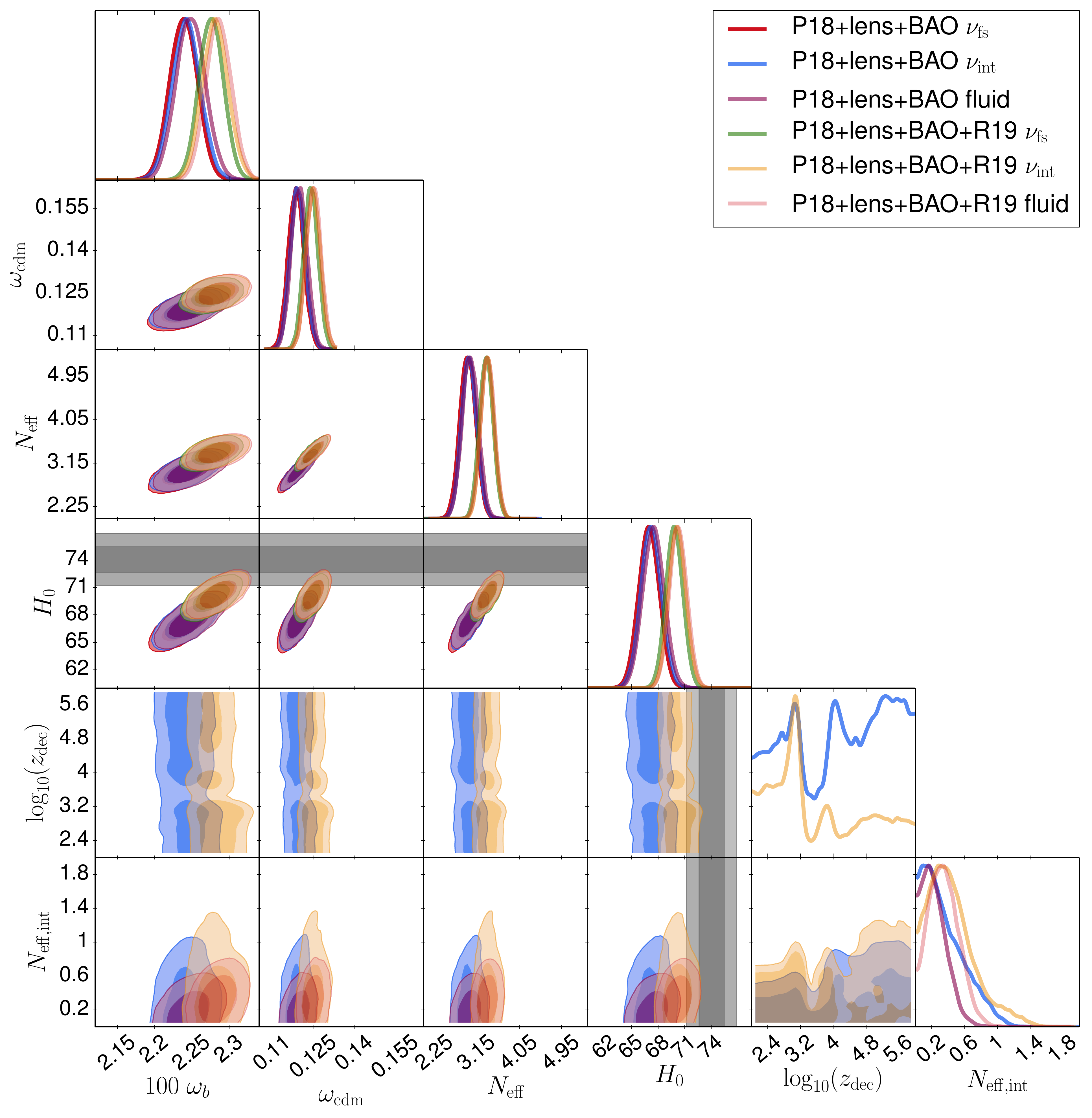}
	\hspace{0.1cm}\includegraphics[width=8.4cm]{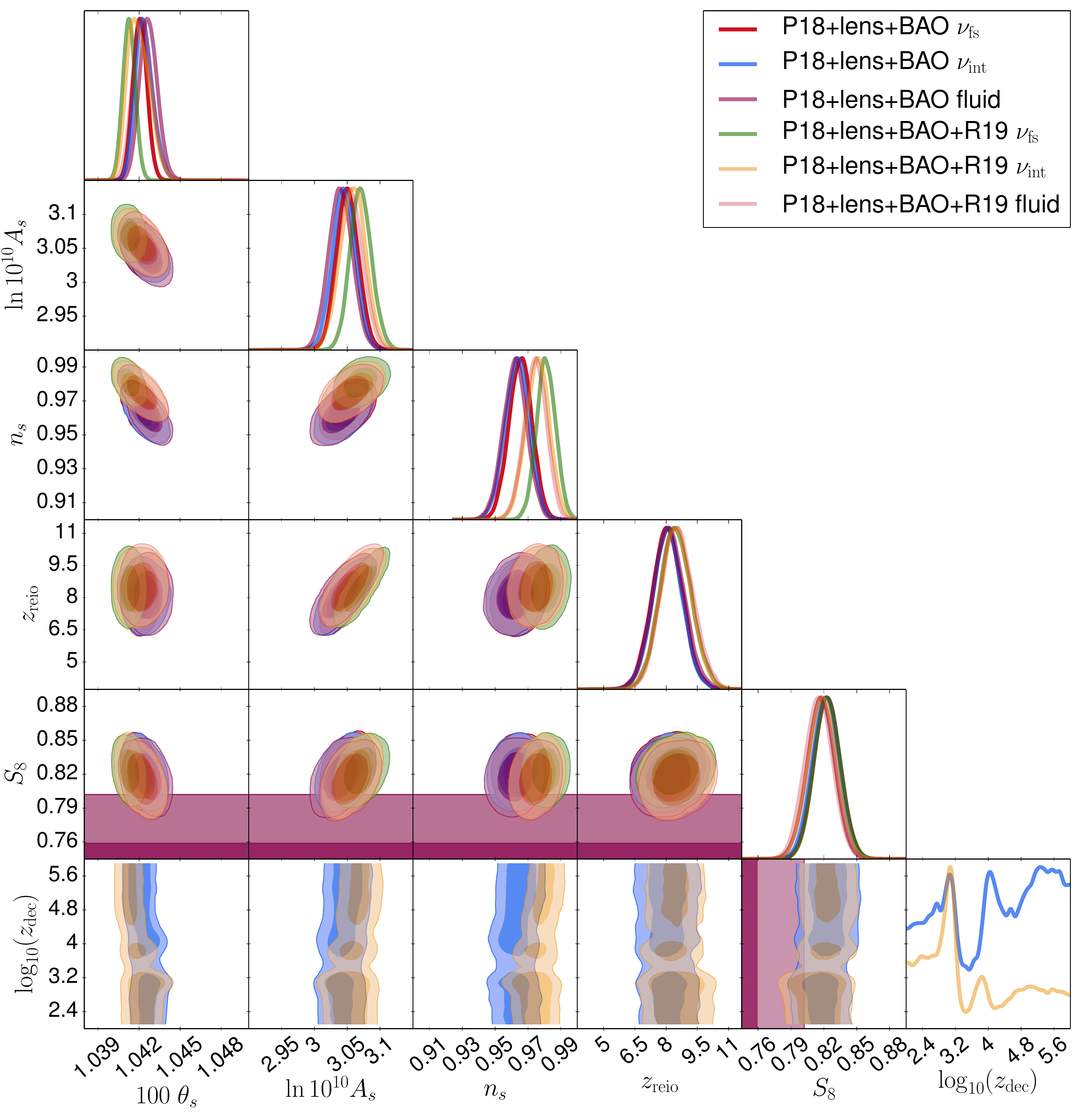}
	\caption{\textbf{Case 4: varying fraction of self-interacting relativistic species, including fluid.} All free-streaming (red and green) vs fraction of self-interacting species interacting (blue and yellow) and fraction of fluid-like (purple and pink) for Planck + BAO (red, blue and purple) and Planck + BAO + $H_0$ prior (green, yellow and pink). The grey bands correspond to the $H_0$ measurement from Riess et al.~\cite{Riess:2019cxk}, while the purple band is the $S_8$ measurement from KiDS+VIKING-450~\cite{Wright:2020ppw}. See Appendix \ref{app:full_plots}, Figure \ref{fig:app:urfl} for the full parameter space plot. \textbf{Left and right panel:} both show the same cases, but different cosmological parameters. \textbf{Left panel}: energy density parameters. \textbf{Right panel:} power spectrum shape and amplitude parameters. Note that beyond $\zdec > 10^6$ up to standard neutrino decoupling and below $\zdec < 10^2$ the 1-d posterior is approximately flat. The fluid-like case only slightly alleviates the $H_0$ and $S_8$ tensions. Although the low $\zdec$ decoupling case at $1000 < \zdec < 2000$ (so essentially fluid-like) does slightly better at the price of being somewhat fine-tuned, it still requires an external solution to the Hubble tension making the combination of Planck with the $H_0$ prior somewhat suspect.}
	\vspace{5mm}
	\label{fig:results:urfl}
\end{figure}

\noindent In this section, we refer to these figures, datasets, and configurations:
\begin{itemize}
	\item Figure \ref{fig:results:urz6i}. P18 +lens. Red (all free-streaming) and blue (partially interacting).
	\item Figure \ref{fig:results:urz6i}. P18 +lens +BAO. Green (all free-streaming) and yellow (partially interacting).
	\item Figure \ref{fig:results:urz6i}. P18 +lens +BAO +R19. Light green (all free-streaming) and cyan (partially interacting).
	\item Figure \ref{fig:results:urfl}. P18 +lens +BAO. Red (all free-streaming), blue (partially interacting), and purple (partially fluid-like).
	\item Figure \ref{fig:results:urfl}. P18 +lens +BAO +R19. Green (all free-streaming), yellow (partially interacting), and pink (partially fluid-like).
\end{itemize}
\newpage
\noindent For this case, almost all $\zdec$ values are allowed and parameter space can roughly be split into three regions with slightly disfavored regions inbetween:
\begin{enumerate}
	\item Nearly free-streaming at $\zdec \gtrsim \textrm{60,000}$ extending up to standard neutrino decoupling (note that beyond $\zdec > 10^6$ the 1-d posterior is approximately flat up to standard neutrino decoupling).
	\item A strongly interacting mode at around $\zdec \sim \textrm{10,000}$.
	\item A fluid-like region from just before recombination until today, with a peak in the posterior shortly before recombination.
\end{enumerate}
The $\zdec \sim \textrm{10,000}$ mode allows for up to $\Neffint \lesssim 1$ at $2\sigma$, while the fluid-like regime is restricted to about $\Neffint \lesssim 0.5$ unless the $H_0$ prior is added, in which case a larger number of extra relativistic species is allowed, up to around $\Neffint \lesssim 1$ at $1\sigma$. We see there is a slight alleviation of the $S_8$ (Figure \ref{fig:results:urz6i}, right panel) and $H_0$ (Figure \ref{fig:results:urz6i}, left panel) tensions compared to the free-streaming case for low values of $\zdec \lesssim 2000$ (see $\zdec$-$H_0$ and $\zdec$-$S_8$ 2-d plots for $\zdec \lesssim 2000$), with the change in $S_8$ being the most significant. Given the relatively minor improvement in the $H_0$ tension compared to the free-streaming case the combination with the $H_0$ prior remains somewhat suspect.

However, if the results including $H_0$ were to be trusted, we intriguingly find a preference for a fluid-like component, with a sharp peak in the posterior at around $1000 < \zdec < 2000$ (with a best fit of $\zdec=1200$, see Table~\ref{results:case4}) and a flattening of the posterior at low $\zdec$ values (continuing until a decoupling today) at a higher level than the free-streaming part of parameter space, with a comparable parameter space volume. However, this would require a novel resolution to the $H_0$ tension that fixes the value to a high number irrespective of the number of extra relativistic species, without affecting these bounds.

In order to isolate the fluid-like region of parameter space, we include on Figure \ref{fig:results:urfl} a fluid-like case with the decoupling redshift set to $\zdec=1$. This makes it easier to derive numerical constraints, i.e. for the data combination P18 +lens +BAO +R19, the parameter constraints shift from $S_8 = 0.823 \pm 0.012$ (free-streaming) to $S_8 = 0.815 \pm 0.013$ (fluid) and from $H_0 = 69.8 \pm 1.0$ (free-streaming) to $H_0 = 70.3 \pm 1.0$ (fluid). It is clear that this is only a marginal improvement in both cases. In order for a fluid-like species to help more with these tensions we need a decoupling redshift in the relatively narrow interval of $1000 < \zdec < 2000$, which is a somewhat fine tuned solution (and still requires additional help in order to reach the $H_0$ value from e.g. Riess et al.~\cite{Riess:2019cxk}).

\begin{table}[tb]
	\tiny
	\centering
	\begin{tabular}{ l | c c c | c c c | c c }
		& \multicolumn{3}{ c | }{Free-streaming} & \multicolumn{3}{ c }{Self-interacting (Case 4)} & \multicolumn{2}{ c }{Self-interacting (fluid-like)} \\
		& P18 +lens & +BAO & +R19 & P18 +lens & +BAO & +R19 & P18 +lens +BAO & +R19 \\ \hline
		$\omega_{b}$ & $0.02217 \pm 0.00022$ & $0.02239 \pm 0.00019$ & $0.02276 \pm 0.00017$ & $0.02221 \pm 0.00022$ & $0.02242^{+0.00018}_{-0.00020}$ & $0.02281^{+0.00017}_{-0.00018}$ & $0.02248 \pm 0.00020$ & $0.02286 \pm 0.00017$\\
		$\omega_{cdm}$ & $0.1178 \pm 0.0028$ & $0.1186 \pm 0.0029$ & $0.1240 \pm 0.0026$ & $0.1183^{+0.0027}_{-0.0030}$ & $0.1191 \pm 0.0028$ & $0.1247^{+0.0026}_{-0.0027}$ & $0.1196 \pm 0.0029$ & $0.1249 \pm 0.0027$\\
		$100 \times \theta_s$ & $1.04226 \pm 0.00054$ & $1.04207 \pm 0.00052$ & $1.04127 \pm 0.00046$ & $1.04244^{+0.00054}_{-0.00066}$ & $1.04238^{+0.00055}_{-0.00075}$ & $1.04198^{+0.00068}_{-0.00094}$ & $1.04269 \pm 0.00066$ & $1.04226^{+0.00066}_{-0.00075}$ \\
		$\ln({10}^{10} A_s)$ & $3.038 \pm 0.017$ & $3.050 \pm 0.017$ & $3.071 \pm 0.017$ & $3.036^{+0.017}_{-0.018}$ & $3.044^{+0.018}_{-0.017}$ & $3.058^{+0.020}_{-0.019}$ & $3.040 \pm 0.018$ & $3.054 \pm 0.019$ \\
		$n_s$ & $0.9567 \pm 0.0085$ & $0.9654 \pm 0.0070$ & $0.9810 \pm 0.0059$ & $0.9561^{+0.0086}_{-0.0082} $ & $0.9633^{+0.0076}_{-0.0071}$ & $0.9757^{+0.0073}_{-0.0070}$ & $0.9622 \pm 0.0073$ & $0.9745 \pm 0.0068$ \\
		$z_{reio}$ & $7.63 \pm 0.75$ & $8.06 \pm 0.73$ & $8.45 \pm 0.76$ & $7.69^{+0.72}_{-0.74}$ & $8.07^{+0.70}_{-0.72}$ & $8.46 \pm 0.75$ & $8.13 \pm 0.75$ & $8.53 \pm 0.77$ \\
		$\Neff$ & $2.81 \pm 0.18$ & $2.95 \pm 0.17$ & $3.35 \pm 0.15$ & $2.85^{+0.18}_{-0.19}$ & $2.98^{+0.16}_{-0.17}$ & $3.37^{+0.14}_{-0.15}$ & $3.00 \pm 0.17$ & $3.37 \pm 0.15$ \\
		$\Neffint$ & --- & --- & --- & $< 0.74$ (95\%CL) & $< 0.86$ (95\%CL) & $0.44^{+0.14}_{-0.37}$ & $< 0.51$ (95\%CL) & $0.35^{+0.15}_{-0.22}$ \\
		$H_0$ $\left[\frac{\textrm{(km/s)}}{\textrm{Mpc}}\right]$ & $65.4 \pm 1.4$ & $67.0 \pm 1.2$ & $69.8 \pm 1.0$ & $65.7^{+1.4}_{-1.5}$ & $67.2^{+1.1}_{-1.2}$ & $70.1^{+0.9}_{-1.0}$ & $67.5 \pm 1.2$ & $70.3 \pm 1.0$ \\
		$S_8$ & $0.835 \pm 0.014$ & $0.823 \pm 0.012$ & $0.823 \pm 0.012$ & $0.833 \pm 0.014$ & $0.821 \pm 0.012$ & $0.818 \pm 0.013$ & $0.818 \pm 0.013$ & $0.815 \pm 0.013$ \\ [0.5mm] \hline
		ln($E$) & $-0.5275 \times 10^{3}$ & $-0.5322 \times 10^{3}$ & $-0.5399 \times 10^{3}$ & $-0.5273 \times 10^{3}$ & $-0.5317 \times 10^{3}$ & $-0.5385 \times 10^{3} $ & $-0.5325 \times 10^{3}$ & $-0.5391 \times 10^{3}$ \\
		$E_{\rm int}/E_{\rm fs}$ & --- & --- & --- & $1.18$ & $1.57$ & $3.94$ & $0.71$ & $2.19$ \\ [0.5mm] \hline
		\multicolumn{9}{ c }{Best fit (corresponding to low $\zdec$ mode)} \\ [0.5mm] \hline
		$\Neff$ & 2.798 & 2.937 & 3.321 & 2.807 & 2.924 & 3.376 & 2.982 & 3.365 \\
		$\Neffint$ & --- & --- & --- & 0.030 & 0.193 & 0.564 & 0.168 & 0.312 \\
		$\log_{10}$$(\zdec)$ & --- & --- & --- & 3.038 & 3.077 & 3.078 & --- & --- \\
		$\chi^2_{\rm{eff}}$ & 1012.85 & 1021.61 & 1036.65 & 1012.79 & 1021.01 & 1032.73 & 1021.22 & 1034.32  \\
		$\Delta \chi^2_{\rm{eff}}$ & --- & --- & --- & -0.06 & -0.60 & -3.91 & -0.39 & -2.32 \\ [0.5mm] \hline
		\multicolumn{9}{ c }{Intermediate $\zdec$ mode best fit} \\ [0.5mm] \hline
		$\Neff$ & --- & --- & --- & 2.768 & 2.930 & 3.463 & --- & --- \\
		$\Neffint$ & --- & --- & --- & 0.002 & 0.297 & 0.448 & --- & --- \\
		$\log_{10}$$(\zdec)$ & --- & --- & --- & 3.849 & 4.004 & 3.773 & --- & --- \\
		$\chi^2_{\rm{eff}}$ & --- & --- & --- & 1012.93 & 1021.46 & 1034.96 & --- & --- \\
		$\Delta \chi^2_{\rm{eff}}$ & --- & --- & --- & +0.07 & -0.15 & -1.68 & --- & --- \\ [0.5mm] \hline
		\multicolumn{9}{ c }{High $\zdec$ mode best fit} \\ [0.5mm] \hline
		$\Neff$ & --- & --- & --- & 2.954 & 2.924 & 3.321 & --- & --- \\
		$\Neffint$ & --- & --- & --- & 0.012 & 0.028 & 0.305 & --- & --- \\
		$\log_{10}$$(\zdec)$ & --- & --- & --- & 5.542 & 5.860 & 5.180 & --- & --- \\
		$\chi^2_{\rm{eff}}$ & --- & --- & --- & 1013.07 & 1021.74 & 1037.36 & --- & --- \\
		$\Delta \chi^2_{\rm{eff}}$ & --- & --- & --- & +0.22 & +0.13 & +0.71 & --- & ---
	\end{tabular}
	\caption{Case 4: statistical information for our baseline configurations and the one including a prior on $H_0$. $\chi^2_{\rm{eff}} = -2 \ln \mathcal{L}$ is the effective chi square, $\Delta \chi^2_{\rm{eff}}$ is with regards to the corresponding free-streaming case, ln($E$) is the Bayesian evidence, and $E_{\rm int}/E_{\rm fs}$ is the Bayesian evidence ratio with regards to the corresponding free-streaming case. $\chi^2_{\rm{eff}}$ values are presented for the best fit cosmology for each of the three modes, where the low $\zdec$ mode is the overall best fit of the run. All credibility intervals are 68\%CL centered around the mean unless otherwise noted. For $\zdec$ bounds we refer to Figure~\ref{fig:results:2FSz6m}, as this case is obviously multimodal and all of parameter space is allowed to some degree.}
	\label{results:case4}
\end{table}

%% file: conclusions.tex
\newpage
\section{Discussion and conclusions}
\label{sec:discussion}
In this paper we have produced comprehensive constraints on new neutrino self-interactions from CMB and BAO data. In comparing with data, we study a range of scenarios for neutrino self-interactions. While neutrino self-interactions are strongly constrained by terrestrial experiments, these constraints can be weakened by imposing the self-interactions on the neutrino mass eigenstates and limiting the number of states that participate (Fig.~\ref{fig:app:gphiconst}). We therefore consider cosmological constraints on self-interactions among all neutrino states as well as a variable fraction of the neutrino states. In the following, we summarize the main results of our analysis and discuss implications for particle physics.

We consider several different scenarios for neutrino interactions, dubbed {\em Cases}, and summarized in Sec. \ref{ssec:parameters}, which are discussed case by case in the following. As a reference, our dataset choices are outlined in Sec. \ref{sec:method}. 
\begin{itemize}
\item{\em Case 1: All species interacting} 
\begin{itemize}
\item P18 +lens and P18 +lens +BAO analyses:\\
If we assume all neutrino species are interacting and allow for a free total amount of neutrinos ($\Neff = \Neffint$) we find two modes. The first mode (dubbed the ``high-$\zdec$ mode") has similar cosmological parameters to the free-streaming case, but allows values of $\zdec$ lower than the standard neutrino decoupling time, we find $\zdec > 10^{5.1}$ at $95\%$ confidence for P18 +lens +BAO. The second mode has a low value of $\zdec$, $\zdec = 10^{4.14\pm 0.056}$ for P18 +lens +BAO. The low-$\zdec$ mode also has a lower value of $n_s$ ($0.9226 \pm 0.0055$ versus $0.9613 \pm 0.0071$) and a lower value of $\Neff$ ($2.57\pm 0.14$ versus $2.92 \pm 0.17$), here both values are quoted for P18 +lens +BAO, but the trend is the same for P18 +lens alone. All other parameters remain similar between the two cases. The low-$\zdec$ mode is not an improved fit to the data, increasing the $\chi_\text{eff}^2$ by $\approx 7$. See Fig. \ref{fig:results:P_Pb_R1z6m} and Table \ref{results:tab:P_Pb_R1z6m} for complete parameter results. 

\item  P18 +lens +BAO +R19 analysis:\\
Adding the local measurement of $H_0$ from R19 increases the best fit value of $\Neff$ to $\Neff = 3.30 \pm 0.15$ and eliminates the low-$\zdec$ mode found above. The inferred cosmological parameters, including the value of $\Neff$, are nearly identical for the interacting neutrino case and our free-streaming control. The neutrino decoupling epoch is bound at $\zdec > 10^{5.3}$ at $68\%$ confidence. See Fig.~\ref{fig:results:Pbr_R1z6m} and Table \ref{results:tab:Pbr_R1z6m} for complete results. 

\item P18 +lens +BAO +R19, omitting high-$\ell$ polarization data:\\
For comparison with \cite{Kreisch:2019yzn} we also also try eliminating high-$\ell$ CMB polarization data. In this case, the low-$\zdec$ mode found with P18 +lens and P18 +lens +BAO mode reappears, but with shifted values of nearly all other cosmological parameters. Notably, the best fit value of $H_0$ is completely consistent with that from R19 alone, appearing to eliminate the Hubble tension. The value of $S_8$ is also reduced slightly, somewhat alleviating the tension with low-redshift data. On the other hand, the low-$\zdec$ mode is not actually an improved fit to the data ($\chi_{eff}^2$ increases by $\approx 4$). Separately, we have no {\em a priori} reason to eliminate the polarization data that forbids the existence of this mode. See Fig.~\ref{fig:results:Pbnhr_R1z6m} and Table \ref{results:tab:Pbnhr_R1z6n} for complete results. 
\end{itemize}
\item{\em Case 2: Two free-streaming species plus free amount of interacting species}\\
If we force two neutrino states to be free-streaming to alleviate non-cosmological constraints on neutrino interactions, P18 +lens +BAO analyses find bounds on $\Neff = 2.91^{+0.18}_{-0.17}$, nearly unchanged from the control case in which all contributions to $\Neff$ are free-streaming, $\Neff = 2.92 \pm 0.17$. A hint of a second low-$\zdec$ mode appears at $\zdec \approx 10^{4.09}$. This mode has a larger value of $\theta_s$, lower $A_s$ and $n_s$, but does not significantly shift other cosmological parameters (including $H_0$ and $S_8$). The hint of the low-$\zdec$ mode remains when adding R19 but the $H_0$ tension is not relieved (new value of $H_0$ is the same as for free-streaming). The low-$\zdec$ mode is not isolated so it is not straightforward to derive a bound on $\zdec$ in this case. Complete results are presented in Fig.~\ref{fig:results:2FSz6m} and Table \ref{results:tab:2FSz6m}. For this scenario, we also provide constraints under the assumption that $\Neffint$ never decouples (that is, it is a fluid). For the fluid case, the values of $n_s$ and $\Neff$ are lowered by approximately $1-2$ and $0.5\sigma$, respectively. The value of $S_8$ is also lowered, somewhat reducing the tension with low-redshift data. The fluid model is, however, a worse fit to the data overall. Complete results are in Fig.~\ref{fig:results:2FSfl} and  Table \ref{results:tab:2FSz6m}. Figures \ref{fig:2FSb_spectra} and \ref{fig:varying_zdec} illustrate how changes to the CMB power spectra induced by the lower $\zdec$ value are compensated by shifts in other cosmological parameters.

\item {\em Case 3: $\Neff = 3.046$, varying fraction of interacting species}\\
If we fix the total relativistic degrees of freedom to the Standard Model value of $\Neff = 3.046$, but allow the interacting fraction to vary we find upper bounds on the interacting component of $\Neffint < 0.79$ for P18 +lens +BAO at $68\%$ confidence. The $\zdec$ posterior shows a small local maximum at around $\zdec \sim \textrm{10,000}$. The region $\zdec \gtrsim 10^{5}$ is, however, preferred by the data (with $\Neffint \lesssim 1$, at about $1\sigma$). The fluid case ($\zdec< 0$) further limits the interacting component to $\Neffint < 0.28$ at $95\%$ confidence. The rest of the cosmological parameters are virtually unchanged between the free-streaming control case, the fit allowing some fraction of neutrinos to self-interact and decouple, and the fit allowing interacting neutrinos that never decouple. Complete results are given in Fig.~\ref{fig:results:PSIz6m} and Table \ref{results:tab:PSI}. 

\item {\em Case 4: Free total $\Neff$ and varying fraction of interacting species}\\
Finally, if we allow $\Nefffs$ and  $\Neffint$ to vary independently the values of $A_s$ and $n_s$ shift towards slightly lower values for all dataset combinations in comparison to the free-streaming control case. But other parameters, including $\Neff$ and $H_0$ are virtually unchanged. The upper bounds on $\Neffint$ are $0.74$ and $0.86$ for P18 +lens and P18 +lens +BAO, respectively, both at $95\%$ confidence. The interacting case is not in any less tension with R19 than the free-streaming control case. Intriguingly, there does appear to be hints of additional modes in $\zdec$ at low, intermediate and high values of $\zdec$ (appearing at $\zdec \approx 10^{3}$, $10^4$, and $10^{5-6}$). The significance of the low-$\zdec$ mode increases dramatically with the inclusion of R19 data and we find a substantial Bayesian evidence ratio of 3.94, yet the $H_0$ tension is not resolved in this case making the combination of discrepant datasets questionable. Complete results are given in Fig.~\ref{fig:results:urz6i} and Table~\ref{results:case4}. Additionally, in Fig.~\ref{fig:results:urfl} and Table~\ref{results:case4} we also present results for a fluid case that never decouples. In this case, we find a tighter bound on $\Neffint$ of $0.51$ for P18 +lens +BAO at $95\%$ confidence, while $S_8$ is slightly lowered compared to the free-streaming case, somewhat alleviating the tension with low-redshift data. When adding R19 a non-zero amount of interacting fluid is preferred at about $1.5\sigma$, but since the $H_0$ tension is not alleviated compared to the free-streaming case the combination of discrepant datasets is questionable.
\end{itemize}

While we find some hints of additional strongly-interacting neutrino modes these are disfavored by the data overall. We conclude that self-interacting neutrinos are not cosmologically favored over free-streaming species, but have not been ruled out for large enough $\zdec$ or small enough $\Neffint$. For large $\zdec$ or small $\Neffint$, self-interacting species are indistinguishable from free-streaming species as the earlier the decoupling or the lower the abundance, the less the self-interacting species has an effect on the CMB.

For the uniform and universal coupling, $g_{ij} = g_\phi \delta_{ij}$ ({\em Case 1}), we find the low-$\zdec$ mode is not only disfavored by the cosmological data, but is also ruled out experimentally (see Section~\ref{subsec:expconstraint}). The high-$\zdec$ mode places a bound on $\zdec$, or equivalently on $G_\nu$, of $\log_{10} (\zdec) > 5.1$ (P18 +lens +BAO, 95\%CL). This can be translated to $G_\nu < 10^{-3.2} \MeV^{-2}$. Note that this translation is computed under the assumption of the standard cosmology, but it is still valid as cosmological parameters for the high $\zdec$ mode agree with those for the standard cosmology with free-streaming neutrinos. 

Partially interacting neutrinos are discussed in {\em Case 2-4}. For these cases, we focus only on the model where $\nu_3$ is the only interacting neutrinos species, $g_{ij}=g_\phi \delta_{3i}\delta_{3j}$, as it has the weakest experimental bounds. The discussion of extra dark radiation will be considered in follow-up work. For $\nu_3$ to have self-interactions without beyond the Standard Model physics other than the Majoron, we need to have $\Neffint \sim 1$. From the results for {\em Case 2} in Table \ref{results:tab:2FSz6m}, the best fit value of $\Neffint = 0.787$ with $\log_{10} (\zdec) = 4.085$ ($G_\nu = 10^{-1.283} \MeV^{-2}$) for the lower-z decoupling mode or $\Neffint = 0.822$ with $\log_{10} (\zdec) = 5.456$ ($G_\nu = 10^{-3.381} \MeV^{-2}$) for the higher-z decoupling mode. Both modes have $\chi_{eff}^2$ values comparable to the free-streaming neutrino case. The bounds on $\zdec$ depend on the prior choice as stated in the caption of Table~\ref{results:tab:2FSz6m}, but we can take the most conservative one, $\log_{10} (\zdec) > 3.9$ (95\%CL) ($G_\nu < 10^{-0.99}\MeV^{-2}$). We can consider this as the bound for interacting $\nu_3$ because the mean value of $\Neff$ does not vary significantly with the value of $\zdec$, as we can see in Figure~\ref{fig:results:2FSfl}. Note that the bound on $\zdec$ is relaxed by several orders of magnitude compared to the universal coupling case, as expected. From {\em Case 3} and {\em Case 4}, we can deduce the bounds on $\Neffint$. Compared to the fluid-like case, allowing decoupling of self-interactions relaxes the bounds significantly. For {\em Case 3}, where $\Neff$ is fixed to $3.046$, we have $\Neffint < 0.50$ for the fluid-like case and $\Neffint < 2.34$ with decoupling (95\%CL). For {\em Case 4}, where $\meff$ is fixed, we have $\Neffint < 0.51$ and $\Neffint < 0.86$ (95\%CL) for the fluid-like and the decoupling cases, respectively.

Finally, we note that in {\em Cases 1} and {\em 2} the modes with self-interacting neutrinos have mean values of the spectral index that are lower than the mean value for the free-streaming case. The allowed inflationary models could then be different in these cosmologies. For instance, if the constraints on the tensor-to-scalar ratio were unchanged in the interacting neutrino cosmology, then the lower value of $n_s$ would favor natural inflation \cite{Ade:2018gkx,Akrami:2018odb,Freese:1990rb}. Neutrino free-streaming is, however, known to affect the tensor power spectrum (e.g. \cite{Weinberg:2003ur}) so constraints on  the tensor-to-scalar ratio should differ somewhat in cosmologies where neutrinos self-interact until late times. We leave a study of this to a future work.

In summary, at present CMB and BAO data exhibit no preference for non-standard neutrino interactions. The cosmological constraints we have produced, though generally weaker, are complementary to laboratory constraints on neutrino interactions. On the other hand, the hints of additional modes with low values of neutrino decoupling demonstrate the potential for cosmological data to uncover new physics of neutrinos or other light relic particles. Considering an expanded suite of datasets, e.g. including galaxy survey data or other late time probes, may shed further light on these scenarios and possibly rule out or strengthen the evidence for these hints. This goes beyond the scope of this paper and is something we leave to future work. Additionally, future CMB surveys are likely to improve constraints on these models, but we also leave a study of the constraining power of future CMB datasets to a later work.\\

%% file: appendix.tex
\appendix
\section{Opacity Function}
\label{app:opacity}
\begin{figure}[H]
	\centering
	\includegraphics[width=12cm]{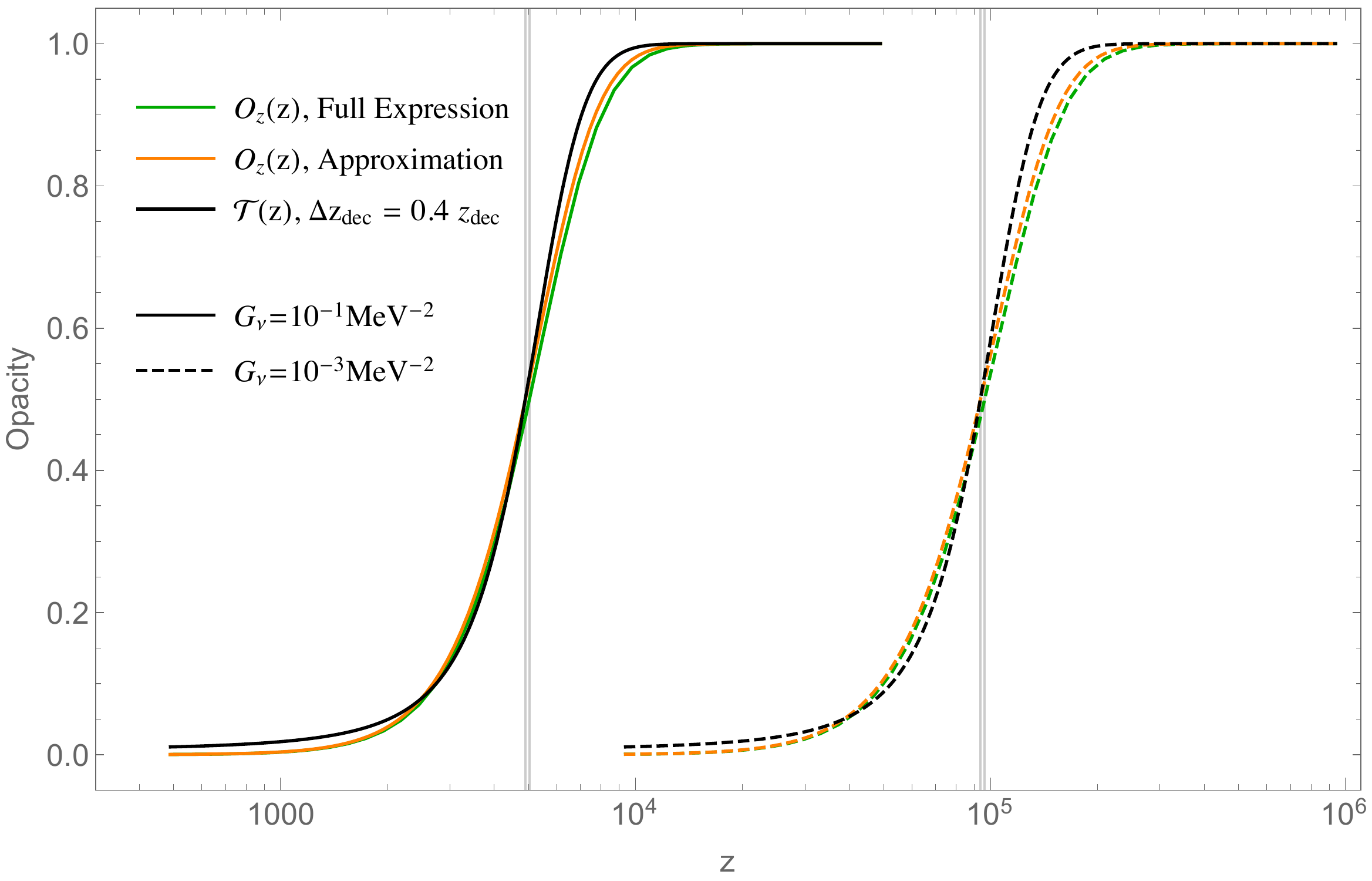}
	\caption{Opacity in terms of $z$ for $G_\nu = 10^{-1} \MeV^{-2}$ (solid) and $G_\nu = 10^{-3} \MeV^{-2}$ (dashed). Grid lines are $z$ where opacity is $0.5$. We also show the transition function $\mathcal{T}(z)$ with corresponding $\zdec$ and $\Delta \zdec = 0.4 \zdec$.}
	\label{fig:opacity}
\end{figure}

We compare the opacity functions with and without the Pauli-blocking factors in Sec.~\ref{sec:SInu}. The approximation ignoring the Pauli-blocking factors gives $\mathcal{O}(10\%)$ larger interaction rates $\Gamma_\nu$, which provides $\mathcal{O}(1\%)$ smaller decoupling redshifts, $z_\text{dec}$. In Figure~\ref{fig:opacity}, we show the opacity functions for $G_\nu = 10^{-1} \MeV^{-2}$ and $G_\nu = 10^{-3} \MeV^{-2}$ as examples. For $G_\nu = 10^{-1} \MeV^{-2}$, we find $\zdec = 5034$ for the full expression and $\zdec = 4902$ for the approximation. For $G_\nu = 10^{-3} \MeV^{-2}$, we have $\zdec = \textrm{96,367}$ and $\zdec = \textrm{93,859}$. The value of $\zdec$ is $\mathcal{O}(1\%)$ smaller with the approximation. In addition, we show the corresponding transition functions (Eq.~\ref{eq:transitionfunc}) with $\dzdec = 0.4\zdec$, which we use in \texttt{CLASS} to implement the effects of neutrino self-interactions. We have checked that using the transition function to suppress higher ($l \geq 2$) moments of neutrino perturbations instead of using the opacity function yields only $\mathcal{O}(0.1\%)$ difference in the power spectrum.

\section{Decoupling width}
\label{app:decoupling_width}
\begin{figure}[H]
	\centering
	\includegraphics[width=\textwidth]{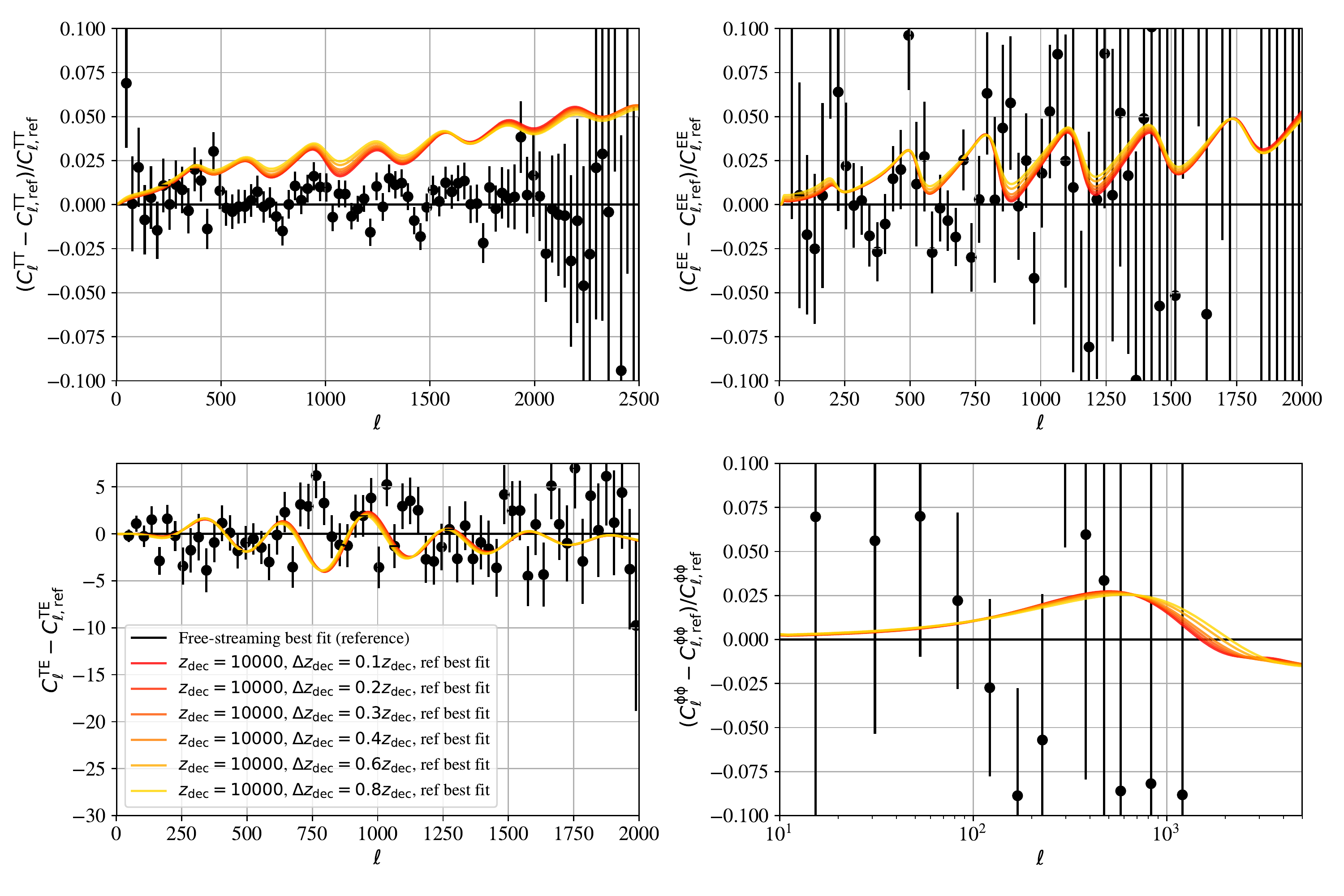}
	\caption{Fractional differences of the CMB temperature (top left) and polarisation (top right) auto-correlation and cross-correlation (bottom left, showing difference rather than fractional difference) angular power spectra, and the CMB lensing power spectrum (bottom right). All cases are compared to a free-streaming neutrinos comparison case with varying width of decoupling $\Delta\zdec$ of 10\% (more red) to 80\% (more yellow) of the decoupling redshift of $\zdec=\textrm{10,000}$.}
	\label{fig:varying_dzdec}
\end{figure}

\begin{figure}[H]
	\centering
	\includegraphics[width=\textwidth]{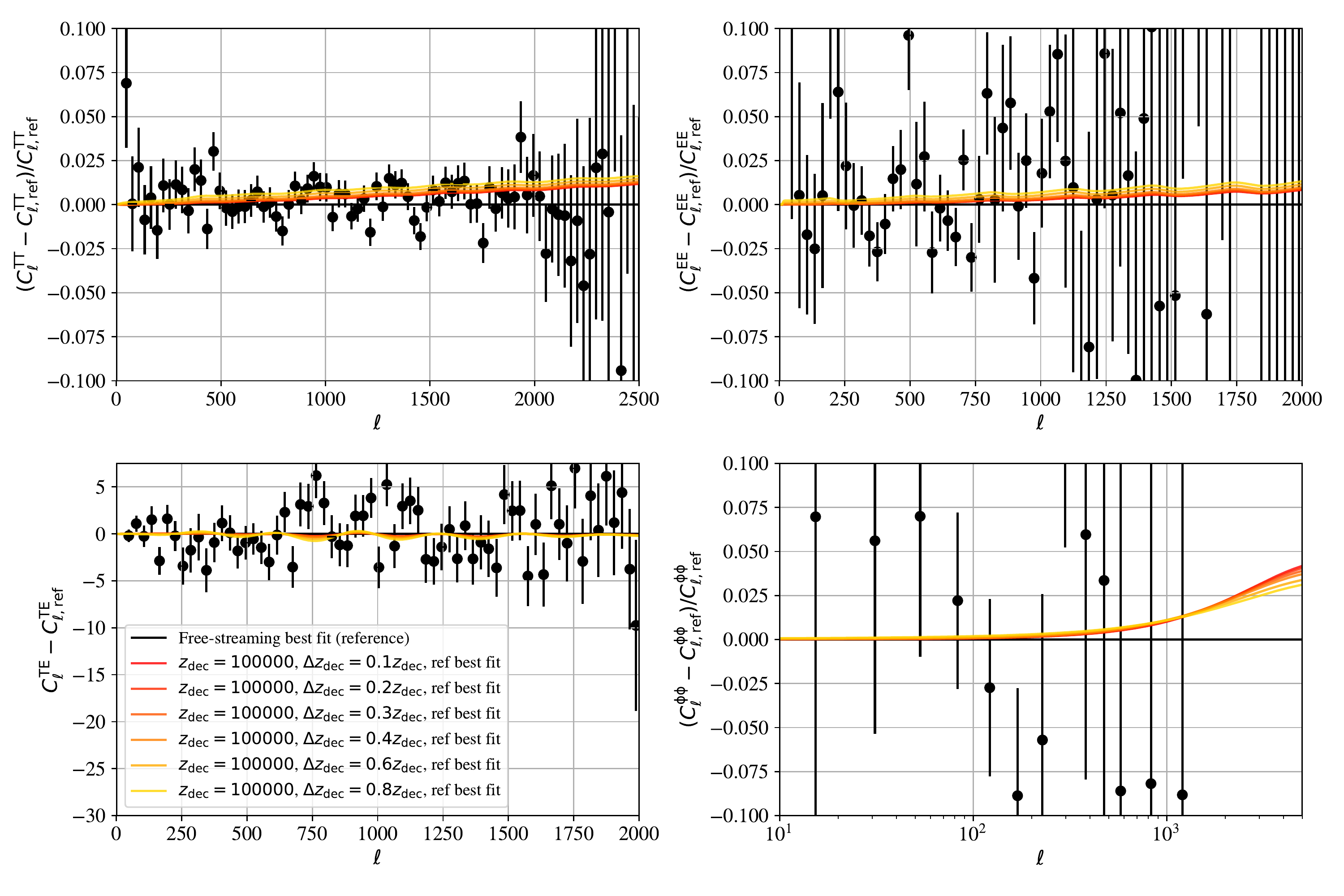}
	\caption{Fractional differences of the CMB temperature (top left) and polarisation (top right) auto-correlation and cross-correlation (bottom left, showing difference rather than fractional difference) angular power spectra, and the CMB lensing power spectrum (bottom right). All cases are compared to a free-streaming neutrinos comparison case with varying width of decoupling $\Delta\zdec$ of 10\% (more red) to 80\% (more yellow) of the decoupling redshift of $\zdec=\textrm{100,000}$.}
	\label{fig:varying_dzdec2}
\end{figure}

In this Appendix, we test the impact of the choice in decoupling width on the spectra and the MCMC-derived posterior distribution for one of our cases (the one referred to as Case 2) for the dataset combination P18 +BAO. In Figures~\ref{fig:varying_dzdec} and \ref{fig:varying_dzdec2}, we show the difference ratio compared to a free-streaming cosmology when we vary the decoupling width from 10\% to 80\% for $\zdec = \textrm{10,000}$ and $\zdec = \textrm{100,000}$, respectively. Qualitatively they are similar and only differ at a sub-percent level. We note that at the spectra level, we found that with our baseline decoupling width of 40\% of $\zdec$ we could reproduce the example spectra of~\cite{Lancaster:2017ksf} Figures 7-9 (but note that our value for $\zdec$ does not map perfectly onto theirs).

\begin{figure}[H]
	\centering
	\includegraphics[width=12cm]{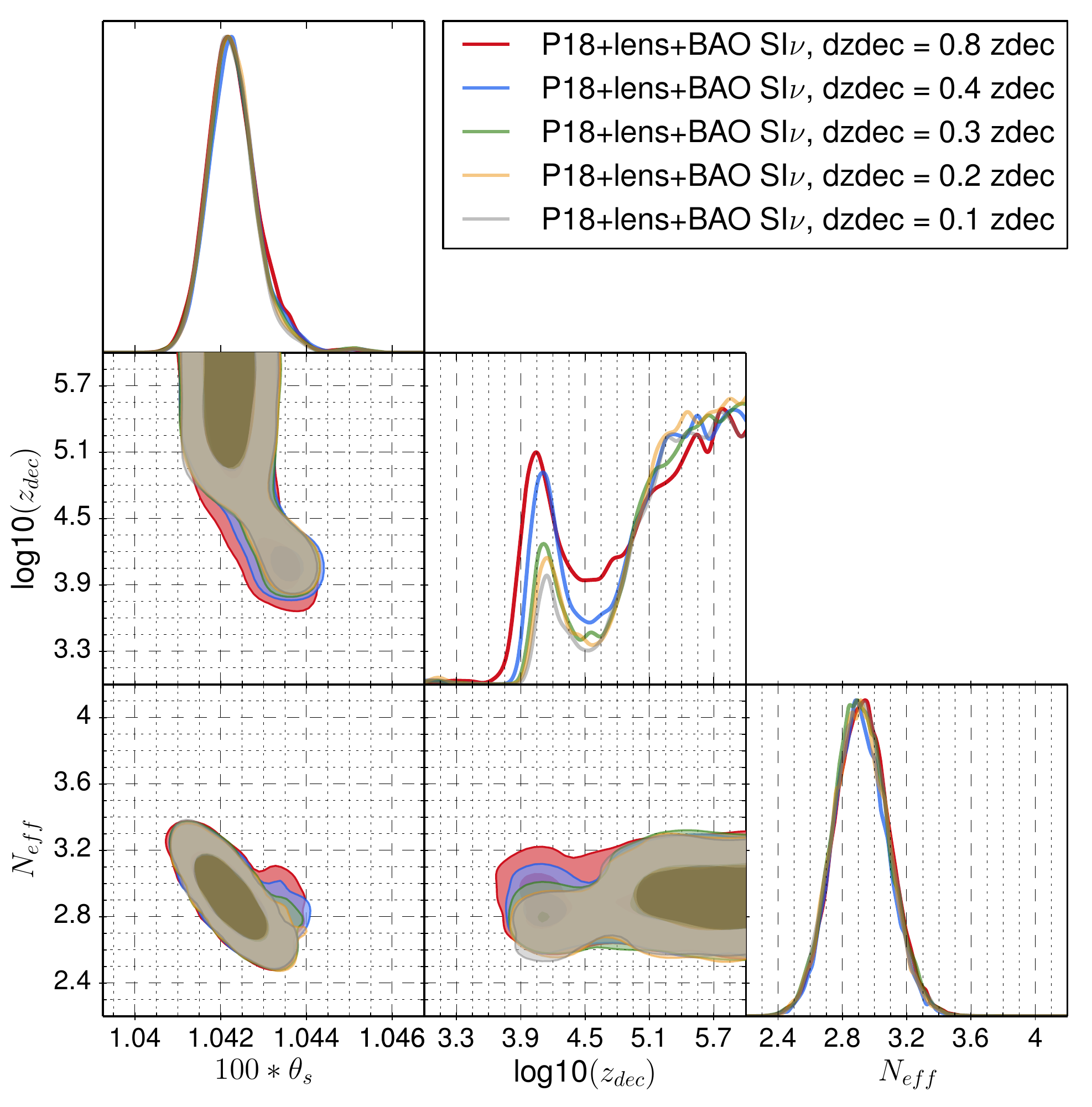}
	\caption{\textbf{Case 2: two free-streaming neutrinos plus self-interacting relativistic species.} The Figure shows select parameters for five runs with different decoupling widths of 10\% (red), 20\% (blue), 30\% (green), 40\% (yellow), and 80\% (grey) of $\zdec$. The MCMC runs otherwise have an identical setup and are for the data combination P18 +lens +BAO.}
	\label{fig:appendix_dzdec}
\end{figure}

This does, however, translate to some difference in the posterior distribution, which is shown in Figure~\ref{fig:appendix_dzdec}. The low $\zdec$ mode persists for all decoupling widths, but a wider decoupling width shifts the mode to slightly lower values and, in general, lower $\zdec$ values are more disfavored for a narrower decoupling width. This extends to the intermediate $\zdec$ region, where a decoupling width of $10-40\%$ means the intermediate region is significantly more disfavored compared to the $80\%$ decoupling width case. This might warrant further study in what kind of models could give rise to such a wide decoupling width, as it is easier to accommodate with cosmological data, but we leave this to future work.

\section{Full parameter space plots}
\label{app:full_plots}
In this Appendix, we show the full parameter space plots corresponding to the Figures in Section~\ref{sec:results}.
\begin{figure}[H]
	\centering
	\includegraphics[width=15cm]{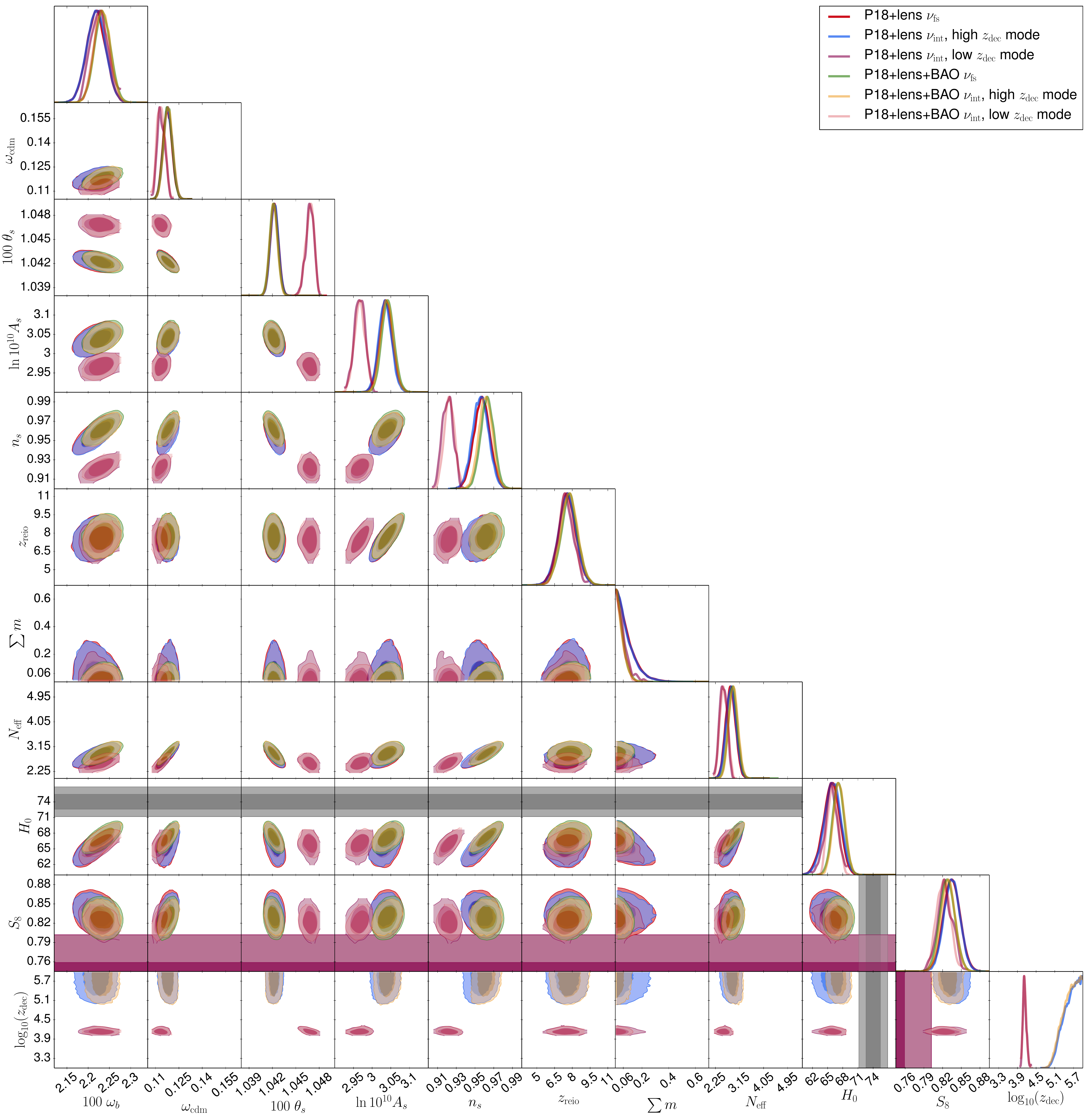}
	\caption{\textbf{Case 1: all species interacting.} Full parameter space corresponding to Figure~\ref{fig:results:P_Pb_R1z6m}.}
	\label{fig:app:P_Pb_R1z6m}
\end{figure}

\begin{figure}[H]
	\centering
	\includegraphics[width=15cm]{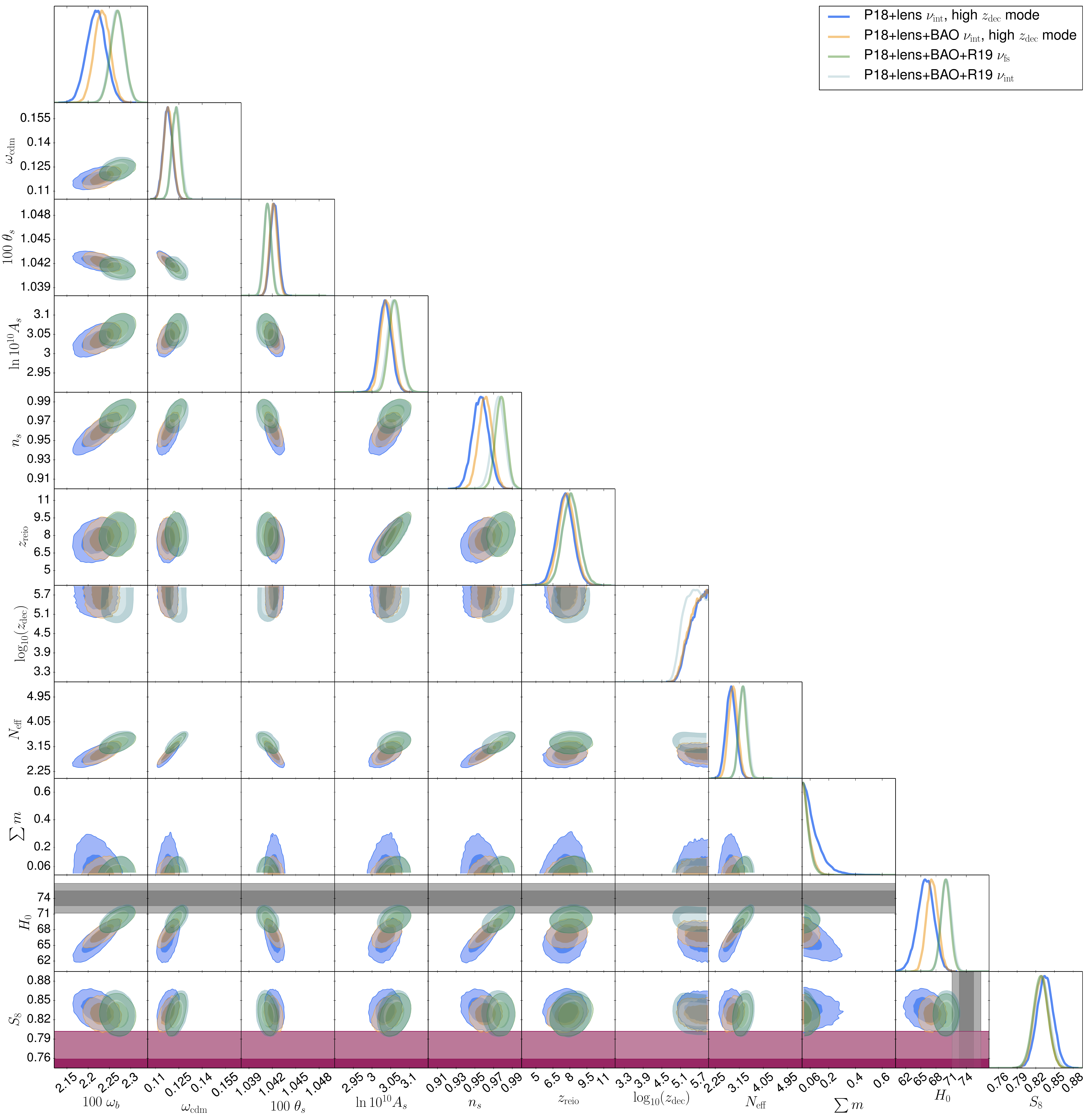}
	\caption{\textbf{Case 1: all species interacting, including $H_0$ prior.} Full parameter space corresponding to Figure~\ref{fig:results:Pbr_R1z6m}.}
	\label{fig:app:Pbr_R1z6m}
\end{figure}

\begin{figure}[H]
	\centering
	\includegraphics[width=15cm]{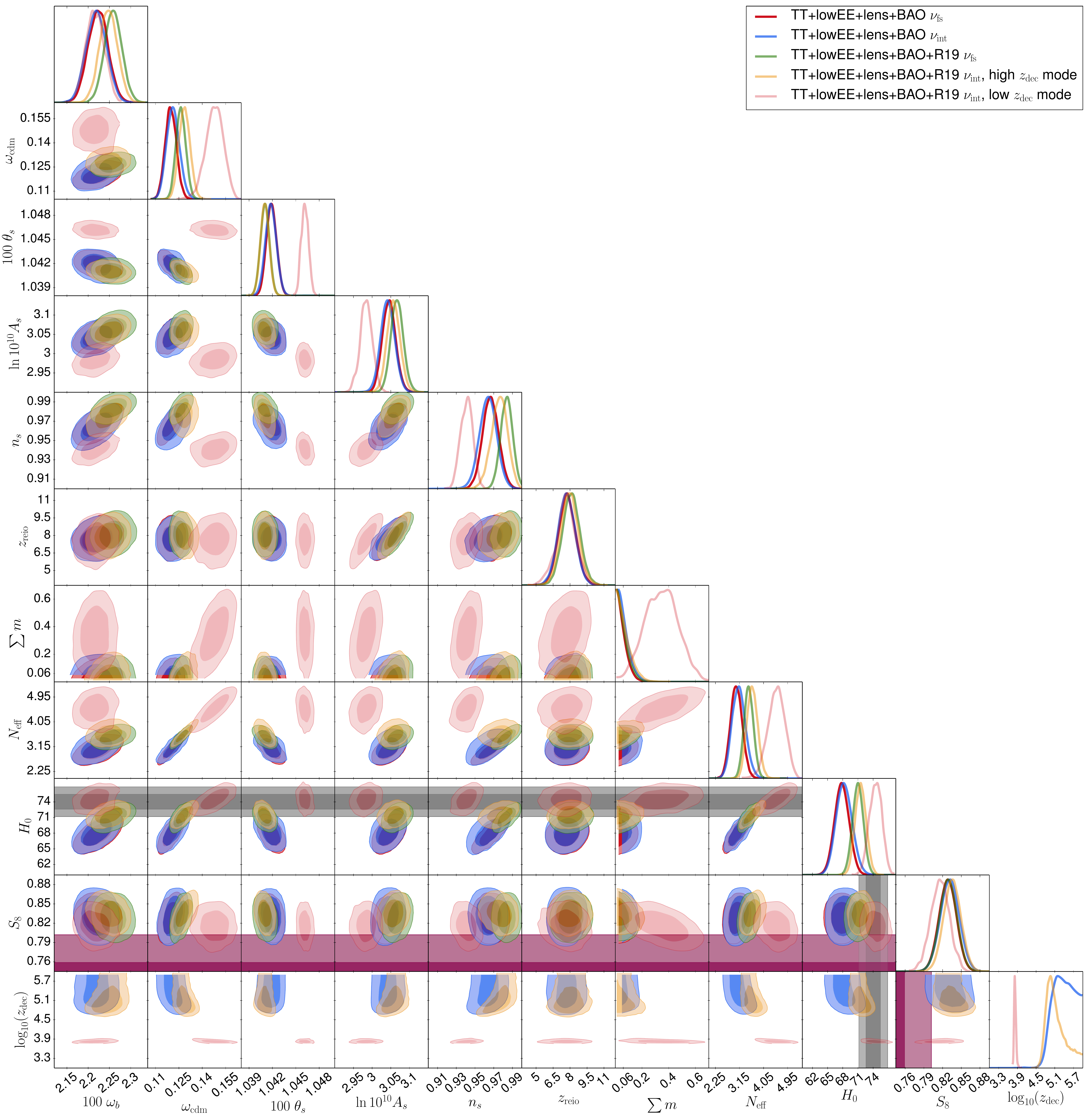}
	\caption{\textbf{Case 1: all species interacting, no high-$\ell$ polarization.} Full parameter space corresponding to Figure~\ref{fig:results:Pbnhr_R1z6m}.}
	\label{fig:app:Pbnhr_R1z6m}
\end{figure}

\begin{figure}[H]
	\centering
	\includegraphics[width=15cm]{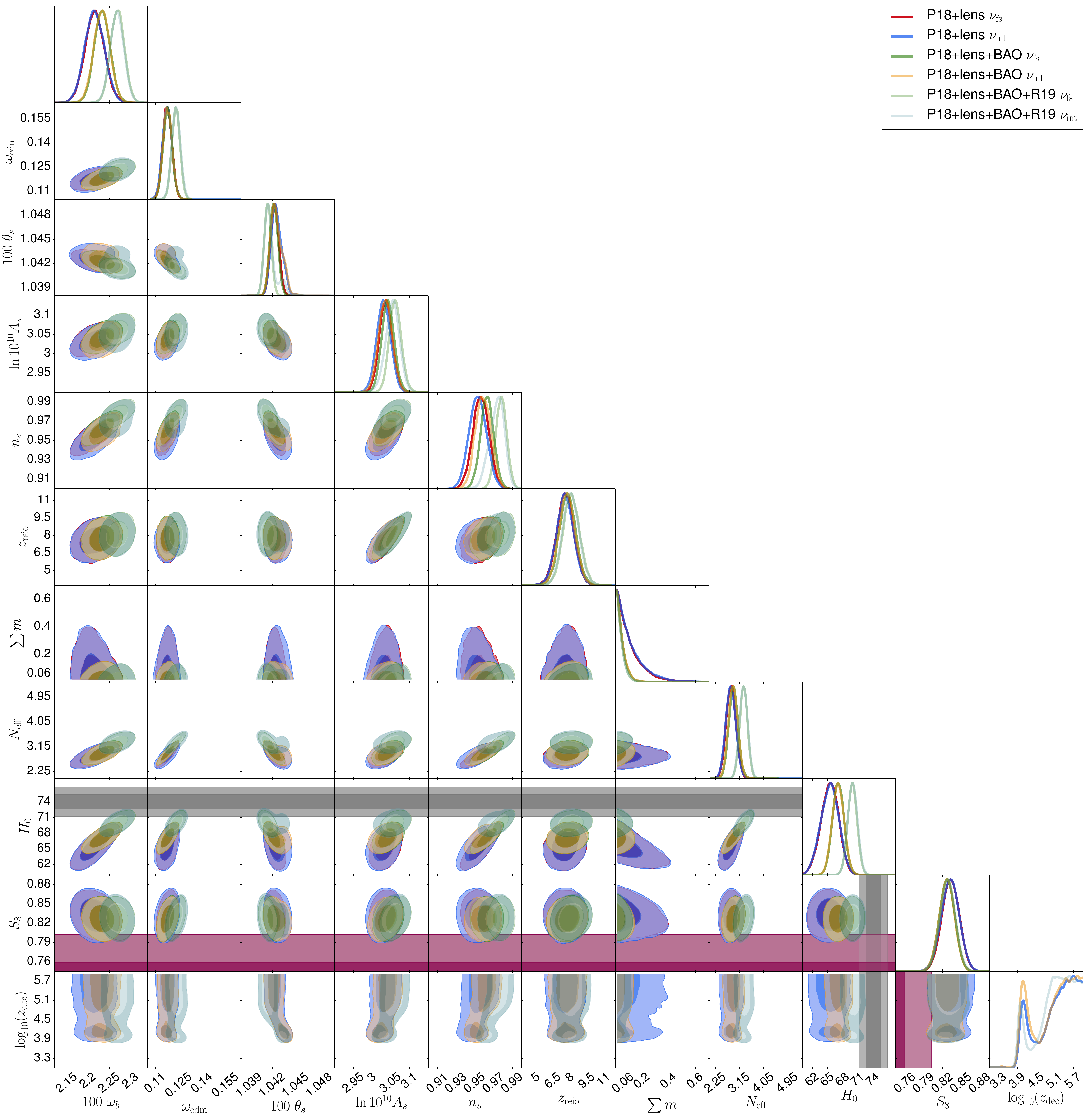}
	\caption{\textbf{Case 2: two free-streaming neutrinos plus self-interacting relativistic species.} Full parameter space corresponding to Figure~\ref{fig:results:2FSz6m}.}
	\label{fig:app:2FSz6m}
\end{figure}

\begin{figure}[H]
	\centering
	\includegraphics[width=15cm]{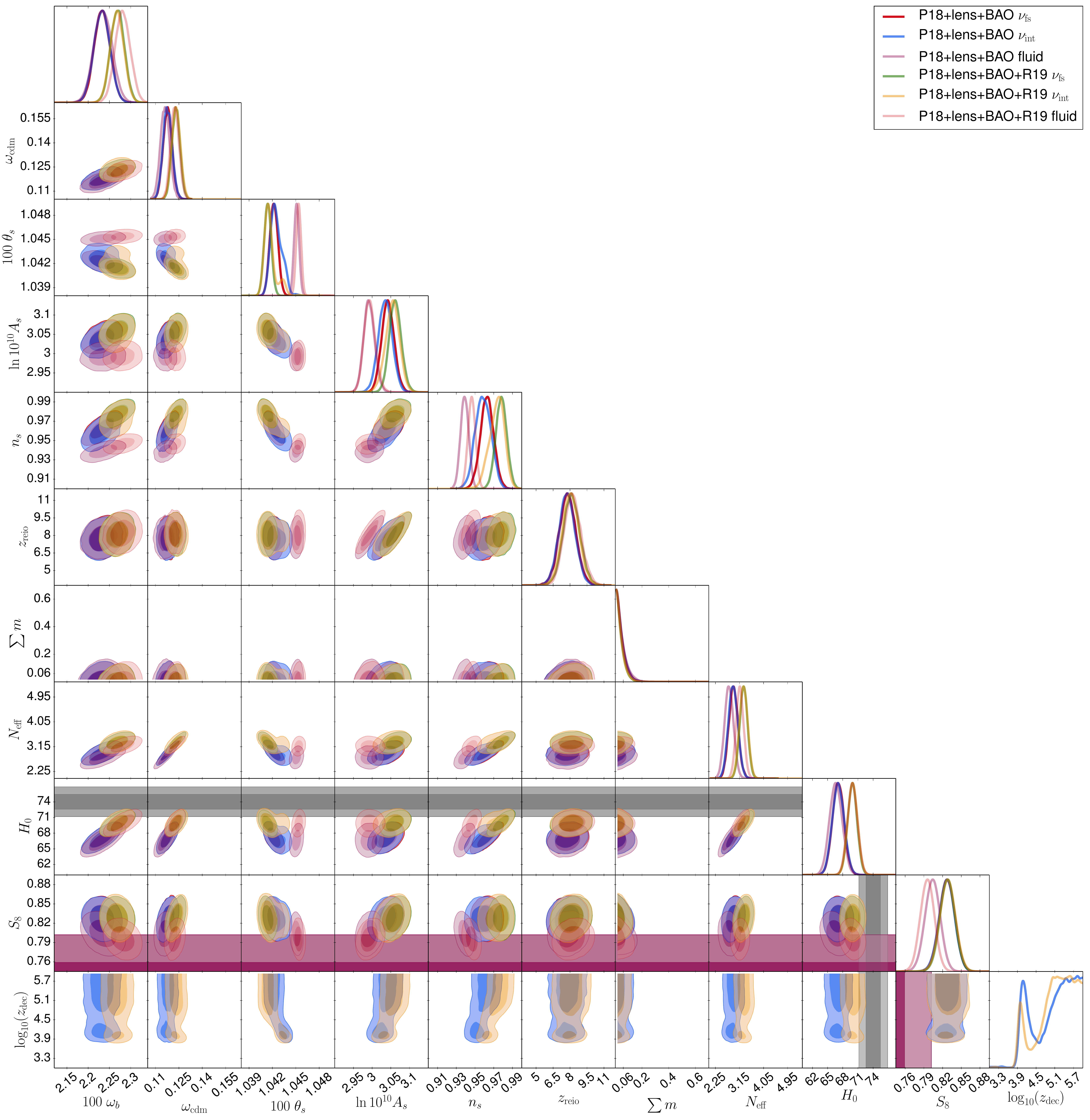}
	\caption{\textbf{Case 2: two free-streaming neutrinos plus self-interacting relativistic species, including fluid-like.} Full parameter space corresponding to Figure~\ref{fig:results:2FSfl}.}
	\label{fig:app:2FSfl}
\end{figure}

\begin{figure}[H]
	\centering
	\includegraphics[width=15cm]{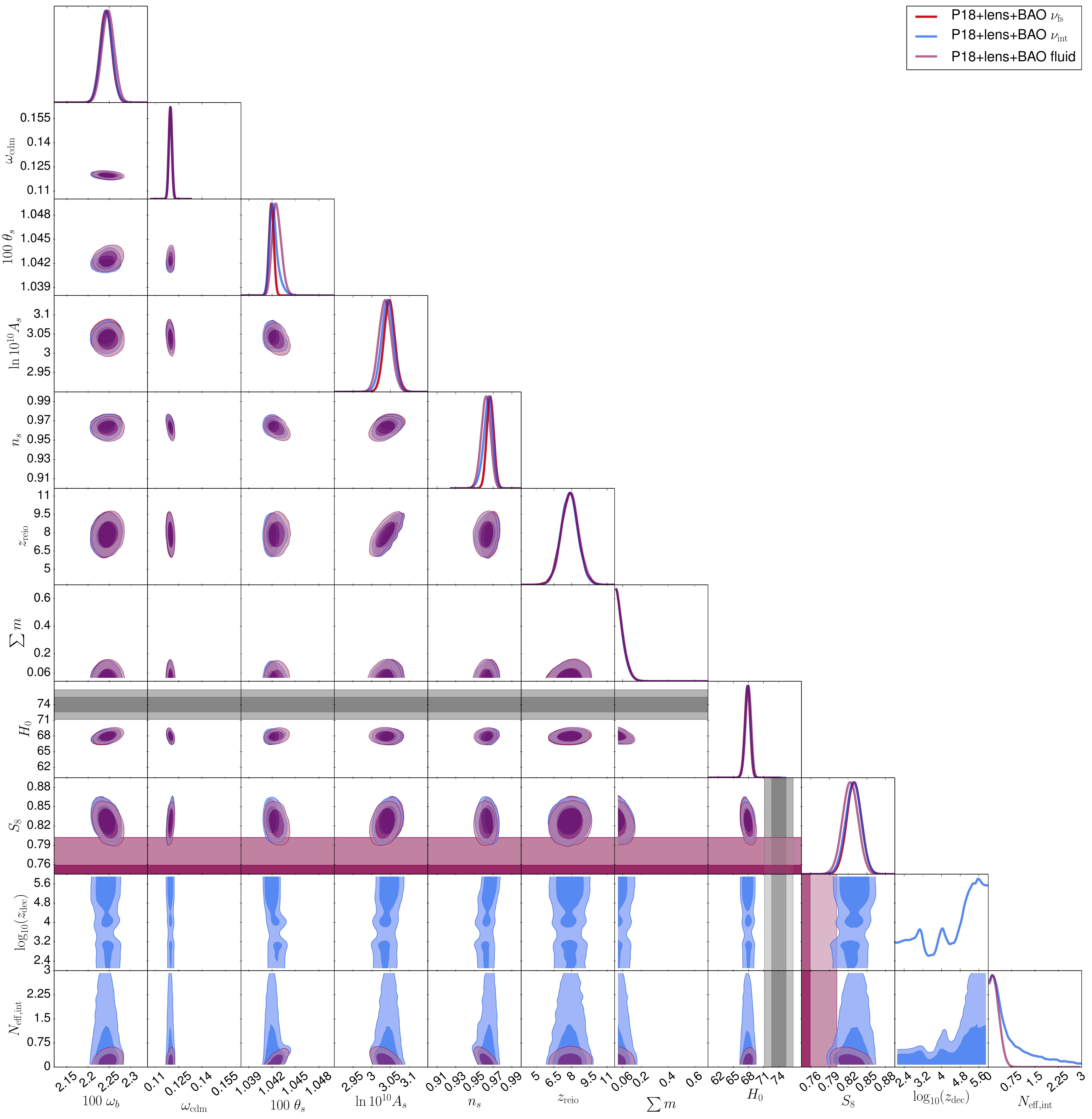}
	\caption{\textbf{Case 3: varying fraction of self-interacting relativistic species, including fluid-like, with fixed $\Neff=3.046$.} Full parameter space corresponding to Figure~\ref{fig:results:PSIz6m}.}
	\label{fig:app:PSIz6m}
\end{figure}

\begin{figure}[H]
	\centering
	\includegraphics[width=15cm]{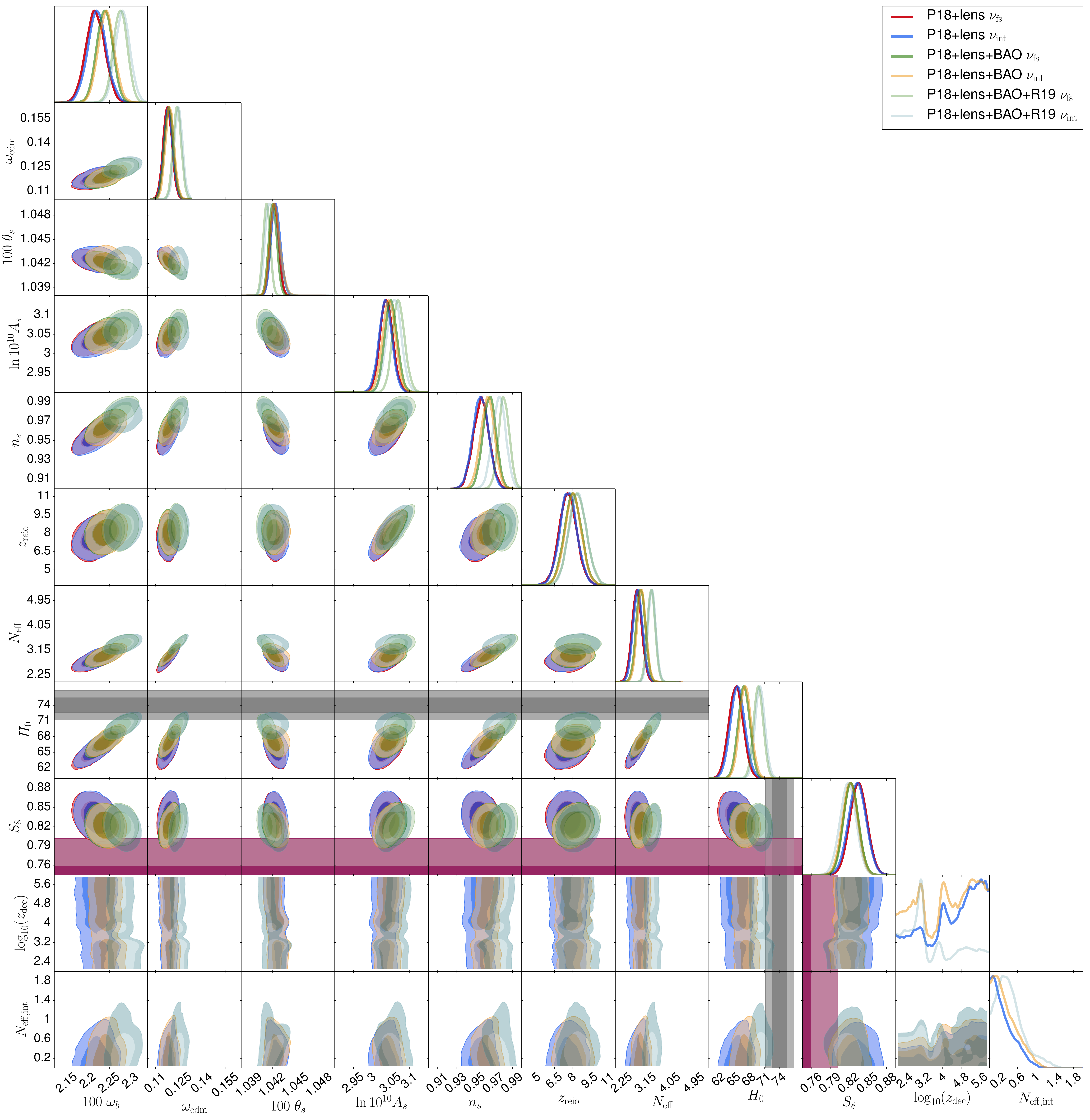}
	\caption{\textbf{Case 4: varying fraction of self-interacting relativistic species.} Full parameter space corresponding to Figure~\ref{fig:results:urz6i}.}
	\label{fig:app:urz6i}
\end{figure}

\begin{figure}[H]
	\centering
	\includegraphics[width=15cm]{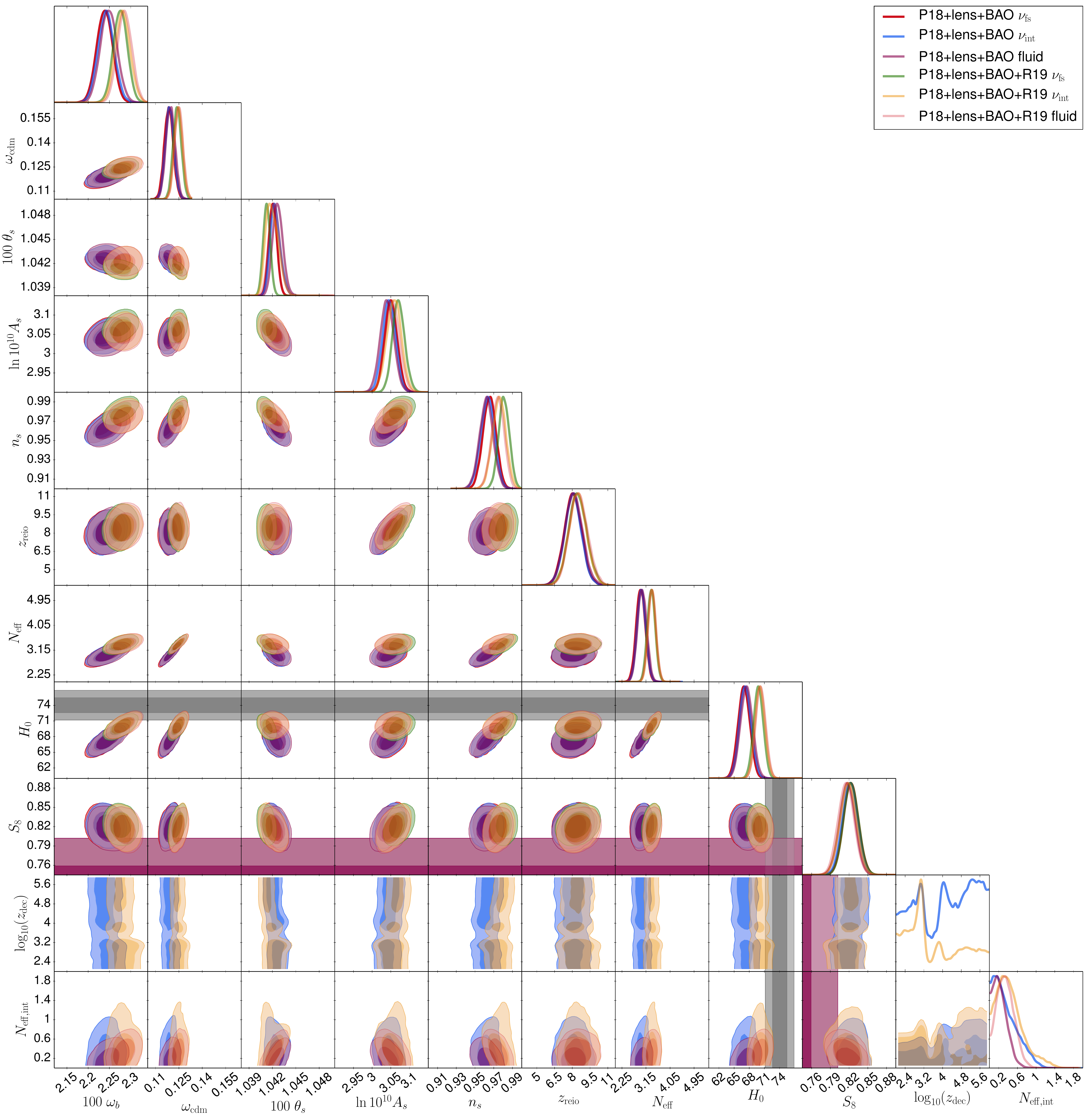}
	\caption{\textbf{Case 4: varying fraction of self-interacting relativistic species, including fluid-like.} Full parameter space corresponding to Figure~\ref{fig:results:urfl}.}
	\label{fig:app:urfl}
\end{figure}

\section{CLASS precision parameters and MultiNest sampling settings}
\label{app:precision_settings}
Although we added a new species with a different name, we will provide the settings used if someone were to add an interaction to the ncdm species as they are the same. In order to accurately compute the effect of self-interactions on a massive relativistic species using the ncdm framwork in CLASS, it is crucial to turn off the fluid approximation. This can be done by ensuring it never kicks in (alternately could use the \texttt{ncdm\_fluid\_approximation} flag)\\
\texttt{ncdm\_fluid\_trigger\_tau\_over\_tau\_k = 1e8}\\
Aside from that we set most of the other precision settings to an arbitrarily small value, although this could be tuned for greater efficiency\\
\texttt{tol\_M\_ncdm = 1e-10}\\
\texttt{tol\_ncdm = 1e-10}\\
\texttt{tol\_ncdm\_synchronous = 1e-10}\\
\texttt{tol\_ncdm\_newtonian = 1e-10}\\
\texttt{tol\_ncdm\_bg = 1e-10}\\
\texttt{tol\_ncdm\_initial\_w = 1e-10}\\
Please see the file include/precisions.h (within newer CLASS versions) for details on what these parameters do and their default values.\\

In order to find all modes we needed to increase the precision settings of the MultiNest sampler beyond commonly used values. We attribute this need to the many orders of magnitude covered by the parameters space in combination with the narrowness of some of the modes compared to the wide allowed parameter space as the interactions asymptote to a free-streaming or fluid-like case. The settings we used were\\
\texttt{evidence\_tolerance = 0.005}\\
\texttt{n\_live\_points = 4000}
Please see the MultiNest documentation for details on what these parameters do.